\documentclass[useAMS,usenatbib]{mn2e}
\usepackage{amssymb,amstext,amsmath,commath}
\usepackage{times,float,graphicx,paralist}
\usepackage{epsfig,epstopdf,url,rotating}
\DeclareGraphicsExtensions{.pdf,.ps,.eps,.png,.jpg,.mps} 
\thinspace
\begin{document}
\title[Environmental Effects on the Molecular Gas of Spirals] {The JCMT Nearby Galaxies Legacy Survey -- X. Environmental Effects on the Molecular Gas and Star Formation Properties of Spiral Galaxies}
\author[A. Mok et al.] {Angus Mok $^1$, C. D. Wilson$^1$, J. Golding$^1$, B. E. Warren$^{2}$, F. P. Israel$^3$, S. Serjeant$^4$,
\newauthor J. H. Knapen$^{5,6}$, J. R. S\'anchez-Gallego$^{5,6,7}$, P. Barmby$^{8}$, G. J. Bendo$^{9}$,
\newauthor E. Rosolowsky$^{10}$, and P. van der Werf$^{11}$\\
\\
$^1$ Department of Physics \& Astronomy, McMaster University, Hamilton, Ontario L8S 4M1, Canada \\
$^2$ International Centre for Radio Astronomy Research, M468, University of Western Australia, 35 Stirling Hwy, Crawley, WA 6009, Australia \\
$^3$ Sterrewacht Leiden, Leiden University, PO Box 9513, 2300 RA Leiden, The Netherlands \\
$^4$ Department of Physics \& Astronomy, The Open University, Milton Keynes MK7 6AA, United Kingdom \\
$^{5}$ Instituto de Astrof\'isica de Canarias, E-38205 La Laguna, Tenerife, Spain \\
$^{6}$ Departamento de Astrof\'\i sica, Universidad de La Laguna, E-38200 La Laguna, Tenerife, Spain \\
$^{7}$ Department of Physics and Astronomy, University of Kentucky, 40506-0055, Lexington, Kentucky, United States \\
$^{8}$ Department of Physics \& Astronomy, University of Western Ontario, London, Ontario N6A 3K7, Canada \\
$^{9}$ UK ALMA Regional Centre Node, Jodrell Bank Centre for Astrophysics, School of Physics and Astronomy, University of Manchester, Oxford Road, \\
Manchester M13 9PL, United Kingdom \\
$^{10}$ Department of Physics, University of Alberta, Edmonton, Ontario T6G 2R3, Canada \\
$^{11}$ Kapteyn Astronomical Institute, Postbus 800, NL-9700 AV Groningen, The Netherlands \\
}
%
%
\def\etal{{ et al.\thinspace}}
\def\gtrsim{\mathrel{\raise0.35ex\hbox{$\scriptstyle >$}\kern-0.6em\lower0.40ex\hbox{{$\scriptstyle \sim$}}}}
\def\lesssim{\mathrel{\raise0.35ex\hbox{$\scriptstyle <$}\kern-0.6em\lower0.40ex\hbox{{$\scriptstyle \sim$}}}}
\def\Msun{\hbox{$\rm\thinspace M_{\odot}$}}
\def\kmsmpc{{\,\rm km\,s^{-1}Mpc^{-1}}}
\def\kms{km~s$^{-1}$}
\def\hi{H\,{\sc i\,\,}}
\def\hino{H\,{\sc i}}
\def\h2{H$_{2}$}
\def\ks{Kolmogorov-Smirnov }
\def\aap{A\&A}
\def\aaps{A\&AS}
\def\araa{ARA\&A}
\def\apjl{ApJ}
\def\apj{ApJ}
\def\apjs{ApJS}
\def\aj{AJ}
\def\mnras{MNRAS}
\def\pasj{PASJ}
\def\pasp{PASP}
\def\nat{Nature}
%
%
\maketitle
\begin{abstract}
We present a study of the molecular gas properties in a sample of 98 \hi - flux selected spiral galaxies within $\sim25$ Mpc, using the CO $J=3-2$ line observed with the James Clerk Maxwell Telescope. We use the technique of survival analysis to incorporate galaxies with CO upper limits into our results. Comparing the group and Virgo samples, we find a larger mean \h2 mass in the Virgo galaxies, despite their lower mean \hi mass. This leads to a significantly higher \h2 to \hi ratio for Virgo galaxies. Combining our data with complementary H$\alpha$ star formation rate measurements, Virgo galaxies have longer molecular gas depletion times compared to group galaxies, due to their higher \h2 masses and lower star formation rates. We suggest that the longer depletion times may be a result of heating processes in the cluster environment or differences in the turbulent pressure. From the full sample, we find that the molecular gas depletion time has a positive correlation with the stellar mass, indicative of differences in the star formation process between low and high mass galaxies, and a negative correlation between the molecular gas depletion time and the specific star formation rate.
\end{abstract}
%
%
\section{Introduction}
\par
Star formation in the Universe takes place inside galaxies, where they convert their available gas reservoirs into stars. In particular, stars are born inside cold, dense molecular clouds and so a complete analysis of the star formation process requires the study of a galaxy's molecular gas content. Previous work has shown that star formation is more closely linked to the molecular gas, as compared to either the atomic hydrogen (\hino) mass or the total gas (\hi + \h2) mass \citep{Bigiel08, Leroy08}. These studies of star formation and molecular gas data for nearby galaxies have also revealed short \h2 depletion times compared to the age of the galaxy \citep{Leroy08, Saintonge11b, Bigiel11}, which suggests an ongoing need to replenish their molecular gas reservoir.
\par
Spiral galaxies are the primary targets for this type of analysis, as star formation does not take place equally in all galaxies. For example, the morphological classification between early- and late-type galaxies tends to correspond to their stellar population, such as the well-known 'red sequence' and 'blue cloud' found in colour-magnitude diagrams \citep[e.g.][]{Strateva01, Bell03}. The key difference between these two galaxy populations is their typical star formation state: whether they are undergoing active star formation or if they are mostly quiescent objects. Even for spiral galaxies along the Hubble sequence, the star formation rate has been shown to differ greatly \citep{Kennicutt98B}.
\par
The local environment of these galaxies can also influence galaxy evolution, from isolated galaxies, to groups of tens of galaxies, to clusters of thousands of galaxies. Denser environments in the Universe are dominated by early-type, red sequence galaxies \citep{Baldry06, Blanton09}. For spiral galaxies inside clusters, observations have shown a deficiency of atomic hydrogen \citep{Chamaraux80, Haynes86, Solanes01, Gavazzi05} and a reduction in the scale length of H$\alpha$ emission \citep{Koopmann06}, which is linked to young stars and star formation. On a smaller scale, some studies of galaxies with nearby companions have shown that star formation is generally unaffected, other than for the rare cases where mergers and interactions leads to a statistically significant but moderate enhancement in the star formation rate \citep{Knapen09, Knapen15}. On the other hand, a larger sample of interacting pairs from the SDSS shows signs of a star formation enhancement at the smallest separations \citep{Ellison13}.
\par
Many possible physical processes have been invoked to explain the effects of environment. For example, galaxies may be affected by harassment from other cluster members \citep{Moore96} or be starved from their gas supply \citep{Larson80}. In more extreme environments, ram-pressure stripping of a galaxy's gas content can take place \citep{Gunn72}, and interactions and mergers can directly increase the gas content of galaxies or change the distribution of their gas. However, it remains unclear whether these processes can directly affect the molecular gas content, with its high surface density and its location deep inside the galaxy's potential well. 
\par
A good place to study the effects of environment on galaxy evolution is the Virgo Cluster, due to its close proximity and the large numbers of infalling spiral galaxies. One of the initial studies of molecular gas in Virgo spirals using the CO tracer was performed with the 7 metre Bell Laboratories antenna \citep{Stark86}, finding correlations between the radio continuum and far-infrared data. Further studies helped determine the conversion ratio between the CO $J=1-0$ luminosity and the molecular gas mass \citep{Knapp87}. Subsequent CO observations with the Five College Radio Astronomy Observatory (FCRAO) have shown that the molecular content of Virgo spirals may be quite similar to that of field galaxies \citep{Kenney89}, but other groups have found hints of a reduction in the size of the molecular gas disk in Virgo spirals \citep{Fumagalli08}. Recent results from the HeViCS survey indicate a reduction in the amount of molecular gas per unit stellar mass in \hi deficient galaxies \citep{Corbelli12}. Finally, \citet{Vollmer12} have looked at resolved measurements of 12 Virgo spiral galaxies and found a relationship between molecular gas mass and star formation, but no evidence for environmental differences in the star formation efficiency. A complete picture of the effects of environment on star formation and the molecular gas content of galaxies remains elusive.
\par 
This paper uses a large sample of 98 \hi - flux selected spiral galaxies, with a majority coming from the original Nearby Galaxies Legacy Survey \citep{Wilson12}. One of the main objectives of the NGLS is to study the effect of environment on a galaxy's molecular gas content by using CO $J=3-2$ observations with the James Clerk Maxwell Telescope (JCMT) \citep{Wilson09, Wilson12}. The CO $J=3-2$ line is a tracer for warmer and denser gas than the CO $J=1-0$ line and appears to be well correlated with the far-infrared luminosity, a proxy for the star formation rate \citep{Iono09, Wilson12}. We supplement the NGLS data with follow up JCMT surveys in the Virgo cluster and some galaxies from the Herschel Reference Survey (HRS) sample. Our analysis also includes data from other sources, including stellar masses and star formation rates.
\par
In \S~2, we present our observations and the general properties of our sample. In \S~3, we discuss the use of survival analysis for datasets containing censored data (such as the \h2 gas mass) and present our analysis of galaxy properties. We also include a detailed comparison of the group and Virgo populations, as well as the CO detected and non-detected samples. In \S~4, we present some specific tests of our results and the correlations found between various properties of the galaxies in our sample.
%
%
\section{Sample Selection, Observations, and Data Processing}
\subsection{Sample Selection}\label{subsec-pro_sample}
\par
The objective of our analysis is to create a large sample of nearby, gas-rich spiral galaxies. First, we select only spiral galaxies with \hi flux $>6.3$ Jy km s$^{-1}$, which corresponds to a log \hi mass of 8.61 (in solar units) at the sample's median distance of 16.7 Mpc. The \hi flux is identified using the HyperLeda database \citep{Paturel03, Makarov14}, which complies different survey results and produces a weighted average of the results. The database can be found online\footnote{http://leda.univ-lyon1.fr/}. We also use the HyperLeda database to select only spiral galaxies using the morphological type code, removing from the sample galaxies that are ellipticals or lenticulars. We identify Virgo galaxies in our sample using the catalog from \citet{Binggeli85}.
\par
Next, we impose a size limit on our galaxies. The non-Virgo sample is limited to galaxies with $D_{25}<5^\prime$, which at our distance limit of $\sim25$ Mpc corresponds to $D_{25}<36$ kpc. We therefore include only the 39 Virgo spiral galaxies with $D_{25}<7.4^\prime$ in order to match the physical size limits of of the Virgo and non-Virgo samples. Note that for all galaxies in the Virgo sample, we have assumed a standard distance of 16.7 Mpc \citep{Mei07}.
\par
To subdivide our sample further, we use the \citet{Garcia93b} catalog. They used the LEDA data base to identify a robust set of 485 groups out of a sample of 6392 local galaxies by their 3D projection in space. Comparing our galaxy sample to the groups from their catalog yields a total of 40 galaxies. In addition, we place two galaxies, NGC0450 and NGC2146A, which are known from the \citet{Karachentsev72} catalog to be in close pairs, into this category as well. This results in a total of 42 spiral galaxies meeting our \hi flux and size criterion that are members of groups. The 17 remaining galaxies constitute our field (or non-group) galaxy sample.
\par
A summary of the sample sources, subdivided into the field, group, and Virgo populations, are presented in Table~\ref{tab-sample}. There are three main sources of the CO $J=3-2$ data. The first is the Nearby Galaxies Legacy Survey \citep{Wilson12}. The second is a follow on study to complete observations in the Virgo cluster (project code M09AC05), using the same criterion as the parent NGLS survey. As a result, we do not expect significant differences between these two sources. Finally, to increase the number of galaxies in our sample, we also include a subset of Herschel Reference Survey (HRS) galaxies that fulfill the criteria listed above and are not already a member of the NGLS \citep{Boselli10}. One potential different between the three sources of data is that the Herschel Reference Survey also has a K-band (or stellar mass) selection, but we do not expect it to greatly influence the results in this paper. Also, a subset of group galaxies is identified by the HRS as being in the Virgo outskirts, which we have kept in the group category given their local environment is likely more similar to the group than Virgo category. A full discussion of the three sources of our CO $J=3-2$ can be found in Appendix~\ref{app-comp}.
\par
Finally, due to our sample criteria, we are biased towards gas-rich, nearby, moderately-sized galaxies and to the detriment of \hi deficient, far-away, larger, non-spiral galaxies. This is done to ensure a consistent statistical sample and to obtain satisfactory detection rates, as discussed in the observation section below. 
\begin{table}
	\caption{Sample Sources}\label{tab-sample}
	\begin{tabular}{lccccc}
	\hline
	Category & Total & Field & Group & Virgo \\
	\hline
	NGLS & $53$ & $12$ & $21$ & $20$ \\
	Virgo Follow-Up (M09AC05) & $17$ & $0$ & $0$ & $17$ \\
	HRS (M14AC03) & $28$ & $5$ & $21$ & $2$\\
	\hline
	Total & $98$ & $17$ & $42$ & $39$ \\
	\end{tabular}
\end{table}

\subsection{Observations}\label{subsec-obs}
\par
The observations and data processing for the $155$ galaxies in the NGLS are described in detail in \citet{Wilson12}, and so only a basic summary is given here. We observed the CO $J=3-2$ line with the JCMT's HARP instrument \citep{Buckle09}, which has $\eta_{MB}=0.6$ and an angular resolution of 14.5$^{\prime\prime}$. All galaxies were mapped out to at least $D_{25}/2$ with a 1 sigma sensitivity of better than 19 mK ($\rm T_A^*$; 32 mK $\rm T_{MB}$) at a spectral resolution of 20 km s$^{-1}$. The CO luminosities are measured on the $\rm T_{MB}$ scale. The internal calibration uncertainty is 10\%. The reduced images, noise maps, and spectral cubes are available via the survey website\footnote{http://www.physics.mcmaster.ca/$\sim$wilson/www\_xfer/NGLS/} and also via the Canadian Astronomical Data Centre\footnote{DOI 10.1111/j.1365-2966.2012.21453.x}.
\par
For galaxies with CO detections, we used the zeroth moment maps made with a noise cutoff of 2$\sigma$ to measure the CO $J=3-2$ luminosity (see \citet{Wilson12} for further details). We use an aperture chosen by eye to capture all of the real emission from the galaxy. For galaxies without detections, 2$\sigma$ upper limits were calculated from the noise maps assuming a line width of 100 km s$^{-1}$ and using an aperture with a diameter of 1$^\prime$. The full sample includes 57 spiral galaxies from the NGLS, 14 spiral galaxies observed with the JCMT in 2009 February-May (JCMT program M09AC05), and 27 spiral galaxies from the Herschel Reference Survey (JCMT program M14AC03). The galaxies were processed using the same methods adopted for the NGLS and the calibration uncertainty for these data is also 10\%. 
\par
We convert the CO $J=3-2$ luminosities to molecular hydrogen mass adopting a CO-to-\h2 conversion factor of $X_{CO} = 2\times 10^{20}$ cm$^{-2}$ (K km s$^{-1}$)$^{-1}$ \citep{Strong88} and a CO $J=3-2/J=1-0$ line ratio of 0.18 \citep{Wilson12}. This measurement is similar to the CO $J=3-2/J=1-0$ line ratio of Virgo spiral galaxies obtained by other groups \citep{Hafok03}. With these assumptions, the molecular hydrogen mass is given by:
\begin{equation}\label{eqn-mh2}
 M_{H_2} = 17.8 (R_{31}/0.18)^{-1} L_{CO(3-2)}
\end{equation}
where $R_{31}$ is the CO $J=3-2/J=1-0$ line ratio, $M_{H_2}$ is in $M_\odot$, and $L_{CO(3-2)}$ is in units of K km s$^{-1}$ pc$^2$.
\par
We note that the assumption of a constant conversion factor may not be correct in all cases. For example, the conversion factor can be affected by the ambient radiation field and the metallicity \citep{Israel97}. Adopting a single value for $X_{CO}$ will produce an underestimate of the molecular gas mass in galaxies where the metallicity is more than about a factor of two below solar \citep{Wilson95, Arimoto96, Bolatto08, Leroy12}. In a recent review paper on the conversion factor, \citet{Bolatto13} suggested a possible prescription that is based on the gas surface density and metallicity. However, given the large scatter in their observational results and the fact that our galaxies are all relatively `normal' spiral galaxies, we have decided to maintain the constant conversion factor used in the previous papers in this series.
\par
We also use data sourced from other surveys. As stated in the section on our sample selections, the \hi fluxes and morphological types for all the galaxies are taken from the most recent values of the HyperLeda database. We also use the database for measurements of redshift distances and $D_{25}$ sizes. Stellar masses for individual galaxies are taken from the S$^4$G survey \citep{Sheth10} using the NASA/IPAC Infrared Science Archive, for all the galaxies where they are available. The S$^4$G survey have calibrated their stellar masses using the two IRAC bands, 3.6 and 4.5 $\mu$m, using the prescription from \citet{Eskew12} and assuming a Salpeter IMF. To convert to a Kroupa IMF, we multiply by a factor of $0.7$, which has been used in previous work \citep[e.g.][]{Elbaz07}. We have also used the distance measurements in the S$^4$G catalog to adjust the stellar masses to correspond to the distances used in this analysis.
\par
For the 8 galaxies where the stellar mass is unavailable, we substitute K-band luminosities from the 2MASS database using the extended source catalogue \citep{Skrutskie06}. We have assumed a total mass-to-light ratio of 0.533 in solar units for Sbc/Sc-type spiral galaxies from Table 7 in \citet{Portinari04} with the Kroupa IMF, as they comprise a majority of the galaxies without masses from the S$^4$G survey. For comparison, the corresponding value for Sa/Sab-type spiral galaxies is 0.698. The average stellar mass for these galaxies is lower than that of the sample as a whole, but since they only comprise a small percentage of the total sample, we do not expect this difference to have a significant effect on our results.
\par
In addition, H$\alpha$-derived star formation rates are taken from \citet{SanchezGallego12}, with typical uncertainties of 18 per cent. We note that the formulas from \citet{Kennicutt09} assume a stellar initial mass function (IMF) from \citet{Kroupa03}. For the additional spiral galaxies from the Virgo cluster observed in the M09AC05 program, we use H$\alpha$ fluxes from the GOLDmine database \citep{Gavazzi03}. For the galaxies from the HRS sample, we use the data from a new H$\alpha$ study of these galaxies \citep{Boselli15}. The H$\alpha$ data from all three sources are corrected for extinction and converted into star formation rates using the same procedure as \citet{SanchezGallego12}. We also investigate the effects of including a mid-IR star formation tracer using data from the S$^4$G survey and find there is moderate scatter between the two measurements. However, we have chosen to present the extinction-corrected H$\alpha$ data in order to maintain continuity with the previous papers in this series.
\par
Detailed properties for the individual galaxies are presented for the field galaxies in Table~\ref{tbl-list_isolated}, the group galaxies in Table~\ref{tbl-list_group}, and the Virgo sample in Table~\ref{tbl-list_virgo}, located in Appendix~\ref{app-prop}. Maps of the CO detected galaxies can be found in Appendix~\ref{app-map}. The field sample is shown in Figure~\ref{fig-det_isolated}, the group sample in Figures~\ref{fig-det_group1} and \ref{fig-det_group2}, and the Virgo sample in Figure~\ref{fig-det_virgo}. Virgo galaxies observed with two overlapping fields are presented in Figure~\ref{fig-det_virgo_raster}, while NGC 4303, observed using the raster method, is presented in Figure~\ref{fig-det_virgo_double}. Note that images of NGC4254 and NGC4579 can be found in \citet{Wilson12} and are not repeated here.
\subsection{Survival Analysis and the Kaplan-Meier Estimator}\label{subsec-res_kaplan}
\begin{figure*}
	\includegraphics[width=8.0cm]{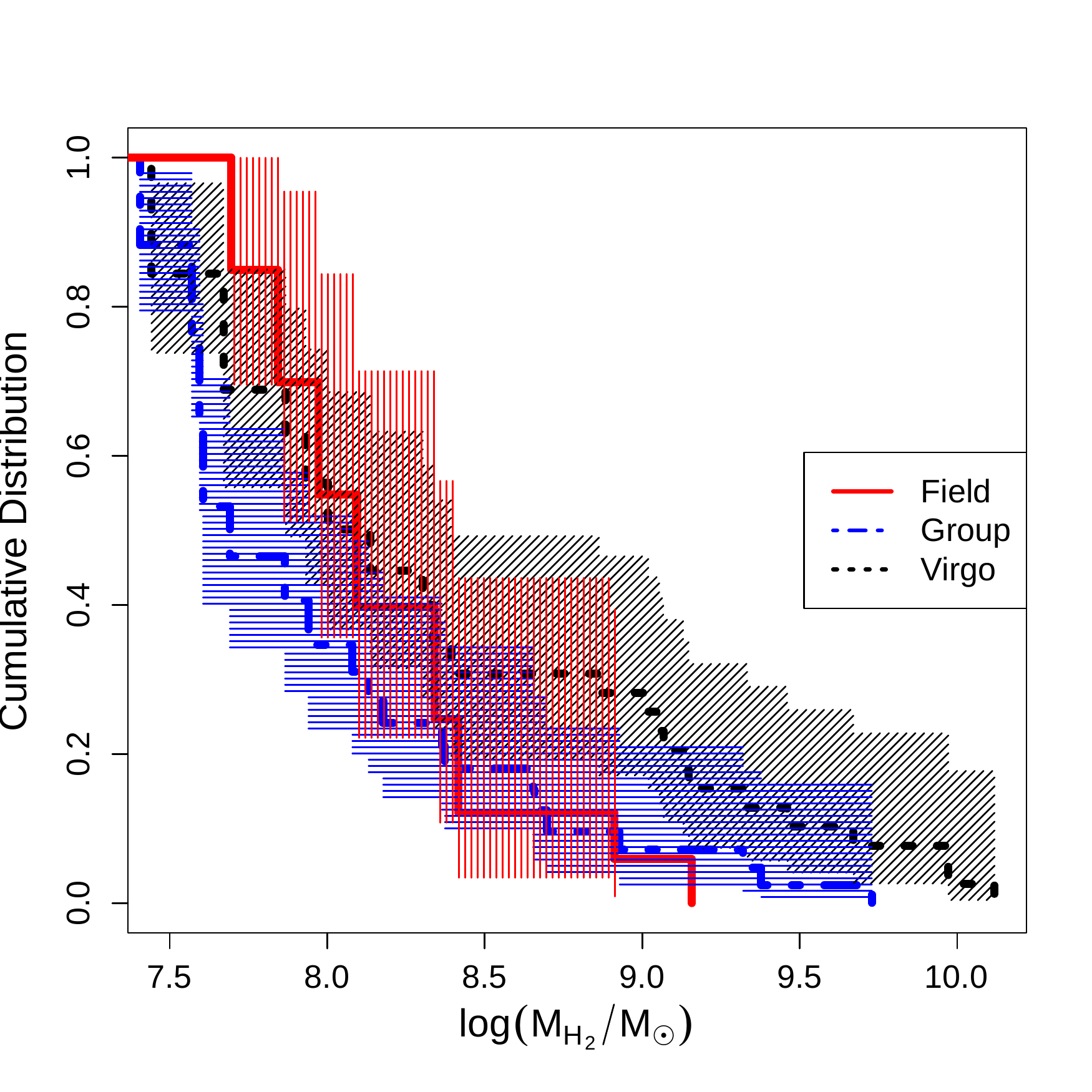}
	\includegraphics[width=8.0cm]{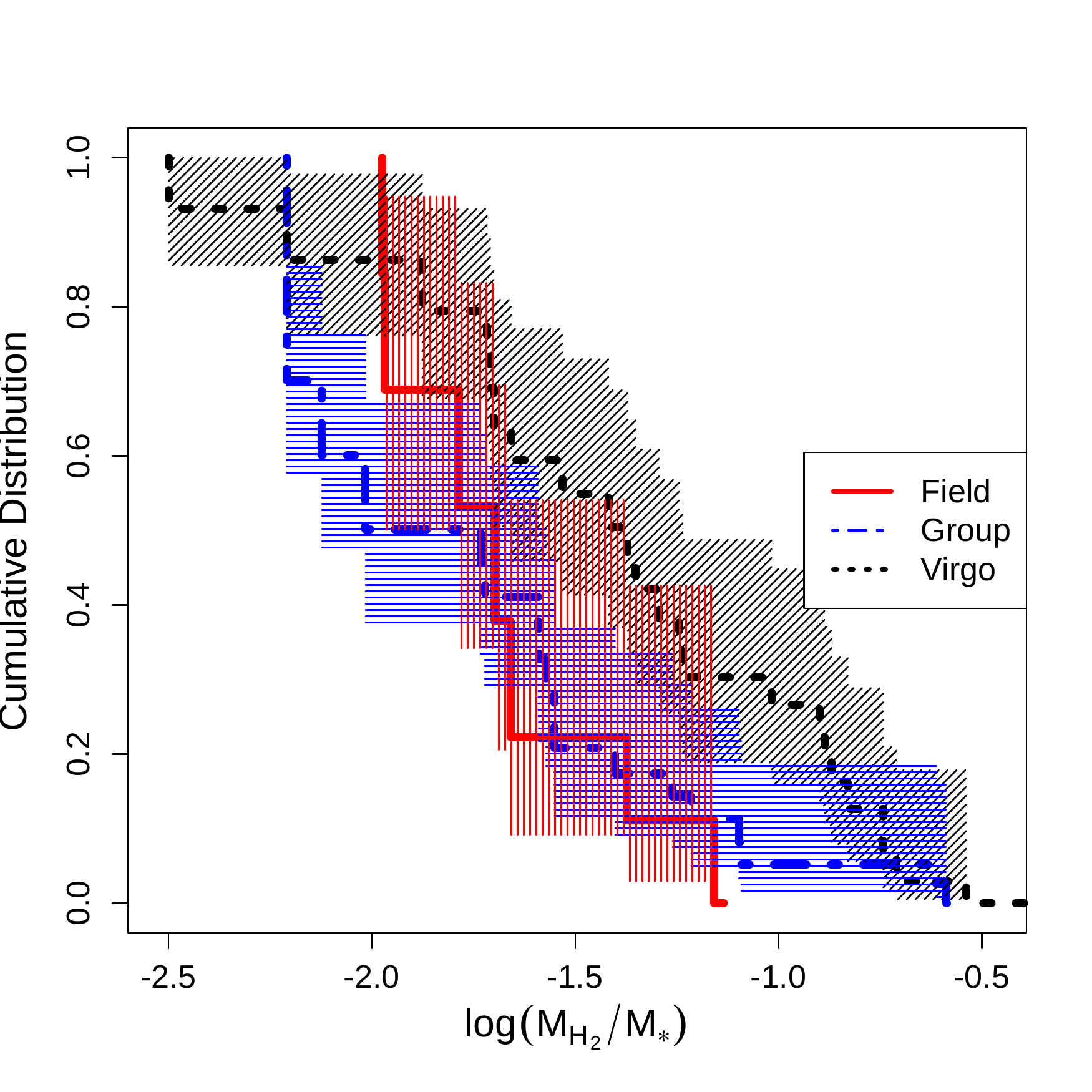}
	\caption{{\bf Left:} Survival functions for the molecular hydrogen gas mass in the field (red solid line), group (blue dot-dash line), and Virgo (black dashed line) sample galaxies using the Kaplan-Meier estimator. The `steps' in the distribution correspond to detections. The $95\%$ confidence intervals are overlaid for the three distributions.{\bf Right:} Survival functions for the molecular hydrogen gas mass divided by the stellar mass in the field, group and Virgo sample galaxies.}
	\label{fig-01C_H2}
\end{figure*}
\par
In statistics, survival analysis is often used with censored datasets, i.e. datasets that include upper or lower limits. The original purpose of survival analysis is in medicine, where data censoring is important in clinical trials because patients can either die or potentially leave the trial. This procedure has subsequently been applied to datasets in other scientific fields, such as astronomy, where `deaths' are replaced with measurements and patients who leave the trial are replaced with censored data, such as upper or lower limits \citep[e.g.][]{Young11}. This method of survival analysis allows us to determine statistical properties that are difficult to ascertain using classical methods when many of the measurements are censored.
\par
For our sample of galaxies, we first use the Kaplan-Meier estimator to fit survival functions to our data. It is a non-parametric, maximum likelihood statistical estimator \citep{Kaplan58}. In this study, we used the statistical package called {\sc Survival}, which is written in {\sc R} and can be found at the standard {\sc R} repository\footnote{http://cran.r-project.org/web/packages/survival/index.html}. Once we have fit a survival function to our dataset, we can then determine the modified versions of important statistical quantities. For example, we can find the `median' value of the dataset by finding the point where the survival function is equal to 0.5 and the `restricted mean' of the dataset by integrating the survival function to the last detected point.
\par
To differentiate between survival functions and determine if they are significantly different from one another, we have used the log-rank test. The log-rank test was first introduced in \cite{Mantel66} and is often used to compare the effectiveness of new treatments in clinical trials. In our paper, we have applied survival analysis to the \h2 mass and its other derived quantities, such as the molecular gas depletion time (the \h2 gas mass divided by the star formation rate) and the stellar mass normalized \h2 mass. The use of survival analysis and the log-rank test for these cases has the primary advantage of incorporating all the data collected, instead of discarding galaxies without CO detections. For non-censored datasets, we will use the standard \ks test for distinguishing between two distributions. An example of the application of the Kaplan-Meier estimator to the \h2 masses and the stellar mass normalized \h2 masses is presented in Figure~\ref{fig-01C_H2}, where the cumulative distribution functions are plotted. 
\par
The application of survival analysis to the total gas mass and its related quantities, such as the total gas depletion time (total gas mass divided by the star formation rate), presents one important difference compared to the \h2 gas mass alone. Since the galaxies without CO detections do have \hi gas masses, it is not proper to treat galaxies without CO detections as pure upper limits for their total gas mass. Rather, they should be considered `interval censored', where the true value is in between two values. In this case, the total gas mass for these galaxies would lie between their measured \hi gas masses and the sum of their \hi gas mass and the \h2 upper limit. For these datasets, we use the statistical package called {\sc Interval}, which can be found at the standard {\sc R} repository\footnote{http://cran.r-project.org/web/packages/interval/index.html}. The package also includes an implementation of the log-rank test routine {\sc ictest} for interval censored data. In order to be consistent, we also use this particular {\sc Interval} routine for our left and right censored data as well. For those cases, we substitute in appropriate interval limits, such as -99 or 99 for the log upper and lower limits.
%
%
\section{Results}
\subsection{Overview of Galaxy Properties in the Three Environments}\label{subsec-res_overview}
\begin{table*}
	\begin{minipage}{160mm}
	\caption{Galaxy Properties as a Function of Environment\label{tbl-env}}
	\begin{tabular}{lccccccccccc}
	\hline
	Type & $\overline{\log M_{H{\sc I}}}$\footnotemark[1] & $\overline{D_{25}}$\footnotemark[1] & $\overline{\rm Distance}$\footnotemark[1] & CO detection rate\footnotemark[2] & $\overline{\log M_{H_2}}$\footnotemark[3] & $\overline{\log M_{H_2 + H{\sc I}}}$\footnotemark[3] & $\overline{\log M_{*}}$\footnotemark[1] \\
(\# of galaxies) & ($M_\odot$) & (kpc) & (Mpc) & (\%) & ($M_\odot$) & ($M_\odot$) & ($M_\odot$) \\
	\hline
Field (17) & $9.23\pm0.07$ & $16.1\pm1.8$ & $20.1\pm0.8$ & $35\pm9$ & $8.17\pm0.07$ & $9.30\pm0.06$ & $9.58\pm0.14$ \\
Group (42) & $9.17\pm0.04$ & $13.1\pm0.8$ & $18.6\pm0.7$ & $40\pm6$ & $7.98\pm0.08$ & $9.24\pm0.04$ & $9.65\pm0.07$ \\
Virgo (39) & $8.99\pm0.05$ & $18.6\pm2.0$ & $16.7$ & $54\pm9$ & $8.34\pm0.13$ & $9.17\pm0.06$ & $9.65\pm0.11$ \\
	\hline
CO Detected (44) & $9.15\pm0.05$ & $19.9\pm1.7$ & $17.5\pm0.5$ & $100$ & $8.61\pm0.11$ & $9.35\pm0.05$ & $9.97\pm0.06$ \\
CO Non-Detected (54) & $9.07\pm0.04$ & $12.5\pm0.8$ & $18.6\pm0.5$ & $0$ & ($8.28\pm0.06$)\footnotemark[4] & ($9.16\pm0.03$)\footnotemark[4] & $9.36\pm0.07$ \\
	\hline
	\end{tabular}
	\begin{tabular}{l}
	\footnotemark[1] Standard error of the means\\
	\footnotemark[2] Binomial confidence intervals\\
	\footnotemark[3] Restricted mean and standard errors from the Kaplan-Meier estimator of the survival functions\\
	\footnotemark[4] Mean of the $2\sigma$ upper limits for the CO Non-Detected sample\\
	\end{tabular}
	\end{minipage}
\end{table*}
\par
The general properties of the galaxies in our sample, separated into field, group, and Virgo subsets, are presented in Table~\ref{tbl-env}. The first three columns are mean \hi mass, $D_{25}$ sizes, and distances taken from the HyperLeda database, using velocities corrected for Virgo infall and assuming a cosmology of $H_o = 70$ km s$^{-1}$ Mpc$^{-1}$, $\Omega_M = 0.27$, $\Omega_\Lambda = 0.73$. The next column is the CO detection rate for each individual sample, followed by the mean \h2 and \h2 + \hi masses, with the calculations outlined in the previous section. Finally, we present the mean stellar mass for each of the samples.
\par
The mean log \h2 mass, calculated from the Kaplan-Meier estimator of the survival function, is highest in the Virgo sample, followed by the field and group samples. The log-rank test on the group and Virgo samples shows a p-value of $0.0279$, which suggests that there is a difference in the \h2 mass distributions between the two populations, when we take into account the censored data. For the case of atomic hydrogen, the mean \hi mass is lowest for the Virgo galaxies, followed by the group and field samples. Relative \hi deficiency of Virgo galaxies has been reported in previous studies of Virgo galaxies \citep[e.g.][]{Chamaraux80, Haynes86, Solanes01, Gavazzi05}, so it is not surprising to see this difference in our sample, even when we have selected based on a \hi flux limit. The \ks test, performed for datasets that do not contain censored data, shows that the \hi masses of the Virgo and group galaxies are not drawn from the same distribution, with a p-value of $4\times10^{-4}$.
\par
The Virgo galaxies are all at the same assumed distance of 16.7 Mpc, while the group and field samples have larger mean distances. As a result, the \ks test reveals that the distribution in the distances of Virgo galaxies can be distinguished from both field and the group galaxies. Replacing the constant 16.7 Mpc with a Gaussian distributed sample with reasonable scatter does not change this result. Although the mean log total gas mass (\hi + \h2) is slightly higher for the field galaxies, followed by the group, and Virgo samples, the difference is not statistically significant. The stellar masses of the Virgo and group samples are also quite similar, while the field galaxies have a lower mean value. The \ks test shows that the stellar mass distributions of the group and Virgo samples cannot be distinguished ($p=0.4031$).
\subsection{Group/Virgo Comparison}\label{subsec-environment}
\begin{table*}
	\begin{minipage}{120mm}
	\caption{Selected Properties of the Group, Virgo, CO Detected, and CO Non-Detected Populations}\label{tab-sum_environment}
	\begin{tabular}{lcccc}
	\hline
	Mean Quantity & Group & Virgo & CO Detected & CO Non-Detected \\
	 & ($42$) & ($39$) & ($44$) & ($54$) \\
	\hline
	$\log M_{H{\sc I}}$ [$M_\odot$] & $9.17\pm0.04$ & $8.99\pm0.05$ & $9.15\pm0.05$ & $9.07\pm0.04$\\
	$\log M_{H_2}$ [$M_\odot$] \footnotemark[1] & $7.98\pm0.08$ & $8.34\pm0.13$ & $8.61\pm0.01$ & ($8.28\pm0.06$)\footnotemark[2]\\
	$\log M_{*}$ [$M_\odot$] & $9.65\pm0.07$ & $9.65\pm0.11$ & $9.97\pm0.06$ & $9.36\pm0.07$\\
	$\log M_{H_2 + H{\sc I}}$ [$M_\odot$]\footnotemark[1] & $9.24\pm0.04$ & $9.17\pm0.06$ & $9.35\pm0.05$ & ($9.16\pm0.03$)\footnotemark[2]\\
	\hline
	$M_{H_2}/M_{H{\sc I}}$\footnotemark[1] & $0.23\pm0.09$ & $0.75\pm0.20$ & $0.87\pm0.19$ & ($0.26\pm0.03$)\footnotemark[2]\\
	$M_{H_2}/M_{*}$\footnotemark[1] & $0.033\pm0.008$ & $0.068\pm0.010$ & $0.07\pm0.01$ & ($0.26\pm0.07$)\footnotemark[2]\\
	$M_{H{\sc I}}/M_{*}$ & $0.52\pm0.11$ & $0.50\pm0.12$ & $0.22\pm0.03$ & $0.82\pm0.12$\\
	$M_{H_2 + H{\sc I}}/M_{*}$\footnotemark[1] & $0.57\pm0.12$ & $0.63\pm0.14$ & $0.29\pm0.03$ & ($1.08\pm0.18$)\footnotemark[2]\\
	\hline
	$\log \rm SFR$ [$M_\odot$ yr$^{-1}$] & $-0.49\pm0.07$ & $-0.69\pm0.11$ & $-0.35\pm0.09$ & $-0.72\pm0.07$\\
	$\log \rm sSFR$ [yr$^{-1}$] & $-10.14\pm0.06$ & $-10.34\pm0.06$ & $-10.32\pm0.06$ & $-10.08\pm0.06$\\
	$\log M_{H_2}/{\rm SFR}$ [yr]\footnotemark[1] & $8.44\pm0.07$ & $8.97\pm0.06$ & $8.96\pm0.08$ & ($9.00\pm0.08$)\footnotemark[2]\\
	$\log M_{H{\sc I}}/{\rm SFR}$ [yr] & $9.71\pm0.07$ & $9.67\pm0.08$ & $9.50\pm0.07$ & $9.79\pm0.05$\\
	$\log M_{H_2 + H{\sc I}}/{\rm SFR}$ [yr]\footnotemark[1] & $9.73\pm0.05$ & $9.86\pm0.07$ & $9.70\pm0.06$ & ($9.88\pm0.05$)\footnotemark[2]\\
	\hline
	\end{tabular}
	\begin{tabular}{l}
	\footnotemark[1] Restricted mean and standard errors from the Kaplan-Meier estimator of the survival functions\\
	\footnotemark[2] Mean of the $2\sigma$ upper limits for the CO Non-Detected sample\\
	\end{tabular}
	\end{minipage}
\end{table*}
\begin{table}
	\caption{Significance Tests Between Different Samples}\label{tab-sign_environment}
	\begin{tabular}{lcccc}
	\hline
	 Mean Quantity & Group/Virgo & Group/Virgo & CO Det./ \\
	 & & (CO Det. Only) & Non-Det. \\
	\hline
	$\log M_{H{\sc I}}$ [$M_\odot$] & ${\bf 4\times10^{-4}}$ & $\underline{0.0181}$ & $0.7680$ \\
	$\log M_{H_2}$ [$M_\odot$] & $\underline{0.0279}$\footnotemark[1] & $0.1635$ & $-$ \\
	$\log M_{*}$ [$M_\odot$] & $0.6304$ & $0.3090$ & ${\bf 1\times10^{-9}}$ \\
	$\log M_{H_2 + H{\sc I}}$ [$M_\odot$] & $0.7720$ \footnotemark[1] & $0.3653$ & $-$ \\
	\hline
	$M_{H_2}/M_{H{\sc I}}$ & ${\bf 0.0095}$ \footnotemark[1] & $0.0786$ & $-$ \\
	$M_{H_2}/M_{*}$ & $\underline{0.0272}$ \footnotemark[1] & $0.1571$ & $-$ \\
	$M_{H{\sc I}}/M_{*}$ & $0.1177$ & $0.0622$ & ${\bf 4\times10^{-7}}$ \\
	$M_{H_2 + H{\sc I}}/M_{*}$ & $0.7483$ \footnotemark[1] & $0.1571$ & $-$ \\
	\hline
	$\log \rm SFR$ [$M_\odot$ yr$^{-1}$] & $\underline{0.0239}$ & $0.2684$ & $\underline{0.0145}$ \\
	$\log \rm sSFR$ [yr$^{-1}$] & $\underline{0.0495}$ & $0.1843$ & $\bf{0.012}$ \\
	$\log M_{H_2}/{\rm SFR}$ [yr] & ${\bf 0.0034}$ \footnotemark[1] & ${\bf 0.0036}$ & $-$ \\
	$\log M_{H{\sc I}}/{\rm SFR}$ [yr] & $0.2406$ & $0.4149$ & $\underline{0.0130}$ \\
	$\log M_{H_2 + H{\sc I}}/{\rm SFR}$ [yr] & $0.0855$ \footnotemark[1] & $0.3090$ & $-$ \\
	\hline
	\end{tabular}
	\begin{tabular}{l}
	\footnotemark[1] Restricted mean and standard errors from the Kaplan-Meier \\
	 estimator of the survival functions\\
	Note: Underline indicate $p<0.05$ and values are bolded for $p<0.01$ \\
	\end{tabular}
\end{table}
\par
In Table~\ref{tab-sum_environment}, we present a summary of the properties of the group/Virgo, CO detected/CO non-detected samples, and a test case comparing the group and Virgo samples using CO detected galaxies only. In Table~\ref{tab-sign_environment}, we present the results from performing the \ks and log-rank statistical tests in order to determine whether the properties of the galaxies in these samples can be distinguished. We have decided to focus on the comparison between the group and Virgo samples, as they have roughly similar number of galaxies and CO detections, as compared to the field sample, which contains fewer galaxies. Future work will focus on further expanding the field sample and creating a sample of isolated galaxies, in order to incorporate these important objects into the analysis.
\par
First, we consider the gas properties of the group and Virgo samples. The mean ratio of \h2 to \hi gas mass is higher for the Virgo galaxies compared to the group galaxies. The log-rank test shows a significant difference in the distribution of the ratio of \h2 to \hi between the Virgo sample and the field sample. This makes sense given that the Virgo galaxies have a lower mean \hi gas masses and higher \h2 gas masses. This seems to follow the general trends from \citet{Kenney89}, who found that the Virgo galaxies are not as \h2 deficient as expected, given their \hi deficiencies. The stellar mass normalized quantities, including the mean \h2 gas masses, generally follow the same trends as the unnormalized quantities. The difference between the stellar mass normalized \hi mass distribution of the group and Virgo galaxies is not as significant as for the unnormalized case. The higher \h2 to \hi ratio in the Virgo sample may indicate that these galaxies are more efficient at converting available \hi to \h2, perhaps through the various forms of interactions that lead to gas flowing towards the centre of host galaxies and the creation of molecular hydrogen. An alternative explanation is that the cluster environment is more effective at stripping the \hi rather than the \h2 gas and thus increasing the global \h2 to \hi ratio \citep{Pappalardo12}. These ideas will be explored in future studies with the resolved data.
\par
With the H$\alpha$ data available for our galaxies, we can determine the star formation rates and specific star formation rates (star formation rate divided by stellar mass) for our galaxies. We find that both values are lower for Virgo galaxies, compared to the group galaxies. However, the \ks test shows that while the star formation rate distributions can be distinguished between the group and Virgo samples at higher significance than the specific star formation rates.
\par
\begin{figure}
	\includegraphics[width=8.0cm]{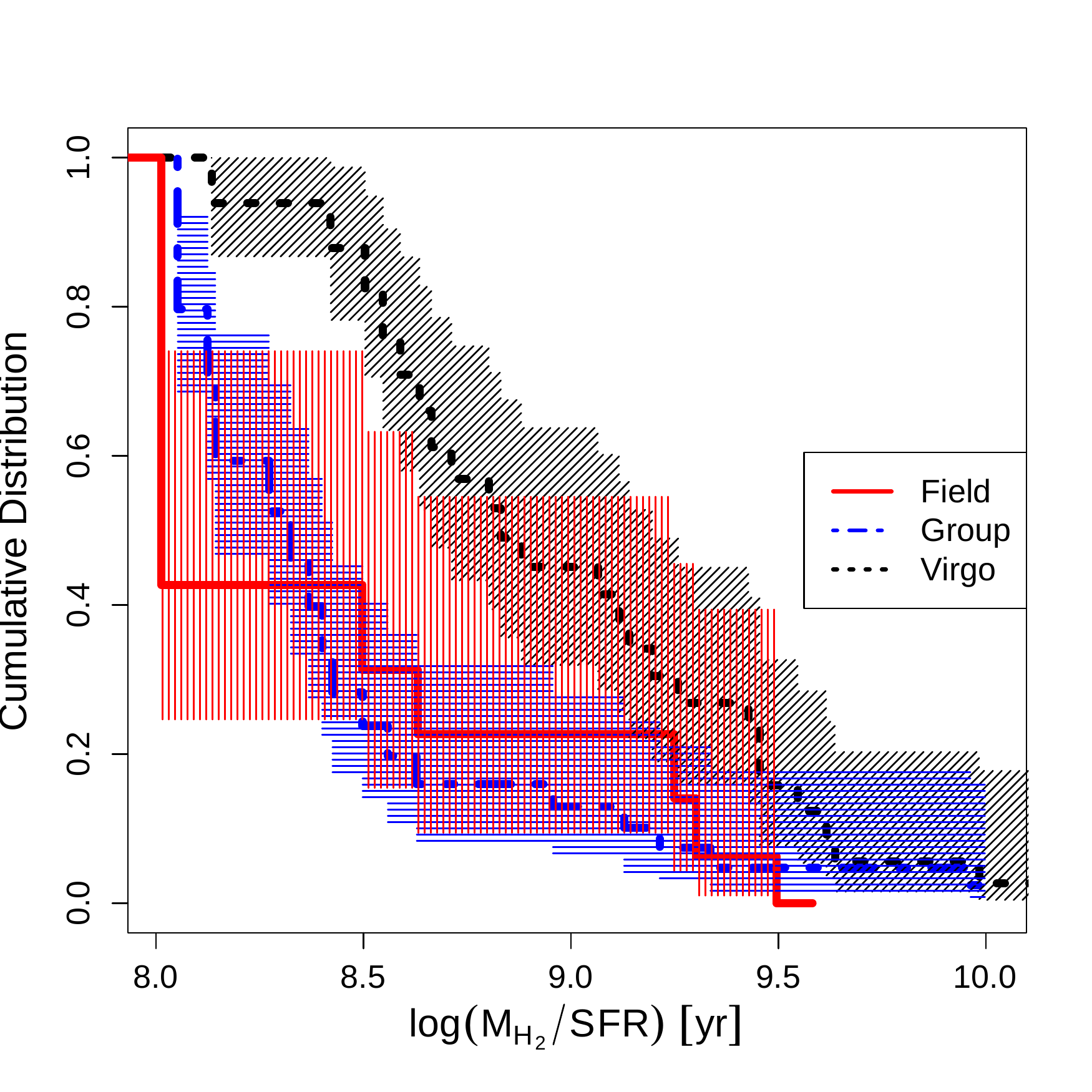}
	\caption{Survival functions for the molecular gas depletion time in the field (red solid line), group (blue dot-dash line), and Virgo (black dashed line) sample galaxies using the Kaplan-Meier estimator. The `steps' in the distribution correspond to detections. The $95\%$ confidence intervals are overlaid for the three distributions.}
	\label{fig-08AC_H2_SFR}
\end{figure}

A method of measuring the relationship between star formation and the gas content of galaxies is through the gas depletion timescale ($M_{H_2}$/SFR [yr]) or its reciprocal, the star formation efficiency (SFR/$M_{H_2}$ [yr$^{-1}$]). The mean log molecular gas depletion timescale is longer for the Virgo sample ($8.97\pm0.06$) than for the group sample ($8.44\pm0.07$), with the log-rank test showing that the Virgo galaxies can be distinguished from the group galaxies at the $p=0.01$ level. This is shown graphically in Figure~\ref{fig-08AC_H2_SFR}, where we see large differences in the shapes of the cumulative distribution functions. For the case where we only consider CO detected galaxies from the group and Virgo sample, the differences are still significant. When we calculate the gas depletion times with respect to the atomic hydrogen gas mass or the total gas mass (\h2 + \hino), the two samples have similar distributions.
\par
One possible explanation for this difference is metallicity effects on the $X_{CO}$ value, as discussed in Section~\ref{subsec-obs}. If Virgo galaxies have systematically different metallicities than group galaxies, then that would have an effect on the molecular gas mass and hence the gas depletion times. The similar mean stellar mass of the two samples suggests any metallicity effects should not be significant. Other differences in the ISM properties, such as temperature or surface density, would also cause changes in the measured molecular gas mass, assuming a variable $X_{CO}$ conversion factor prescription \citep{Bolatto13}. However, it seems unlikely that Virgo spirals would have the extreme temperatures or surface densities required to cause a substantial difference in the $X_{CO}$ factor.
\par
Another explanation for the variations in the molecular gas depletion time is that star formation may be inhibited in the more extreme cluster environment, even in the presence of a comparable or even higher amount of \h2 gas, due to other mechanisms such as increasing thermal support in the gas. Turbulent pressure can play a large role in regulating star formation \citep{Krumholz09} and this pressure is likely to be different between the three environments. For example, studies of the dense gas tracer HCN in nearby disk galaxies from the HERACLES survey found variations in the star formation efficiency with environment and towards the centre of galaxies \citep{Usero15}. The authors suggests models where such variations are caused by differences in the Mach number, which can also apply to the Virgo cluster environment. For example, \citet{Alatalo15} found an increased $^{13}$CO/$^{12}$CO ratio in early-type galaxies in the Virgo cluster, which they attribute to enrichment due to low-mass stars or to variations in the gas pressure in the dense environment. Another possibility is differences in the gravitational stability of spiral disks, which can be parametrized by the Toomre Q parameter \citep{Toomre64, Kennicutt89}. Close neighbors and the additional pressure from the cluster environment may cause changes in the disks of spiral galaxies and its star formation efficiency.
\par
Comparing to previous surveys, we note that \citet{Leroy08} found a constant value of the molecular gas depletion timescale of $(1.90\pm0.4)\times10^9$ (or $\sim10^{9.27}$) years using resolved maps of 23 galaxies from the HERACLES survey with CO $J=2-1$ line, FUV + IR star formation rates, and a Kroupa IMF. The study of a sample of 30 galaxies from the HERACLES survey found a \h2 depletion timescale of $2.2\times10^9$ (or $\sim10^{9.34}$) years \citep{Leroy13}. This value is longer than our integrated values for our group and Virgo samples. The COLD GASS survey \citep{Saintonge11b} found a molecular gas depletion time of $\sim1$ Gyr, varying between $\sim0.5$ Gyr for low-mass galaxies ($\sim10^{10}M_\odot$) to $\sim3$ Gyr for high-mass galaxies ($\sim10^{11}M_\odot$). This is more similar to the results from the analysis of our sample of galaxies.
\par
One reason for the variations in the molecular gas depletion time between our sample and the HERACLES survey may be the difference between integrated and resolved measurements. We have calculated a single value of the molecular gas depletion time for the entire galaxy, as compared to performing pixel by pixel measurements. \citet{Leroy13} found a median gas depletion time of $2.2$ Gyr when weighted by individual measurements, but a lower value of $1.3$ Gyr (or $\sim10^{9.11}$ years) when weighted by galaxy. Second, the presence of lower mass galaxies in our sample could be driving down the mean molecular gas depletion timescale. This is because we are weighting all galaxies equally in this analysis, whereas resolved measurements are dominated by the measurements from larger, higher mass galaxies. Third, our use of the CO $J=3-2$ line may be another reason, since that line traces a smaller fraction of the molecular gas than the CO $J=2-1$ line, mainly the warmer and denser component. Although we have included an average line ratio in our calculations, we may still be missing some of the more diffuse gas in the disk. A fourth possibility is the difference in the star formation rate indicator (H$\alpha$ + extinction correction vs. FUV + IR), though it seems unlikely that the FUV + IR would provide systematically lower star formation rates. This is because the FUV component measures star formation on a longer timescale and both the FUV and IR have contributions not related to recent star formation \citep{Kennicutt12}. Finally, the H$\alpha$ emission may not fully coincide with the results from the CO maps, which for most observations were concentrated in the inner 2 arcminutes of the galaxy. This may lead to a higher star formation rate (measured over a larger portion of the galaxy) as compared to its gas content. A future extension to this project will look at resolved measurements for the galaxies in this sample, which will then create a more consistent comparison and may help resolve any differences.
\subsection{CO Detected/Non-Detected Comparison}\label{subsec_detected}
\par
\begin{figure*}
	\includegraphics[width=8.0cm]{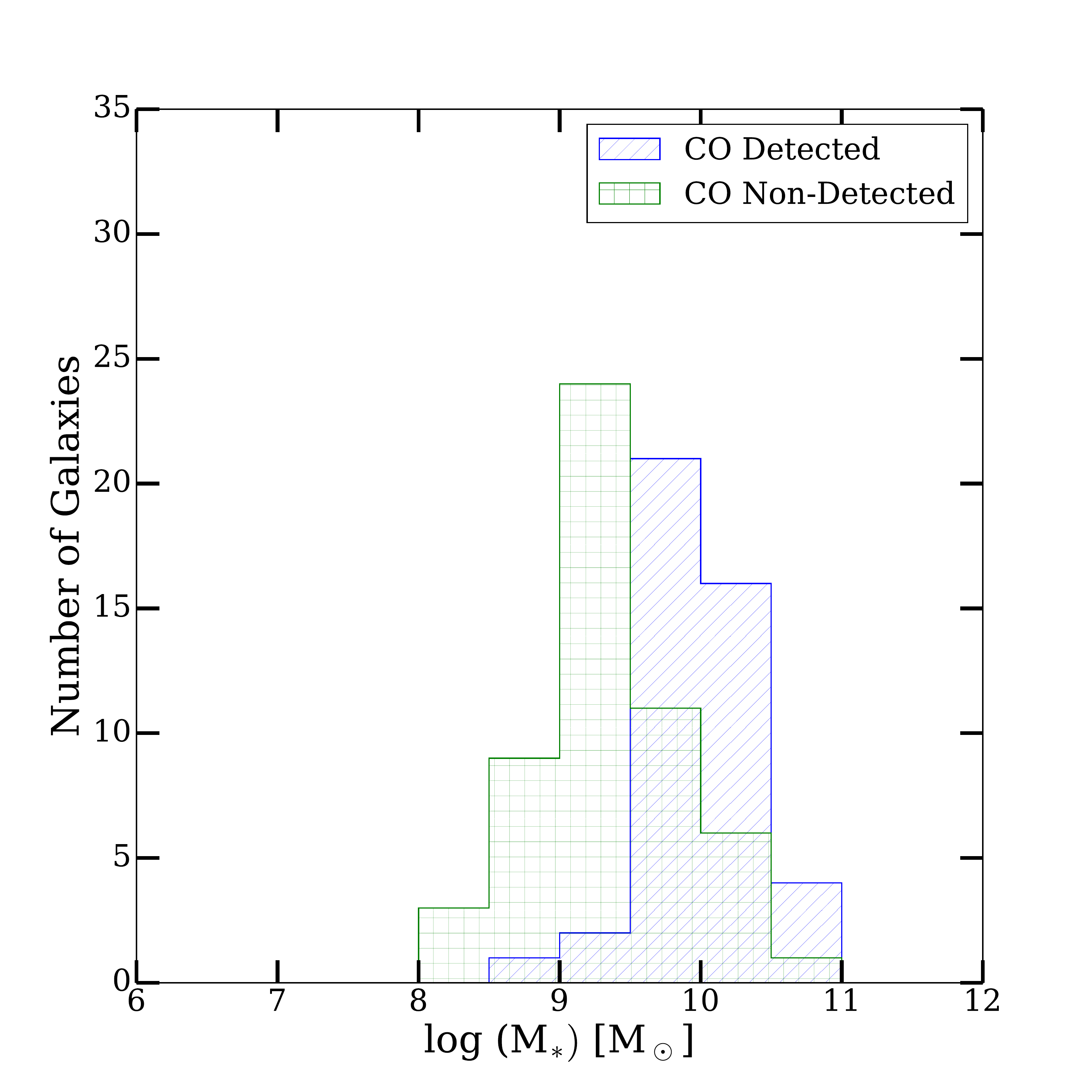}
	\includegraphics[width=8.0cm]{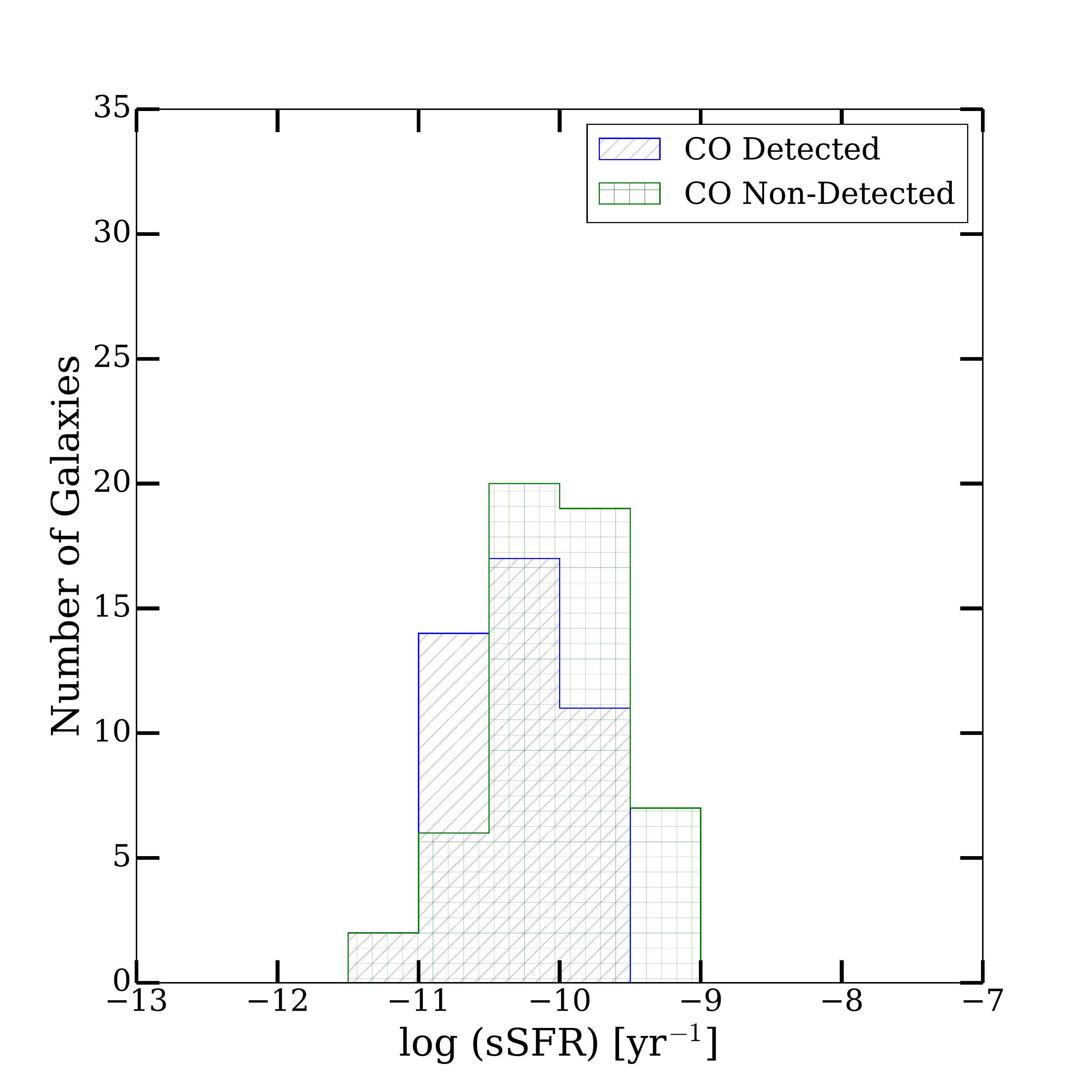}
	\caption{{\bf Left:} A histogram showing the significant difference between the stellar mass distributions of the CO detected and non-detected samples ($p=1\times10^{-9}$). {\bf Right:} A histogram showing the difference in the sSFR distribution between the CO detected and non-detected samples ($p=0.012$).}
	\label{fig-codet}
\end{figure*}

There are 44 CO detected galaxies in our sample, including 6 field galaxies, 17 group galaxies, and 21 Virgo galaxies. There are 54 CO non-detected galaxies, including 11 field galaxies, 25 group galaxies, and 18 Virgo galaxies. Looking at their gas properties, the two samples have roughly the same \hi gas mass, which suggests the lack of a strong correlation between \h2 and \hi gas masses in our \hi - flux selected sample. The CO detected sample has a larger average stellar masses than the CO non-detected sample, as seen in the left panel of Figure~\ref{fig-codet}. There is a very significant difference between their stellar mass distributions, with a corresponding p-value of $1\times10^{-7}$. The relationship between the stellar mass and CO detection would be expected if the \h2 gas mass is well-correlated with the stellar mass, which has also been seen in \citet{Boselli14}.
\par
We also find that the upper limits for the \h2 and total gas mass in CO non-detected galaxies are lower than the corresponding values for the CO detected sample. However, the results are not as conclusive when considering the stellar mass normalized values. This is likely due to the large difference in the stellar mass of the two samples, with the CO detected galaxies being much more massive than the non-detected galaxies. This results in stellar-mass normalized \h2 and total gas masses that are comparable to the upper limits for CO detected sample. We will need more data to determine if there are any systematic differences between the two samples.
\par
The mean log star formation rate is higher in the CO detected galaxies than in the CO non-detected galaxies (log SFR [$M_\odot$ yr$^{-1}$] of $-0.35\pm0.09$ vs. $-0.72\pm0.07$). The corresponding p-value for the \ks test between the two distributions is $0.0145$. This difference in the star formation rates between the two samples is likely caused by the correlation between star formation rate and molecular gas, the material required to form stars. With the stellar masses of the CO detected galaxies substantially higher than those of the CO non-detected galaxies, the \ks test show a difference in the distribution of sSFR between the two samples, which can also be seen in the right panel of Figure~\ref{fig-codet}. This is likely a consequence of the well-known negative correlation between stellar mass and specific star formation rate for star forming galaxies. Finally, the \hi gas depletion times are longer in the CO non-detected galaxies compared to the CO non-detected galaxies. This is likely also due to the significant difference in their star formation rates.
\subsection{CO $J=3-2$ Detection Rates}\label{subsec-dis_det}
\par
The overall detection rate for the sample is 44 per cent. The CO $J=3-2$ detection rate is slightly lower for the field ($35\pm9$ per cent) sample, as compared to the group ($40\pm6$ per cent) and Virgo ($54\pm9$ per cent) samples. Some of the factors that would cause this difference include the Virgo galaxies being on average closer than the group and field galaxies, which can influence the CO detection rates, since closer galaxies would be easier to detect. Another important factor is sample variance and the smaller number of field galaxies in our sample, only 17 in total. As a result, the detection or non-detection of one or two galaxies can have a large influence on the CO detection rate.
\par
We note that the stellar mass for the field sample is on average lower than the Virgo and group samples. Given the correlation between stellar mass and molecular gas mass \citep{Lisenfeld11, Boselli14}, this difference in the stellar mass will contribute to the difference in the detection rates. These low stellar mass galaxies (below $\sim 10^9 M_\odot$) may even be considered dwarf spirals according to the stellar mass classification from other galaxy surveys \citep{Geha12}. In addition, if any galaxies in our sample are metal poor, this would also affect the \h2 mass we estimate via the metallicity dependence of the CO-to-\h2 conversion factor \citep{Wilson95}, as well as the detection rate. The stellar mass distribution of our field, group, and Virgo sample, with the majority of our galaxies in the mass range of $10^9-10^{11} M_\odot$, combined with the observed mass-metallicity relationship fit \citep{Tremonti04}, would lead to ($12$ $+$ log(O/H)) values of between $8.63-9.11$. It is likely that few of these galaxies have metallicities more than a factor of two below solar ($12$ $+$ log(O/H) $=8.69$), the rough limit at which any metallicity effects would become significant \citep{Wilson95, Arimoto96, Israel00, Bolatto08}.
\par
Finally, the redshift-limited (recessional velocities of between 1500 and 5000 km s$^{-1}$) AMIGA survey of isolated galaxies detects $51\pm 5$\% in the CO $J=1-0$ line, a detection rate similar to our Virgo samples and slightly higher than our overall sample \citep{Lisenfeld11}. In comparison to our sample, their stellar masses (measured by $L_K$) are roughly in the range of $10^8 - 10^{11}M_\odot$, similar in mass to than our sample. Their sample of isolated galaxies is based on the catalogue of \citet{Karachentseva73} and is chosen to possess no nearby similarly sized neighbours in the sky.
%
%
\section{Discussion}
\begin{table*}
	\begin{minipage}{120mm}
	\caption{Selected Properties of the Early-type Spiral, Late-Type Spiral, Pair, and Non-Pair Populations}\label{tab-sum_morph}
	\begin{tabular}{lcccc}
	\hline
	Mean Quantity & Early-type Spirals & Late-type Spirals & Pair & Non-Pair \\
	 & ($40$) & ($57$) & ($14$) & ($84$)\\
	\hline
	$\log M_{H{\sc I}}$ [$M_\odot$] & $9.13\pm0.05$ & $9.09\pm0.04$& $9.06\pm0.08$ & $9.11\pm0.03$\\
	$\log M_{H_2}$ [$M_\odot$]\footnotemark[1] & $8.36\pm0.08$ & $8.00\pm0.07$& $8.47\pm0.13$ & $8.06\pm0.07$\\
	$\log M_{*}$ [$M_\odot$] & $9.86\pm0.08$ & $9.49\pm0.07$& $9.79\pm0.11$ & $9.61\pm0.06$\\
	$\log M_{H_2 + H{\sc I}}$\footnotemark[1] & $9.50\pm0.07$ & $9.25\pm0.07$& $9.36\pm0.08$ & $9.35\pm0.06$\\
	\hline
	$M_{H_2}/M_{H{\sc I}}$\footnotemark[1] & $0.63\pm0.18$ & $0.28\pm0.10$ & $1.09\pm0.48$ & $0.31\pm0.07$\\
	$M_{H_2}/M_{*}$\footnotemark[1] & $0.055\pm0.010$ & $0.038\pm0.007$ & $0.066\pm0.016$ & $0.041\pm0.006$\\
	$M_{H{\sc I}}/M_{*}$ & $0.32\pm0.06$ & $0.70\pm0.11$& $0.31\pm0.07$ & $0.59\pm0.08$\\
	$M_{H_2 + H{\sc I}}/M_{*}$\footnotemark[1] & $0.40\pm0.07$ & $0.80\pm0.13$ & $0.38\pm0.07$ & $0.67\pm0.09$\\
	\hline
	$\log SFR$ [$M_\odot$ yr$^{-1}$] & $-0.44\pm0.09$ & $-0.62\pm0.08$ & $-0.46\pm0.11$ & $-0.57\pm0.07$\\
	$\log sSFR$ [yr$^{-1}$] & $-10.30\pm0.07$ & $-10.10\pm0.06$ & $-10.26\pm0.12$ & $-10.18\pm0.05$\\
	$\log M_{H_2}/{\rm SFR}$ [yr]\footnotemark[1] & $8.68\pm0.10$ & $8.55\pm0.06$ & $8.85\pm0.20$ & $8.57\pm0.06$\\
	$\log M_{H{\sc I}}/{\rm SFR}$ [yr] & $9.57\pm0.07$ & $9.70\pm0.05$ & $9.52\pm0.12$ & $9.68\pm0.05$\\
	$\log M_{H_2 + H{\sc I}}/{\rm SFR}$ [yr]\footnotemark[1] & $9.73\pm0.06$ & $9.79\pm0.05$ & $9.73\pm0.11$ & $9.78\pm0.04$\\
	\hline
	\end{tabular}
	\begin{tabular}{l}
	\footnotemark[1] Restricted mean and standard errors from the Kaplan-Meier estimator of the survival functions\\
	\end{tabular}
	\end{minipage}
\end{table*}
\subsection{Effects of Morphology}\label{subsec-dis_morph}
\begin{table}
	\centering
	\caption{Significance Test Between Different Samples}\label{tab-sign_test}
	\begin{tabular}{lccc}
	\hline
	Mean Quantity & Early/Late & Pair/Non-Pair\\
	\hline
	$\log M_{H{\sc I}}$ [$M_\odot$] & $0.6553$ & $0.9524$\\
	$\log M_{H_2}$ [$M_\odot$]\footnotemark[1] & $0.0614$ & $0.1799$\\
	$\log M_{*}$ [$M_\odot$] & ${\bf 0.0013}$ & $0.4489$\\
	$\log M_{H_2 + H{\sc I}}$ [$M_\odot$]\footnotemark[1] & $0.0504$ & $0.8574$\\
	\hline
	$M_{H_2}/M_{H{\sc I}}$\footnotemark[1] & $0.0976$ & ${\bf 0.0098}$\\
	$M_{H_2}/M_{*}$\footnotemark[1] & $0.3055$ & $0.1971$\\
	$M_{H{\sc I}}/M_{*}$ & $\bf{0.0014}$ & $\underline{0.0450}$\\
	$M_{H_2 + H{\sc I}}/M_{*}$\footnotemark[1] & $\underline{0.0066}$ & $0.2072$\\
	\hline
	$\log SFR$ [$M_\odot$ yr$^{-1}$] & $0.4178$ & $0.4489$\\
	$\log sSFR$ [yr$^{-1}$] & $0.0620$ & $0.5856$\\
	$\log M_{H_2}/{\rm SFR}$ [yr]\footnotemark[1] & $0.1249$ & $\underline{0.0163}$\\
	$\log M_{H{\sc I}}/{\rm SFR}$ [yr] & $0.2626$ & $0.5856$\\
	$\log M_{H_2 + H{\sc I}}/{\rm SFR}$ [yr]\footnotemark[1] & $0.4890$ & $0.8262$\\
	\hline
	\end{tabular}
	\begin{tabular}{l}
	\footnotemark[1] Restricted mean and standard errors from the Kaplan-Meier\\
	 estimator of the survival functions\\
	Note: Underline indicate $p<0.05$ and values are bolded for $p<0.01$ \\
	\end{tabular}
\end{table}
\par
Morphology can have a large effect on the star formation \citep{Kennicutt98B, Bendo07} and the molecular gas properties \citep{Kuno07} of spiral galaxies. For our sample of galaxies, we have performed a simple comparison of the early-type spirals (a, ab, b, and bc) with the late-type spirals (c, cd, d, and m). As stated in the discussion of our sample selection in Section~\ref{subsec-pro_sample}, we used the HyperLeda morphological codes for this classification, which employs a weighted average of multiple measurements. One galaxy classified as S?, NGC3077, is excluded from this analysis. In total, there are 40 early-type spirals and 57 late-type spirals. Selected properties for the two samples are presented in Table~\ref{tab-sum_morph}, as well as the results from the significance tests in Table~\ref{tab-sign_test}.
\par
The \ks test shows a significant difference in the stellar mass ($p=0.0013$), with the mean value being higher for early-type spirals. This results in significantly lower stellar mass normalized \hi gas masses and total gas masses for the sample. On the other hand, the \ks test shows that the distributions of \hi mass, star formation rates, and the specific star formation rates are not significantly different between the two samples. Furthermore, the mean molecular gas mass and the molecular gas depletion times are not significantly different between the two samples. This suggests that variations between the early- and late-type spirals in this study should not be an important contributor to the differences observed between the group and Virgo samples.
\subsection{Effects of Close Pairs}\label{subsec-dis_pairs}
\par
Interacting galaxies and mergers have been linked to more active star formation in their nucleus \citep{Keel85} and can lead to inflows of gas towards the centre \citep{Mihos96}, increased cooling, and greater fragmentation \citep{Teyssier10}. From a large sample of SDSS galaxies, \citet{Ellison08} found that there is a slight statistical enhancement in the star formation rate for close pairs. This enhancement is also seen for cases where these pairs actually undergo mergers and interactions \citep{Knapen09, Knapen15}. However, this effect may be less apparent for galaxies that are inside denser environments \citep{Ellison10}. Therefore, we decided to investigate the effects on our results of removing any close pairs from the sample. Once again, we use the catalog of \citet{Karachentsev72} to compare galaxies known to be in pairs with their non-pair counterparts. For the group sample, there are 5 galaxies in pairs (NGC0450, NGC2146A, NGC3507, NGC3455, NGC4123) . For the Virgo sample, there are 9 galaxies in pairs (NGC4294, NGC4298, NGC4302, NGC4411A, NGC4430, NGC4561, NGC4567, NGC4568, NGC4647). Of these, 3 out of the 5 group galaxies and 5 out of 9 of the Virgo galaxies are CO detected. Note that for the galaxies in the Virgo cluster, close pairs may not be true interacting galaxies, due to the close proximity of these galaxies in the sky.
\par
For most of the galaxy properties in this study, such as the stellar mass and atomic gas properties, the pair and non-pair samples are not significantly different. However, the \h2 to \hi gas mass ratio is higher and the \h2 gas depletion time is longer in the pair sample. From the significance tests in Table~\ref{tab-sign_test}, the p-values are indicative of a significant difference between the two samples. These differences in the \h2 gas depletion time and the \h2 to \hi gas mass ratio are similar to those found when comparing between the group and Virgo galaxies. The various environmental effects, such as stripping of the atomic hydrogen in the outskirts and the interaction effects on the molecular gas, may also occur for the more extreme cases of close pairs.
\par
We have also tested removing these pairs from our group and Virgo samples. Most of the results from our comparison of group and Virgo samples remain the same, such as the stellar masses and atomic gas properties. Virgo galaxies still possess a slightly higher mean molecular gas mass. On the other hand, while the \h2 to \hi ratio is higher for Virgo galaxies at $0.48\pm0.14$ compared to $0.25\pm0.10$ for the group galaxies, with the pairs removed the log-rank no longer shows a significant difference between the two distributions ($p=0.1136$). Similarly, the mean log \h2 gas depletion times [yr] for the Virgo galaxies is longer at $8.97\pm0.06$ compared to $8.44\pm0.07$ for the group sample, but now with a log-rank test value of $p=0.079$.
\par
These results suggest that these environmental trends in \h2 to \hi ratio and \h2 gas depletion times are similar when we make the comparison between the group/Virgo and between the pair/non-pair populations. Removing the presence of the \citet{Karachentsev72} pairs reduces the overall significance of the differences found between the group and Virgo samples. Physically, the environmental effects discussed in the previous section will likely be amplified for the galaxies that are strongly interacting. The question remains whether the observed variations in the molecular gas and star formation properties in the cluster environment affect all galaxies or whether the difference is mainly due to the denser environment producing more close pairs? A more systematic analysis is require to fully disentangle these two effects, such as increasing the number of galaxies in our sample, the use of a more rigorous method of defining pairs, and observing trends with distance to the cluster center or with multiple nearby clusters (such as the Fornax Cluster). 
\subsubsection{\hi Rich Galaxies}\label{subsec-dis_richhi}
\par
We have used the less traditional \hi flux as the primary selection in our sample of galaxies. As a result, our full sample includes many galaxies with normal \hi mass, but low stellar mass. We can see their presence most readily in the CO non-detected sample or by looking at the specific galaxies with high \hi gas mass to stellar mass ratios. In general, these objects will likely be missed by optically-selected surveys and may even exist as an understudied class of galaxies. Similar \hi rich objects have been observed recently by the Bluedisks project \citep{Wang13} and HIghMass survey \citep{HuangS14}. The HIghMass survey galaxies have high \hi gas mass ($M_{H{\sc I}} > 10^{10} M_\odot$) and high \hi fractions compared to galaxies with the same stellar mass. After measuring their star formation rates, they found that the HIghMass galaxies have comparatively high specific star formation rates, which the authors attribute to their more recent formation times. The CO non-detected galaxies in our sample possess similar qualities, with lower stellar masses, high \hi to stellar mass ratios, and higher specific star formation rates compared to the CO detected galaxies.
\subsection{Correlation between Galaxy Properties}\label{subsec-dis_correlation}
\begin{figure}
	\includegraphics[width=8.0cm]{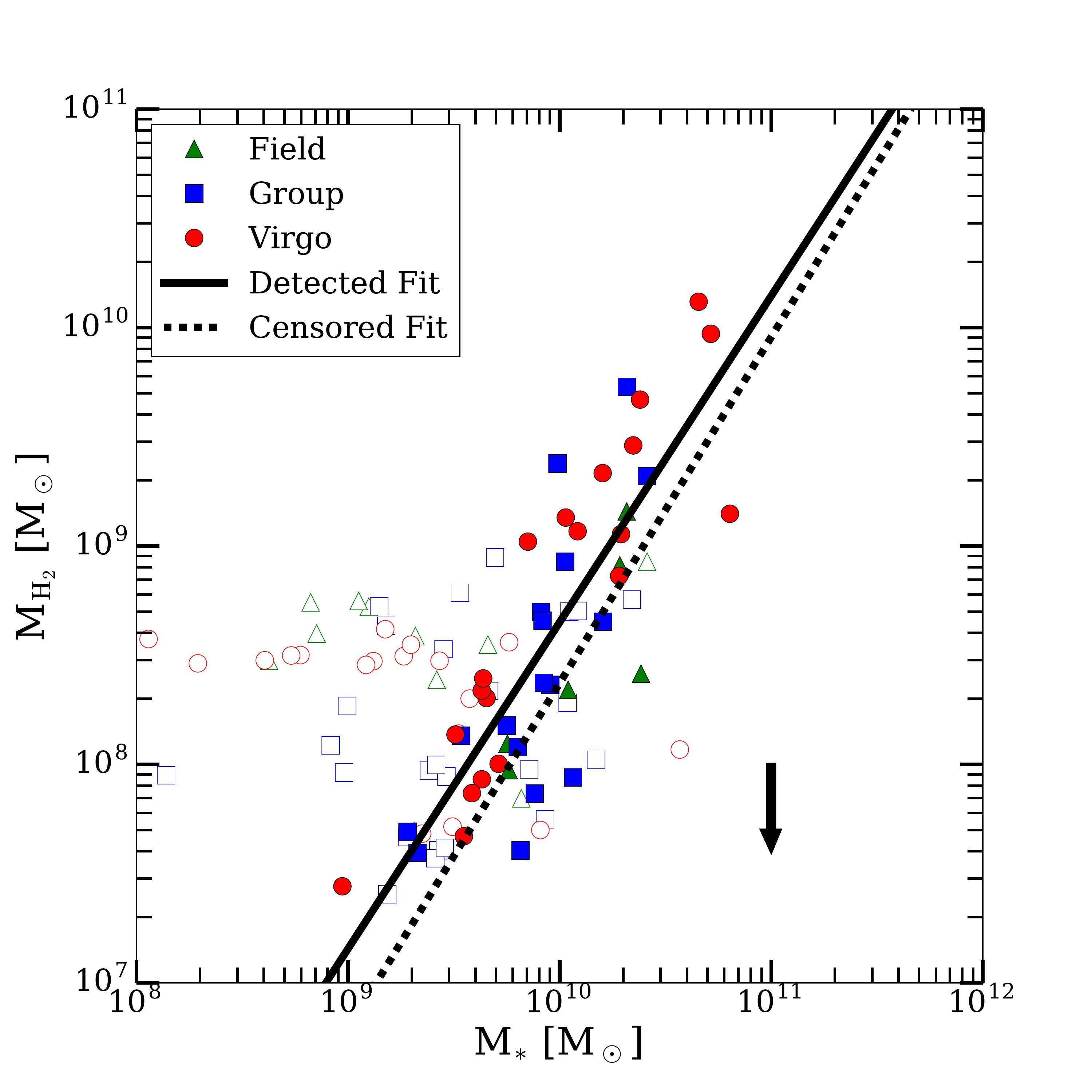}
	\caption{The molecular gas mass as a function of the stellar mass for the field, group, and Virgo sample. Filled points are detections while open points are upper limits, with the direction indicated by the large black arrow and a length of 1$\sigma$. Also plotted are the linear fit to the detected galaxies in the entire sample, with a slope of $1.49\pm0.15$. The Pearson coefficient is $0.84$ ($p=7\times10^{-13}$). The Buckley-James fit produced a slope of $1.58\pm0.15$. This relationship between the molecular gas mass and the stellar mass have been seen in previous survey for spiral galaxies \citep{Lisenfeld11, Boselli14}.}
	\label{fig-002_Mstar_H2}
\end{figure}
\begin{figure*}
	\includegraphics[width=8.0cm]{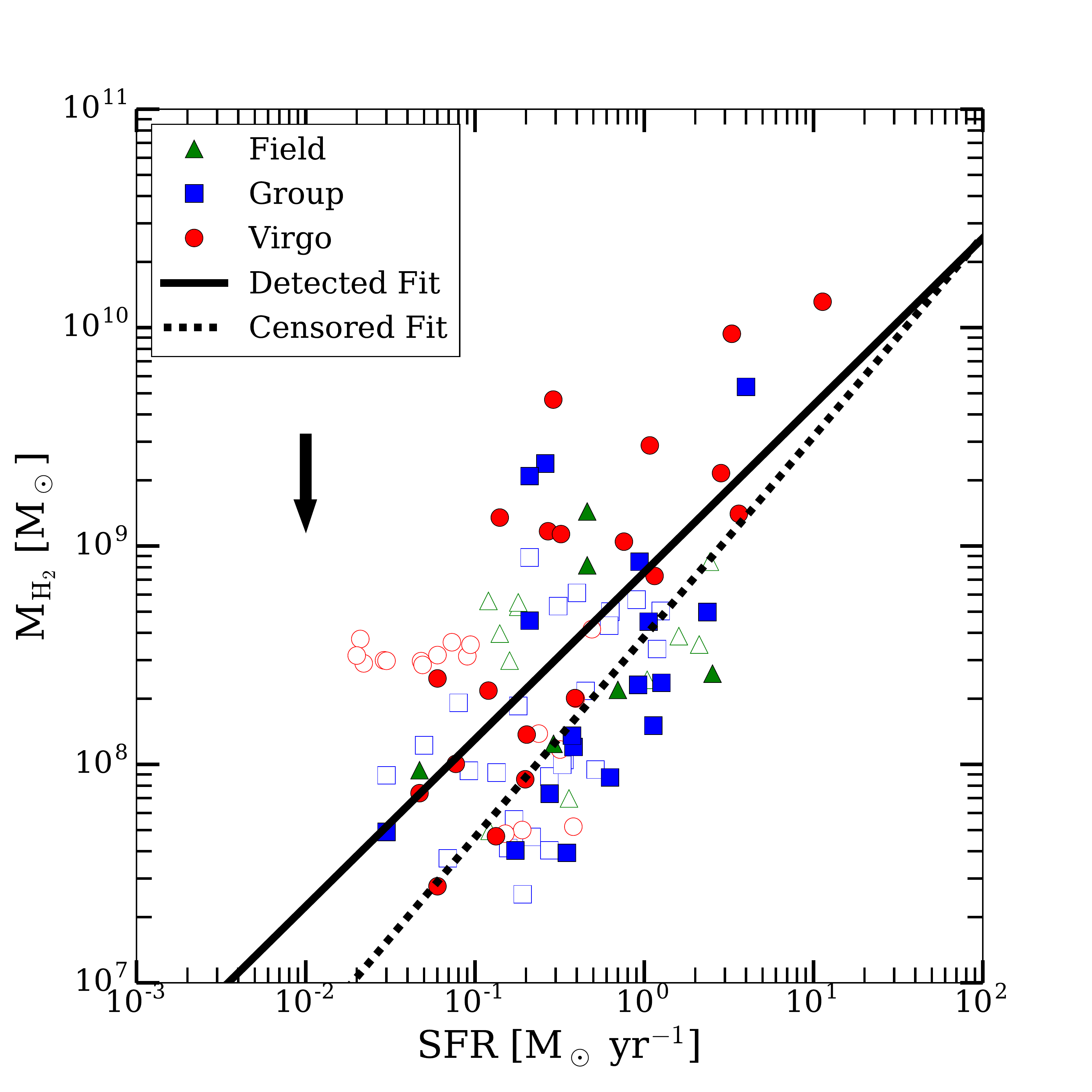}
	\includegraphics[width=8.0cm]{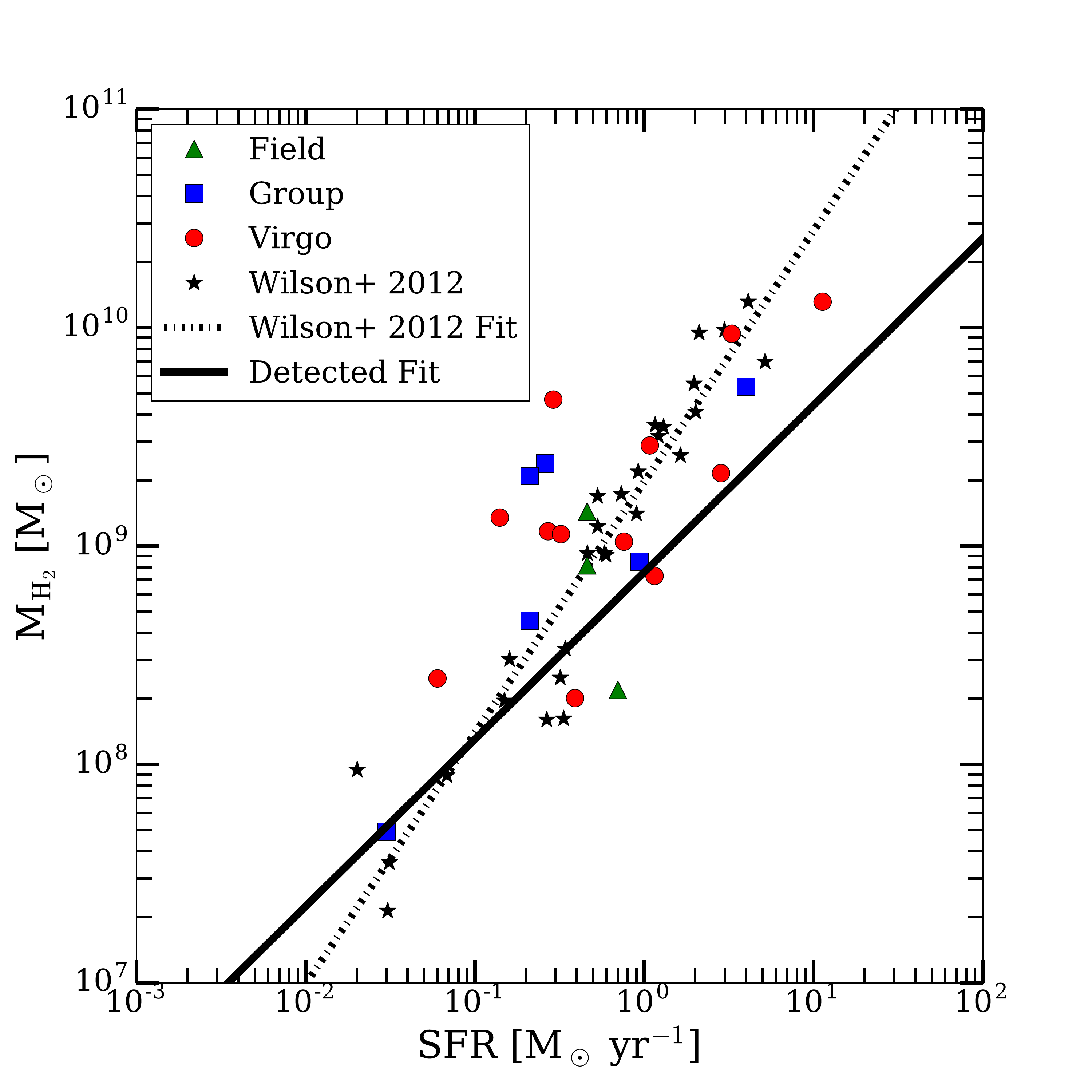}
	\caption{{\bf Left:} The molecular gas mass as a function of the star formation rate for the field, group, and Virgo sample. Filled points are detections while open points are upper limits, with the direction indicated by the large black arrow and a length of 1$\sigma$. Also plotted are the linear fit to the detected galaxies in the entire sample, with a slope of $0.76\pm0.14$. The corresponding Pearson coefficient is $0.64$ ($p=4\times10^{-6}$). The Buckley-James (censored) fit produced a slope of $0.92\pm0.15$. {\bf Right:} The molecular gas mass as a function of the star formation rate for the field, group, and Virgo sample, only including galaxies with a CO S/N ratio greater than $5$. Also plotted in black stars are the smaller sample of galaxies from \citet{Wilson12}, which includes all NGLS galaxies that are also part of the Spitzer Infrared Nearby Galaxy Survey, with star formation rates are measured using FIR fluxes. The dotted line is the linear fit to the galaxies from \citet{Wilson12} ($m=1.149\pm0.005$), with the solid black line from the detected fit to the whole sample plotted for comparison purposes.}
	\label{fig-T002_SFR_H2_Comp}
\end{figure*}
\par
We seek to determine the important scaling relationships among the galaxies in our sample by plotting the different physical properties and identifying any possible correlations. For this analysis, we have chosen to use a simple linear fit to the CO detected galaxies, since the measurement errors on these galaxy properties are small compared to the scatter in the data points. To take into account the effects of data censoring for values along the y-axis, the Buckley-James estimator was used \citep{Buckley79}. To perform the Buckley-James regression, we used the subroutine {\sc bj} in the statistical package called {\sc rms}, which can be found at the standard {\sc R} repository\footnote{http://cran.r-project.org/web/packages/rms/index.html}. We have decided to use the Buckley-James regression in our analysis because of its similarity to survival analysis, with both techniques attempting to incorporate upper limits into the statistical treatment.
\par
First, we present the relationship between stellar mass and molecular gas mass in Figure~\ref{fig-002_Mstar_H2}, which shows a positive correlation between the two parameters, with a slope of $1.49\pm0.15$. This result indicates that the more massive galaxies in our sample contain more molecular gas and has been seen previously in \citet{Lisenfeld11} for late-type galaxies. \citet{Boselli14}, using the Herschel Reference Survey, have also noted relatively constant $M_{H_2}/M_*$ ratios for spiral galaxies. On the whole, this result suggests that those galaxies with more stars have more fuel for future star formation.
\par
Next, we show the relationship between the molecular gas mass and the star formation rate, which is similar to other analyses based on the Kennicutt-Schmidt law \citep{Kennicutt98A}. From the left panel in Figure~\ref{fig-T002_SFR_H2_Comp}, we find a considerable scatter around the fit. One difference from similar studies is the use of the CO $J=3-2$ line, which traces denser and warmer molecular gas when compared to the lower transition CO lines. The use of CO and H$\alpha$ measurements covering different portions of the galaxies, as discussed when comparing integrated \h2 gas depletion times with other surveys, may also contribute to the scatter. Note that we have presented the plot as M$_{H_2}$ vs star formation rate, since our method of calculating the censored data fit only allows for censoring for the variable in the y-axis. Previous studies of resolved molecular gas and star formation rate measurements have found a slope near unity \citep{Bigiel08, Leroy13}, though other groups have found larger values for the slope of the star formation rate vs. molecular gas mass \citep{Kennicutt07}.
\par
To determine the main cause of the large scatter, we have compared our results to an earlier paper in this series in the right panel of Figure~\ref{fig-T002_SFR_H2_Comp}. The previous paper focused on NGLS galaxies that are also found in the SINGS (Spitzer Infrared Nearby Galaxies Survey) sample. The star formation rates in \citet{Wilson12} are measured using FIR fluxes instead of H$\alpha$ measurements, but they have been converted into star formation rates with the SF conversion factor from \citet{Kennicutt12}, also assuming a Kroupa IMF. The galaxies from our larger sample seem to follow the same trend as the galaxies from \citet{Wilson12}, though with a much larger scatter. This scatter could be due to the marginal nature of some of our CO detections, as we have selected all galaxies with a S/N ratio $>3$. When we plot this relationship including only the galaxies in this paper with S/N $>5$, we find that the scatter is reduced, though not to the level from the \citet{Wilson12} paper.
\begin{figure}
	\includegraphics[width=8.0cm]{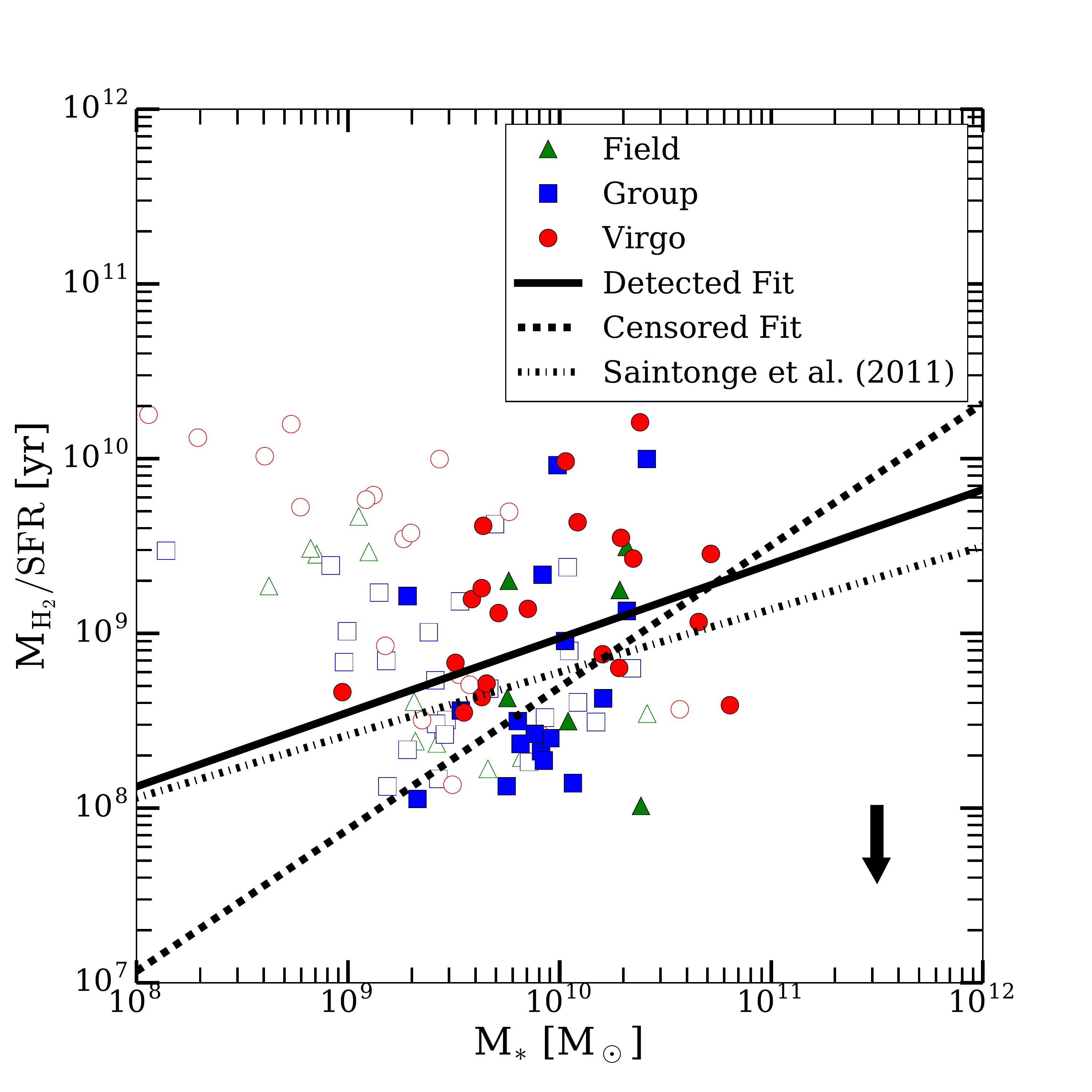}
	\caption{The molecular gas depletion time as a function of stellar mass for the field, group, and Virgo sample. Filled points are detections while open points are upper limits, with the direction indicated by the large black arrow and a length of 1$\sigma$. Also plotted are the linear fit to the detected galaxies in the entire sample, with a slope of $0.42\pm0.21$ and a Pearson coefficient of $0.30$ ($p=0.047$). The Buckley-James fit produced a steeper slope of $0.81\pm0.22$. The results are compared to the fit from \citet{Saintonge11b} for a sample of 222 galaxies.}
	\label{fig-010A_Mstar_TdepH2}
\end{figure}
\begin{figure}
	\includegraphics[width=8.0cm]{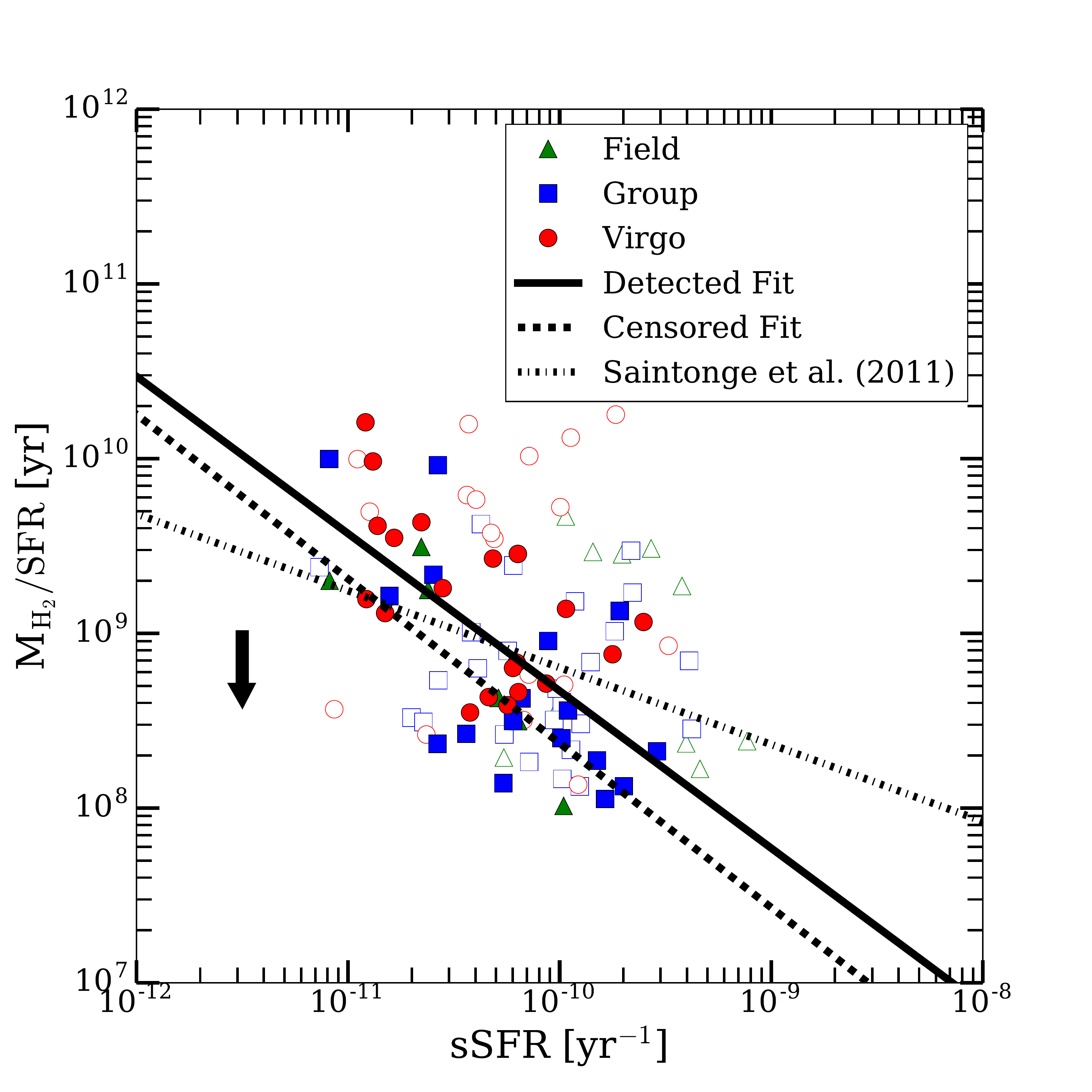}
	\caption{The molecular gas depletion time as a function of the specific star formation rate for the field, group, and Virgo sample. Filled points are detections while open points are upper limits, with the direction indicated by the large black arrow and a length of 1$\sigma$. Also plotted are the linear fit to the detected galaxies in the entire sample, with a slope of $-0.90\pm0.17$ and a Pearson coefficient of $-0.65$ ($p=2\times10^{-6}$). The Buckley-James fit produced a slope of $-0.94\pm0.16$. The results are compared to the fit from \citet{Saintonge11b} for a sample of 222 galaxies.}
	\label{fig-013A_sSFR_TdepH2}
\end{figure}
\par
In addition, we can also look at the trends in the \h2 gas depletion time (molecular gas mass divided by the star formation rate) for the galaxies in our sample. In Figure~\ref{fig-010A_Mstar_TdepH2}, we find that there is a positive correlation between the molecular gas depletion time and the stellar mass, a result also noted in other surveys \citep{Saintonge11b}. The Pearson correlation parameter using the detected galaxies is $0.30$, which is weaker than the other correlations presented in this study. However, the p-value from the correlation is $0.047$, which is a strong hint that a correlation does exist. \citet{Saintonge11b} provide three possible explanations: bursty star formation in low mass galaxies reducing the star formation rates, a quenching mechanism that reduces the star formation efficiency in high mass galaxies, and/or observations not detecting a larger fraction of molecular gas in low mass galaxies. Other groups, such as \citet{Leroy08}, have not found any significant correlations between molecular gas depletion time and stellar mass. Furthermore, it has been suggested that this correlation would disappear with a mass-dependent CO conversion factor \citep{Leroy13}.
\par
Finally, in order to tie together three of the main parameters from our study (star formation, molecular gas mass, stellar mass), we explore the correlation between the specific star formation rate (sSFR) and the molecular gas depletion time. In Figure~\ref{fig-013A_sSFR_TdepH2}, we note a negative correlation between those two parameters, consistent with results from \citet{Saintonge11b}, \citet{Huang14}, and \citet{Boselli14}. This correlation suggests that molecular gas is depleted more quickly in galaxies with high specific star formation rates (a high current star formation compared to their stellar mass). In other words, the molecular gas depletion times inside galaxies are dependent on the fraction of new stars compared to the old stellar population. A related study by \citet{Kannappan13} looked at the tight relationship between the fractional stellar mass growth rate (the mass of new stars formed over the past $\sim1$ Gyr) and the gas to stellar mass ratio, which is suggestive of a link between the amount of new stars formed, the old stellar component, and the amount of gas available.
\par
We note that the correlations found between the different galaxy properties in our sample were largely unchanged after subdividing between the group and Virgo sample, as seen in Figure~\ref{fig-010A_Mstar_TdepH2} and Figure~\ref{fig-013A_sSFR_TdepH2}. The differences in slopes of the total sample and slopes of the group and Virgo sub-samples are within their respective error bars while the differences in the intercepts of these relations are likely related to the variations observed in the properties of the group and Virgo samples in the previous sections.
%
%
\section{Conclusions}
\par
This paper presents the results of an analysis of an \hi flux limited sample of 98 spiral galaxies from the Nearby Galaxies Legacy Survey (NGLS), a Virgo follow-up program, and selected galaxies from the Herschel Reference Survey (HRS). The sample was further subdivided into group and Virgo galaxies in order to determine any possible environmental effects. We studied their molecular gas content through CO $J=3-2$ observations using the James Clerk Maxwell Telescope (JCMT) and star formation properties using H$\alpha$ measurements. We have also used survival analysis in order to incorporate data from galaxies with only upper limits on their CO measurements. 
\begin{itemize}
\item
The overall CO $J=3-2$ detection rate for the galaxies in our sample is 44 per cent. The CO detected galaxies have a larger mean stellar mass and star formation rate compared to the CO non-detected galaxies. On the other hand, the mean specific star formation rates and \hi gas masses are similar between the two samples.
\item
The mean log \hi mass is larger for group galaxies compared to the Virgo galaxies, with the \ks test showing that the distribution of \hi masses in the Virgo and group galaxies are significantly different. Conversely, the \h2 masses are higher in the Virgo compared to the group sample. As a result, the Virgo galaxies possess a significantly higher \h2 to \hi ratio than the group sample. These galaxies inside the cluster may be better at converting their \hi gas into \h2 gas, perhaps due to environmental effects on inflows towards the centre or the \h2 gas not being stripped as efficiently as the \hi gas.
\item
The mean log molecular gas depletion time ($M_{H_2}/{\rm SFR}$ [yr]) is longer in the Virgo sample ($8.97\pm0.06$) compared to the group ($8.44\pm0.07$) sample. This difference in the molecular gas depletion time may be a combination of environmental factors that both increase the \h2 gas mass, as discussed in the previous point, and decrease the star formation rate in the presence of large amounts of molecular gas, such as heating processes in the cluster environment or differences in turbulent pressure \citep{Usero15, Alatalo15}.
\item
The molecular gas depletion time ($M_{H_2}/{\rm SFR}$) depends positively on the stellar mass and negatively on the specific star formation rate, consistent with previous studies on these relationships \citep{Saintonge11b, Boselli14}. Higher mass galaxies have a longer molecular gas depletion time, i.e., they are converting their molecular gas to stars at a slower rate. This may be caused by more bursty star formation in low mass galaxies and/or quenching mechanisms in higher mass galaxies, as suggested by \citet{Saintonge11b}. We find that galaxies with high specific star formation rates have shorter molecular gas depletion times, suggesting that galaxies with high star formation rates relative to their stellar populations would run out of fuel faster and may be undergoing a different and less sustainable star formation process. This is similar to results from other studies, including a large survey with nearby and high redshift galaxies, where \citet{Genzel15} found that the gas depletion time depends most strongly on a galaxy's sSFR relative to the sSFR of the star-formation main sequence.
\end{itemize}
%
%
\section{Acknowledgments}
\par
We wish to thank the referee for their thorough review and useful suggestions. The James Clerk Maxwell Telescope has historically been operated by the Joint Astronomy Centre on behalf of the Science and Technology Facilities Council of the United Kingdom, the National Research Council of Canada and the Netherlands Organisation for Scientific Research. The research of C.D.W. is supported by grants from NSERC (Canada). We acknowledge financial support to the DAGAL network from the People Programme (Marie Curie Actions) of the European Union’s Seventh Framework Programme FP7/2007-2013/ under REA grant agreement number PITN-GA-2011-289313, and from the Spanish Ministry of Economy and Competitiveness (MINECO) under grant number AYA2013-41243-P.
\par
We acknowledge the usage of the HyperLeda database (http://leda.univ-lyon1.fr). This research has made use of the NASA/IPAC Extragalactic Database (NED) which is operated by the Jet Propulsion Laboratory, California Institute of Technology, under contract with the National Aeronautics and Space Administration. This publication makes use of data products from the Wide-field Infrared Survey Explorer, which is a joint project of the University of California, Los Angeles, and the Jet Propulsion Laboratory/California Institute of Technology, funded by the National Aeronautics and Space Administration. This work is based [in part] on archival data obtained with the Spitzer Space Telescope, which is operated by the Jet Propulsion Laboratory, California Institute of Technology under a contract with NASA. Support for this work was provided by an award issued by JPL/Caltech.
\par
The Digitized Sky Surveys were produced at the Space Telescope Science Institute under U.S. Government grant NAG W-2166. The images of these surveys are based on photographic data obtained using the Oschin Schmidt Telescope on Palomar Mountain and the UK Schmidt Telescope. The plates were processed into the present compressed digital form with the permission of these institutions. The National Geographic Society - Palomar Observatory Sky Atlas (POSS-I) was made by the California Institute of Technology with grants from the National Geographic Society. The Second Palomar Observatory Sky Survey (POSS-II) was made by the California Institute of Technology with funds from the National Science Foundation, the National Geographic Society, the Sloan Foundation, the Samuel Oschin Foundation, and the Eastman Kodak Corporation. The Oschin Schmidt Telescope is operated by the California Institute of Technology and Palomar Observatory.The UK Schmidt Telescope was operated by the Royal Observatory Edinburgh, with funding from the UK Science and Engineering Research Council (later the UK Particle Physics and Astronomy Research Council), until 1988 June, and thereafter by the Anglo-Australian Observatory. The blue plates of the southern Sky Atlas and its Equatorial Extension (together known as the SERC-J), as well as the Equatorial Red (ER), and the Second Epoch [red] Survey (SES) were all taken with the UK Schmidt.
%
%
\bibliographystyle{mn2e}
\bibliography{paper3}
%
%
\appendix
\section{Comparison of the Three CO Datasets}\label{app-comp}
\par
\begin{figure*}
	\centering
	\includegraphics[width=8.0cm]{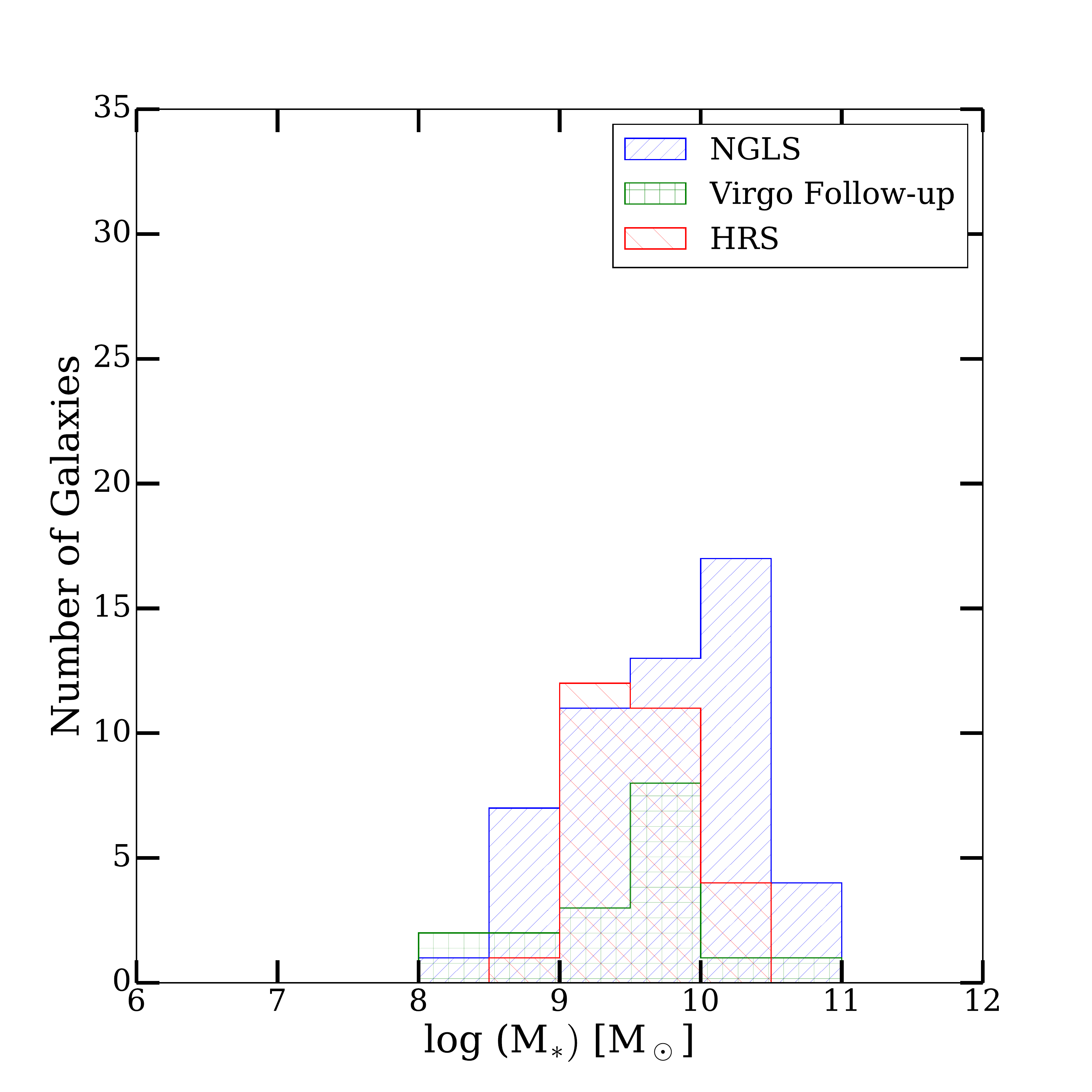}
	\includegraphics[width=8.0cm]{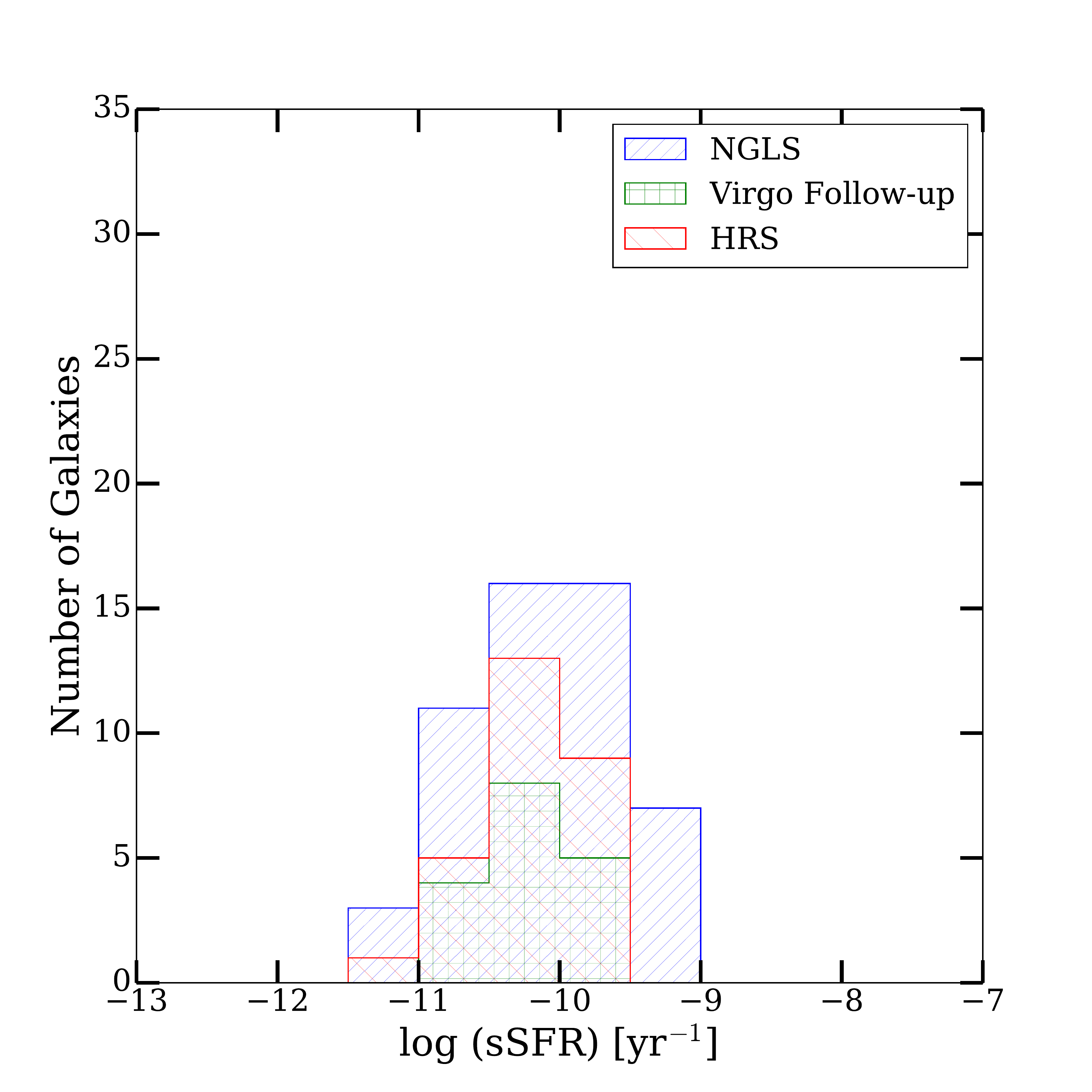}
	\includegraphics[width=8.0cm]{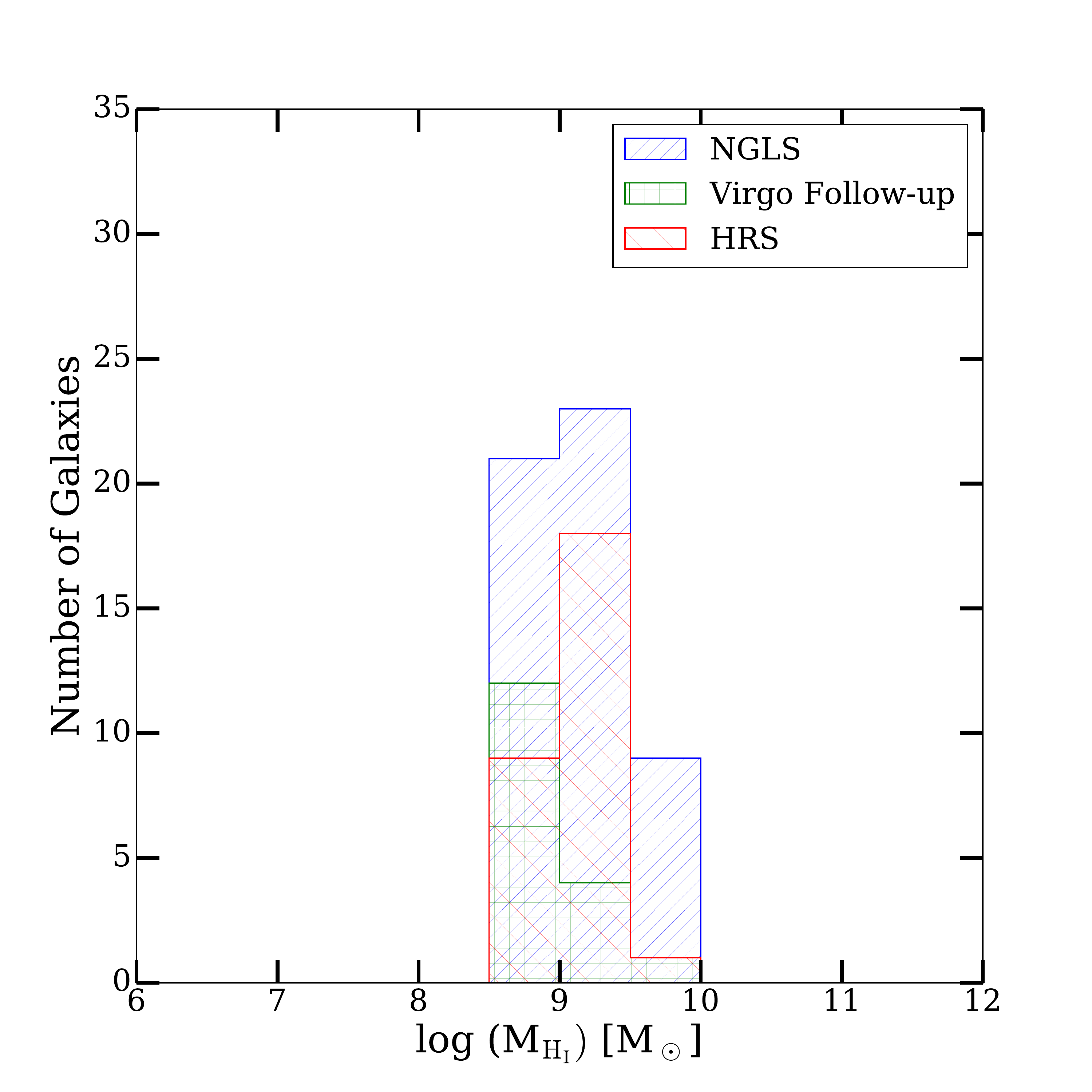}
	\includegraphics[width=8.0cm]{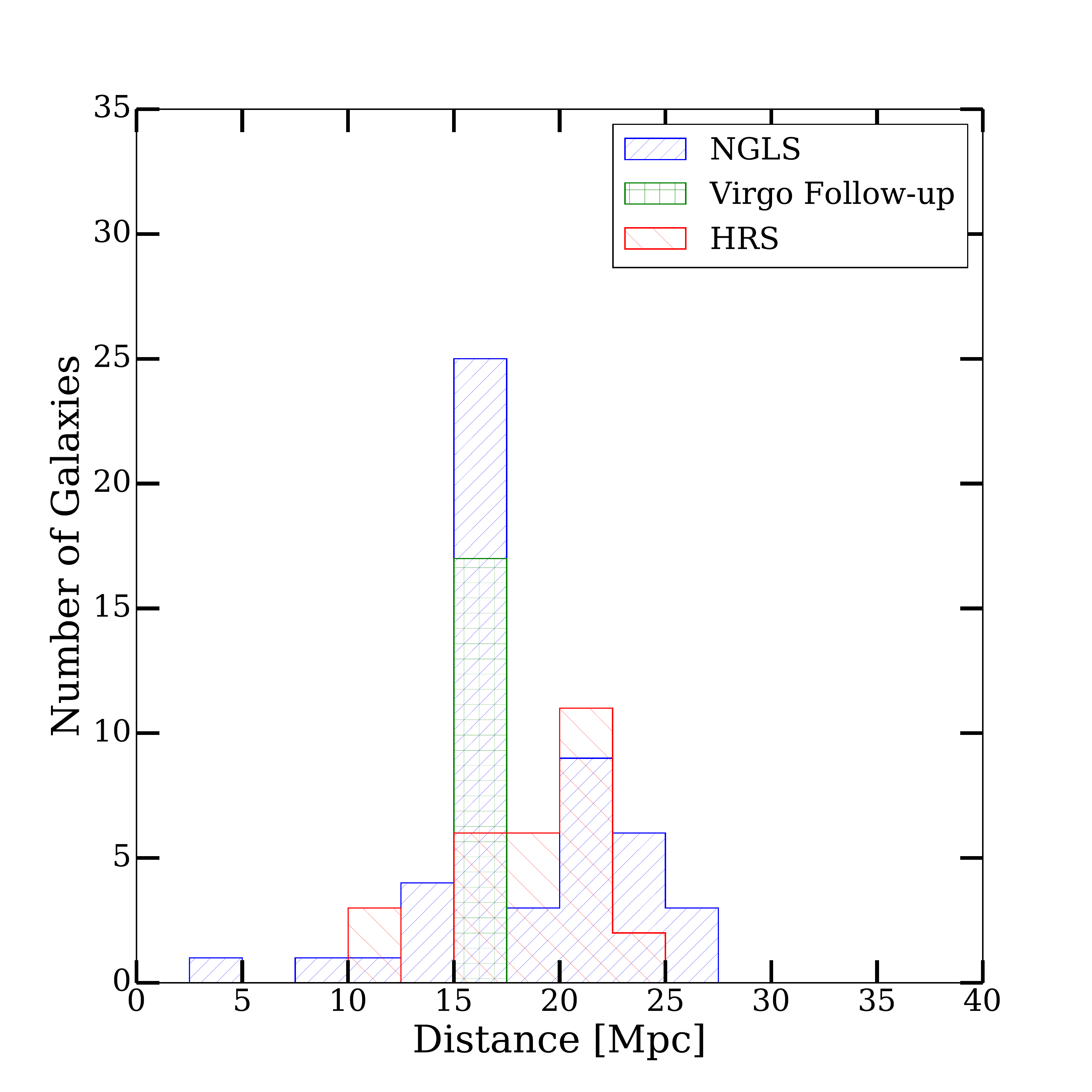}
	\caption{We present histograms of selected properties of the three sample sources, the NGLS, Virgo follow-up sample, and the additional HRS galaxies. The properties presented are stellar mass (top left), sSFR (top right), \hi mass (bottom left), and distance (bottom right).}
	\label{fig-app-comp}
\end{figure*}

We present plots of selected properties from the three observing programs that make up our sample in Figure~\ref{fig-app-comp}. There is a large difference in the distribution of distances between the three datasets, as the NGLS and Virgo follow-up have an obvious peak at our assumed Virgo distance of 16.7 Mpc. For the atomic hydrogen mass, there is only a significant difference between the distributions of the Virgo follow-up and the HRS ($p=0.003$) datasets using the \ks test, likely due to the Virgo follow-up program only containing Virgo galaxies. For the sSFR, we only find a small difference between the distributions for the NGLS and the HRS datasets ($p=0.036$). For the stellar mass distributions, we find no significant differences using the \ks test. Finally, for the molecular gas mass, we use the log-rank test to find the only significant difference is between the NGLS and the HRS datasets ($p=0.003$), where the resulting mean molecular gas mass is lower in the HRS dataset.
\par
Any differences between the samples can be attributed to the small numbers of galaxies in each dataset and to the percentage of galaxies in each environment, since our sample criteria is very similar between the three datasets. For example, the original NGLS dataset contain roughly equal numbers of group and Virgo galaxies, with a smaller number of field galaxies. The Virgo follow-up only contains Virgo galaxies. The HRS, which contains galaxies not already in the sample from the NGLS and Virgo follow-up, is skewed towards group galaxies. Given that the three datasets trace the environment differently, that is likely one of the main causes of the observed variations.
\newpage
\section{Selected Properties of the Galaxies in our Sample}\label{app-prop}
\begin{table*}
	\begin{minipage}{160mm}
	\caption{Selected Properties of Field Galaxies \label{tbl-list_isolated}}
	\begin{tabular}{lccccccccccccc}
	\hline
	Name & Type\footnotemark[1] & $D_{25}$\footnotemark[1] & Distance\footnotemark[2] &$\Delta T$\footnotemark[3] & $L_{CO(3-2)}$\footnotemark[4] & $\log(M_{H{\sc I}})$\footnotemark[5] & $\log(M_{H_2})$\footnotemark[6] & $\log (M_*)$\footnotemark[7] & SFR\footnotemark[8] \\
	& & (kpc) & (Mpc) & (mK) & ($10^7$ K km s$^{-1}$ pc$^2$) & ($M_{\odot}$) & ($M_{\odot}$) & ($M_{\odot}$) & ($M_{\odot}$ yr$^{-1}$) \\
	\hline
ESO477-016 & Sbc & $16.5$ & $24.2$ & $16$ & $<1.48$ & $9.06$ & $<8.72$ & $9.10$ & $0.18$ \\
ESO570-019 & Sc & $7.4$ & $18.4$ & $21$ & $<1.12$ & $9.13$ & $<8.60$ & $8.85$ & $0.14$ \\
IC1254 & SABb & $5.8$ & $21.9$ & $21$ & $<1.57$ & $8.90$ & $<8.75$ & $9.05^\dagger$ & $0.12$ \\
NGC0210 & SABb & $32.7$ & $22.4$ & $25$ & $1.46\pm0.47$ & $9.79$ & $8.42$ & $10.38^\dagger$ & $2.53$ \\
NGC6118 & Sc & $31.4$ & $24.2$ & $26$ & $<2.38$ & $9.54$ & $<8.93$ & $10.41$ & $2.44$ \\
NGC6140 & Sc & $10.8$ & $17.8$ & $20$ & $<0.99$ & $9.67$ & $<8.55$ & $9.66$ & $2.11$ \\
NGC7742 & Sb & $12.3$ & $24.8$ & $20$ & $4.58\pm0.84$ & $9.09$ & $8.91$ & $10.28$ & $0.46$ \\
PGC045195 & Sd & $21.9$ & $20.7$ & $16$ & $<1.09$ & $9.48$ & $<8.59$ & $9.32$ & $1.60$ \\
PGC057723 & SABb & $7.9$ & $14.9$ & $20$ & $<0.69$ & $9.31$ & $<8.39$ & $9.42^\dagger$ & $1.04$ \\
UGC06378 & Sc & $13.0$ & $23.0$ & $19$ & $<1.55$ & $9.12$ & $<8.74$ & $8.82$ & $0.18$ \\
UGC06792 & Sc & $10.1$ & $15.5$ & $23$ & $<0.84$ & $8.79$ & $<8.48$ & $8.63$ & $0.16$ \\
NGC4013 & Sb & $22.1$ & $15.5$ & $19$ & $8.08\pm1.04$ & $9.15$ & $9.16$ & $10.32$ & $0.46$ \\
NGC3437** & SABc & $13.7$ & $20.1$ & $23$ & $1.23\pm0.23$ & $9.11$ & $8.34$ & $10.04$ & $0.70$ \\
NGC3485** & Sb & $13.9$ & $21.9$ & $22$ & $<0.20$ & $9.46$ & $<7.84$ & $9.82$ & $0.36$ \\
NGC3501** & Sc & $22.1$ & $17.8$ & $25$ & $0.53\pm0.15$ & $8.99$ & $7.97$ & $9.76$ & $0.05$ \\
NGC3526** & Sc & $14.9$ & $21.3$ & $16$ & $<0.14$ & $8.95$ & $<7.70$ & $9.31$ & $0.12$ \\
NGC3666** & SBc & $16.8$ & $16.7$ & $17$ & $0.70\pm0.16$ & $9.36$ & $8.09$ & $9.75$ & $0.29$ \\
	\hline
	\end{tabular}
	\begin{tabular}{l}
	Note: ** next to name indicates that the galaxy is from the HRS program\\
	\footnotemark[1] Morphologies and D$_{25}$ extracted from HyperLeda database\\
	\footnotemark[2] Distances extracted from HyperLeda, corrected for Virgo infall and assuming $H_o = 70$ km s$^{-1}$ Mpc$^{-1}$, $\Omega_M = 0.27$, $\Omega_\Lambda = 0.73$\\
	\footnotemark[3] RMS noise in individual spectra in the data cube at $20$ km s$^{-1}$ resolution on $T_{MB}$ scale\\
	\footnotemark[4] Upper limits are $2\sigma$ limits calculated over an area of $1'$ and a line width of $100$ km s$^{-1}$\\
	\footnotemark[5] $M_{H{\sc I}}$ calculated using values for \hi flux from HyperLeda database\\
	\footnotemark[6] $M_{H_2}$ calculated assuming a CO $J=3-2/J=1-0$ line ratio of $0.18$\\
	\footnotemark[7] $\log (M_*)$ from the S$^4$G survey \citep{Sheth10}, except for galaxies with $^\dagger$ symbol, where $\log (M_*)$ is from K-band luminosity assuming \\ stellar mass-to-light ratio of $0.533$ \citep{Portinari04}\\
	\footnotemark[8] Star formation rates from \citet{SanchezGallego12} for NGLS galaxies, \citet{Boselli15} for HRS galaxies \\
	\end{tabular}
	\end{minipage}
\end{table*}
\begin{table*}
	\begin{minipage}{175mm}
	\caption{Selected Properties of Group Galaxies \label{tbl-list_group}}
	\begin{tabular}{lccccccccccccc}
	\hline
	Name & Type\footnotemark[1] & $D_{25}$\footnotemark[1] & Distance\footnotemark[2] &$\Delta T$\footnotemark[3] & $L_{CO(3-2)}$\footnotemark[4] & $\log(M_{H{\sc I}})$\footnotemark[5] & $\log(M_{H_2})$\footnotemark[6] & $\log (M_*)$\footnotemark[7] & SFR\footnotemark[8] & Group ID \footnotemark[9] \\
	& & (kpc) & (Mpc) & (mK) & ($10^7$ K km s$^{-1}$ pc$^2$) & ($M_{\odot}$) & ($M_{\odot}$) & ($M_{\odot}$) & ($M_{\odot}$ yr$^{-1}$) & \\
	\hline
IC0750 & Sab & $8.3$ & $13.7$ & $25$ & $13.41\pm0.97$ & $8.92$ & $9.38$ & $9.99$ & $0.26$ & $269$ \\
IC1066 & Sb & $8.5$ & $24.2$ & $19$ & $<1.71$ & $9.02$ & $<8.79$ & $9.53$ & $0.40$ & $387$ \\
NGC0450 & SABc & $21.8$ & $25.4$ & $20$ & $0.85\pm0.22$ & $9.46$ & $8.18$ & $9.75$ & $1.13$ & P \\
NGC0615 & Sb & $23.8$ & $25.9$ & $15$ & $<1.60$ & $9.42$ & $<8.75$ & $10.34$ & $0.90$ & $27$ \\
NGC1140 & SBm & $11.9$ & $20.1$ & $15$ & $<0.95$ & $9.39$ & $<8.53$ & $9.45$ & $1.19$ & $71$ \\
NGC1325 & SBbc & $26.3$ & $20.7$ & $21$ & $<1.40$ & $9.29$ & $<8.70$ & $10.05$ & $0.63$ & $97$ \\
NGC2146A & SABc & $20.3$ & $25.9$ & $24$ & $<2.49$ & $9.56$ & $<8.95$ & $9.69^\dagger$ & $0.21$ & P \\
NGC2742 & Sc & $18.8$ & $22.4$ & $24$ & $2.53\pm0.83$ & $9.27$ & $8.65$ & $10.21$ & $1.06$ & $167$ \\
NGC3077 & S? & $6.0$ & $3.9$ & $26$ & $0.28\pm0.05$ & $9.12$ & $7.69$ & $9.28$ & $0.03$ & $176$ \\
NGC3162 & SABc & $12.6$ & $20.7$ & $22$ & $2.80\pm0.66$ & $9.41$ & $8.70$ & $9.91$ & $2.36$ & $194$ \\
NGC3227 & SABa & $21.3$ & $18.4$ & $25$ & $11.74\pm1.29$ & $8.99$ & $9.32$ & $10.41$ & $0.21$ & $194$ \\
NGC3254 & Sbc & $14.9$ & $21.9$ & $19$ & $<1.42$ & $9.56$ & $<8.70$ & $10.09$ & $1.25$ & $197$ \\
NGC3353 & Sb & $6.7$ & $17.2$ & $26$ & $<1.21$ & $8.88$ & $<8.64$ & $9.18$ & $0.62$ & $201$ \\
NGC3507 & SBb & $13.3$ & $15.5$ & $21$ & $1.30\pm0.42$ & $8.95$ & $8.36$ & $9.96$ & $0.92$ & $228$ \\
NGC3782 & Scd & $5.1$ & $14.3$ & $21$ & $<0.52$ & $9.01$ & $<8.27$ & $9.00$ & $0.18$ & $258$ \\
NGC4041 & Sbc & $16.4$ & $21.9$ & $22$ & $30.09\pm2.52$ & $9.56$ & $9.73$ & $10.32$ & $3.99$ & $266$ \\
NGC4288 & SBcd & $5.7$ & $11.5$ & $21$ & $<0.34$ & $8.92$ & $<8.09$ & $8.92$ & $0.05$ & $269$ \\
NGC4504 & SABc & $13.7$ & $14.9$ & $18$ & $<0.61$ & $9.51$ & $<8.34$ & $9.67$ & $0.45$ & $293$ \\
NGC4772 & Sa & $19.1$ & $16.1$ & $17$ & $<0.54$ & $8.82$ & $<8.28$ & $10.04$ & $0.08$ & $292$ \\
NGC5477 & Sm & $2.7$ & $8.6$ & $22$ & $<0.25$ & $8.80$ & $<7.95$ & $8.14$ & $0.03$ & $371$ \\
NGC5486 & Sm & $9.5$ & $24.2$ & $16$ & $<1.49$ & $9.18$ & $<8.72$ & $9.15$ & $0.31$ & $373$ \\
IC3908** & SBcd & $13.6$ & $19.0$ & $13$ & $2.56\pm0.25$ & $8.96$ & $8.66$ & $9.92$ & $0.21$ & $314$ \\
NGC3346** & SBc & $14.9$ & $19.5$ & $19$ & $0.41\pm0.13$ & $9.09$ & $7.87$ & $9.88$ & $0.28$ & $214$ \\
NGC3370** & Sc & $14.4$ & $20.1$ & $26$ & $<0.27$ & $9.26$ & $<7.98$ & $9.86$ & $0.52$ & $219$ \\
UGC06023** & SBcd & $10.3$ & $21.3$ & $24$ & $<0.25$ & $8.98$ & $<7.94$ & $9.46$ & $0.28$ & $227$ \\
NGC3455** & SABb & $11.5$ & $17.2$ & $22$ & $<0.13$ & $8.96$ & $<7.67$ & $9.28$ & $0.22$ & $219$ \\
NGC3681** & Sbc & $9.9$ & $19.5$ & $20$ & $<0.16$ & $9.37$ & $<7.75$ & $9.93$ & $0.17$ & $237$ \\
NGC3684** & Sbc & $12.3$ & $18.4$ & $22$ & $0.68\pm0.20$ & $9.37$ & $8.08$ & $9.80$ & $0.38$ & $237$ \\
NGC3756** & SABb & $12.4$ & $22.4$ & $26$ & $<0.29$ & $9.36$ & $<8.02$ & $10.17$ & $0.34$ & $250$ \\
NGC3795** & Sbc & $12.9$ & $21.3$ & $26$ & $<0.26$ & $8.89$ & $<7.97$ & $9.38$ & $0.09$ & $244$ \\
NGC3982** & SABb & $12.5$ & $20.1$ & $26$ & $4.76\pm0.81$ & $9.22$ & $8.93$ & $10.03$ & $0.94$ & $250$ \\
NGC4123** & Sc & $18.5$ & $20.1$ & $28$ & $0.49\pm0.16$ & $9.56$ & $7.94$ & $10.06$ & $0.63$ & $275$ \\
NGC4668** & SBcd & $11.2$ & $24.2$ & $24$ & $<0.28$ & $9.11$ & $<8.00$ & $9.42$ & $0.33$ & $299$ \\
NGC4688** & Sc & $16.8$ & $15.5$ & $29$ & $<0.11$ & $9.15$ & $<7.61$ & $9.43$ & $0.28$ & $292$ \\
NGC4701** & Sc & $5.8$ & $11.5$ & $26$ & $<0.07$ & $9.02$ & $<7.41$ & $9.19$ & $0.19$ & $292$ \\
NGC4713** & Scd & $5.3$ & $10.9$ & $12$ & $0.22\pm0.06$ & $9.03$ & $7.60$ & $9.33$ & $0.35$ & $315$ \\
NGC4771** & Sc & $15.5$ & $17.2$ & $13$ & $0.23\pm0.07$ & $8.94$ & $7.61$ & $9.82$ & $0.17$ & $315$ \\
NGC4775** & Scd & $15.0$ & $23.0$ & $12$ & $1.33\pm0.42$ & $9.46$ & $8.37$ & $9.92$ & $1.26$ & $314$ \\
NGC4808** & Sc & $8.2$ & $12.0$ & $25$ & $0.76\pm0.17$ & $9.28$ & $8.13$ & $9.53$ & $0.37$ & $315$ \\
UGC06575** & Sc & $11.0$ & $21.3$ & $27$ & $<0.26$ & $9.15$ & $<7.96$ & $8.98$ & $0.13$ & $244$ \\
UGC07982** & Sc & $14.9$ & $17.8$ & $13$ & $<0.10$ & $8.69$ & $<7.57$ & $9.41$ & $0.07$ & $315$ \\
UGC08041** & SBcd & $18.5$ & $20.1$ & $12$ & $<0.12$ & $9.19$ & $<7.62$ & $9.46$ & $0.16$ & $315$ \\
	\hline
	\end{tabular}
	\begin{tabular}{l}
	Note: ** next to name indicates that the galaxy is from the HRS program\\
	\footnotemark[1] Morphologies and D$_{25}$ extracted from HyperLeda database\\
	\footnotemark[2] Distances extracted from HyperLeda, corrected for Virgo infall and assuming $H_o = 70$ km s$^{-1}$ Mpc$^{-1}$, $\Omega_M = 0.27$, $\Omega_\Lambda = 0.73$\\
	\footnotemark[3] RMS noise in individual spectra in the data cube at $20$ km s$^{-1}$ resolution on $T_{MB}$ scale\\
	\footnotemark[4] Upper limits are $2\sigma$ limits calculated over an area of $1'$ and a line width of $100$ km s$^{-1}$\\
	\footnotemark[5] $M_{H{\sc I}}$ calculated using values for \hi flux from HyperLeda database\\
	\footnotemark[6] $M_{H_2}$ calculated assuming a CO $J=3-2/J=1-0$ line ratio of $0.18$\\
	\footnotemark[7] $\log (M_*)$ from the S$^4$G survey \citep{Sheth10}, except for galaxies with $^\dagger$ symbol, where $\log (M_*)$ is from K-band luminosity assuming \\ stellar mass-to-light ratio of $0.533$ \citep{Portinari04}\\
	\footnotemark[8] Star formation rates from \citet{SanchezGallego12} for NGLS galaxies, \citet{Boselli15} for HRS galaxies\\
	\footnotemark[9] Group IDs from \citet{Garcia93b}, while P indicates pairs from \citet{Karachentsev72}\\ 
	\end{tabular}
	\end{minipage}
\end{table*}
\begin{table*}
	\begin{minipage}{160mm}
	\caption{Selected Properties of Virgo Galaxies \label{tbl-list_virgo}}
	\begin{tabular}{lccccccccccccccccc}
	\hline
	Name & Type\footnotemark[1] & $D_{25}$\footnotemark[1] & Distance\footnotemark[2] &$\Delta T$\footnotemark[3] & $L_{CO(3-2)}$\footnotemark[4] & $\log(M_{H{\sc I}})$\footnotemark[5] & $\log(M_{H_2})$\footnotemark[6] & $\log (M_*)$\footnotemark[7] & SFR\footnotemark[8] \\
	& & (kpc) & (Mpc) & (mK) & ($10^7$ K km s$^{-1}$ pc$^2$) & ($M_{\odot}$) & ($M_{\odot}$) & ($M_{\odot}$) & ($M_{\odot}$ yr$^{-1}$) \\
	\hline
IC3061* & SBc & $22.9$ & $16.7$ & $22$ & $<0.83$ & $8.70$ & $<8.47$ & $9.12$ & $0.05$ \\
IC3074* & SBd & $2.9$ & $16.7$ & $19$ & $<0.84$ & $8.89$ & $<8.48$ & $8.61^\dagger$ & $0.03$ \\
IC3322A* & SBc & $8.8$ & $16.7$ & $21$ & $0.26\pm0.07$ & $9.09$ & $7.67$ & $9.55$ & $0.13$ \\
IC3371* & Sc & $5.1$ & $16.7$ & $21$ & $<0.81$ & $8.68$ & $<8.46$ & $8.29$ & $0.02$ \\
IC3576* & SBm & $11.1$ & $16.7$ & $21$ & $<0.89$ & $8.95$ & $<8.50$ & $8.77$ & $0.06$ \\
NGC4206 & Sbc & $16.3$ & $16.7$ & $14$ & $0.48\pm0.14$ & $9.38$ & $7.93$ & $9.63$ & $0.20$ \\
NGC4254 & Sc & $52.9$ & $16.7$ & $27$ & $73.80\pm2.51$ & $9.66$ & $10.12$ & $10.66$ & $11.31$ \\
NGC4273* & Sc & $22.4$ & $16.7$ & $23$ & $5.88\pm0.69$ & $8.95$ & $9.02$ & $9.85$ & $0.76$ \\
NGC4298 & Sc & $13.4$ & $16.7$ & $21$ & $6.56\pm1.16$ & $8.84$ & $9.07$ & $10.08$ & $0.27$ \\
NGC4303A & Sbc & $47.5$ & $16.7$ & $26$ & $52.64\pm5.72$ & $9.65$ & $9.97$ & $10.71$ & $3.29$ \\
NGC4302* & Sc & $32.3$ & $16.7$ & $14$ & $6.37\pm0.91$ & $9.28$ & $9.05$ & $10.29$ & $0.32$ \\
NGC4303* & Sbc & $47.5$ & $16.7$ & $26$ & $52.64\pm5.72$ & $9.65$ & $9.97$ & $10.71$ & $3.29$ \\
NGC4316* & Sc & $13.9$ & $16.7$ & $22$ & $0.56\pm0.16$ & $8.70$ & $8.00$ & $9.71$ & $0.08$ \\
NGC4330* & Sc & $16.9$ & $16.7$ & $24$ & $0.41\pm0.14$ & $8.68$ & $7.87$ & $9.59$ & $0.05$ \\
NGC4383 & Sa & $12.8$ & $16.7$ & $20$ & $1.13\pm0.22$ & $9.16$ & $8.30$ & $9.65$ & $0.39$ \\
NGC4390 & SABc & $7.6$ & $16.7$ & $19$ & $<0.88$ & $8.63$ & $<8.50$ & $9.26$ & $0.09$ \\
NGC4411A* & Sc & $9.5$ & $16.7$ & $23$ & $<0.80$ & $8.65$ & $<8.46$ & $9.08$ & $0.05$ \\
NGC4411B* & SABc & $10.8$ & $16.7$ & $22$ & $<0.99$ & $8.91$ & $<8.55$ & $9.30$ & $0.09$ \\
NGC4423 & Sd & $10.2$ & $16.7$ & $18$ & $0.16\pm0.05$ & $8.91$ & $7.44$ & $8.97$ & $0.06$ \\
NGC4430 & Sb & $13.3$ & $16.7$ & $17$ & $1.22\pm0.40$ & $8.62$ & $8.34$ & $9.63$ & $0.12$ \\
NGC4470 & Sa & $13.2$ & $16.7$ & $19$ & $<0.84$ & $8.68$ & $<8.47$ & $9.43$ & $0.03$ \\
NGC4480* & SABc & $20.1$ & $16.7$ & $20$ & $0.77\pm0.24$ & $8.87$ & $8.14$ & $9.51$ & $0.20$ \\
NGC4498* & Sc & $19.8$ & $16.7$ & $24$ & $<0.78$ & $8.77$ & $<8.14$ & $9.52$ & $0.24$ \\
NGC4519* & Scd & $13.0$ & $16.7$ & $22$ & $<1.12$ & $9.49$ & $<8.30$ & $9.57$ & $0.39$ \\
NGC4522 & SBc & $35.5$ & $16.7$ & $13$ & $1.39\pm0.21$ & $8.72$ & $8.39$ & $9.64$ & $0.06$ \\
NGC4548 & Sb & $14.7$ & $16.7$ & $35$ & $<0.33$ & $8.86$ & $<8.07$ & $10.57$ & $0.32$ \\
NGC4561 & SBcd & $8.6$ & $16.7$ & $26$ & $<1.17$ & $9.11$ & $<8.62$ & $9.18$ & $0.49$ \\
NGC4567 & Sbc & $27.2$ & $16.7$ & $20$ & $7.58\pm0.99$ & $9.02$ & $9.13$ & $10.03$ & $0.14$ \\
NGC4568 & Sbc & $42.2$ & $16.7$ & $19$ & $26.30\pm1.98$ & $8.88$ & $9.67$ & $10.38$ & $0.29$ \\
NGC4579 & SABb & $34.4$ & $16.7$ & $20$ & $7.89\pm2.51$ & $8.79$ & $9.15$ & $10.80$ & $3.62$ \\
NGC4639 & Sbc & $13.5$ & $16.7$ & $20$ & $<0.28$ & $8.97$ & $<7.70$ & $9.91$ & $0.19$ \\
NGC4647 & SABc & $17.5$ & $16.7$ & $23$ & $12.11\pm1.86$ & $8.70$ & $9.33$ & $10.20^\dagger$ & $2.84$ \\
NGC4651 & Sc & $15.5$ & $16.7$ & $19$ & $4.10\pm0.79$ & $9.47$ & $8.86$ & $10.28$ & $1.15$ \\
NGC4654 & Sc & $22.7$ & $16.7$ & $12$ & $16.22\pm1.61$ & $9.49$ & $9.46$ & $10.35$ & $1.08$ \\
PGC040604* & SBm & $5.3$ & $16.7$ & $24$ & $<1.05$ & $8.64$ & $<8.57$ & $8.06^\dagger$ & $0.02$ \\
UGC07557* & SABm & $11.9$ & $16.7$ & $24$ & $<1.02$ & $9.03$ & $<8.56$ & $9.76$ & $0.07$ \\
UGC07590 & Sbc & $6.1$ & $16.7$ & $19$ & $<0.89$ & $8.87$ & $<8.50$ & $8.73$ & $0.02$ \\
NGC4294** & SBc & $11.7$ & $16.7$ & $24$ & $<0.15$ & $9.20$ & $<7.71$ & $9.49$ & $0.38$ \\
NGC4396** & Scd & $13.6$ & $16.7$ & $25$ & $<0.14$ & $8.87$ & $<7.68$ & $9.35^\dagger$ & $0.15$ \\
	\hline
	\end{tabular}
	\begin{tabular}{l}
	Note: * indicates galaxy is in the Virgo follow-up program, ** indicates that the galaxy is from the HRS program\\
	\footnotemark[1] Morphologies and D$_{25}$ extracted from HyperLeda database\\
	\footnotemark[2] Distances set to be at 16.7 Mpc \citep{Mei07}\\
	\footnotemark[3] RMS noise in individual spectra in the data cube at $20$ km s$^{-1}$ resolution on $T_{MB}$ scale\\
	\footnotemark[4] Upper limits are $2\sigma$ limits calculated over an area of $1'$ and a line width of $100$ km s$^{-1}$\\
	\footnotemark[5] $M_{H{\sc I}}$ calculated using values for \hi flux from HyperLeda database\\
	\footnotemark[6] $M_{H_2}$ calculated assuming a CO $J=3-2/J=1-0$ line ratio of $0.18$\\
	\footnotemark[7] $\log (M_*)$ from the S$^4$G survey \citep{Sheth10}, except for galaxies with $^\dagger$ symbol, where $\log (M_*)$ is from K-band luminosity assuming \\ stellar mass-to-light ratio of $0.533$ \citep{Portinari04}\\
	\footnotemark[8] Star formation rates from \citet{SanchezGallego12} for NGLS galaxies, GOLDMine database for the Virgo follow-up \citep{Gavazzi03},\\
	 and \citet{Boselli15} for HRS galaxies\\
	\end{tabular}
	\end{minipage}
\end{table*}
\clearpage
\section{CO $J=3-2$ Maps of NGLS Galaxies}\label{app-map}
\par
We present the CO $J=3-2$ maps of our sample of galaxies from the Nearby Galaxies Legacy Survey.
\clearpage
\begin{figure*}
	\includegraphics[height=3.75cm]{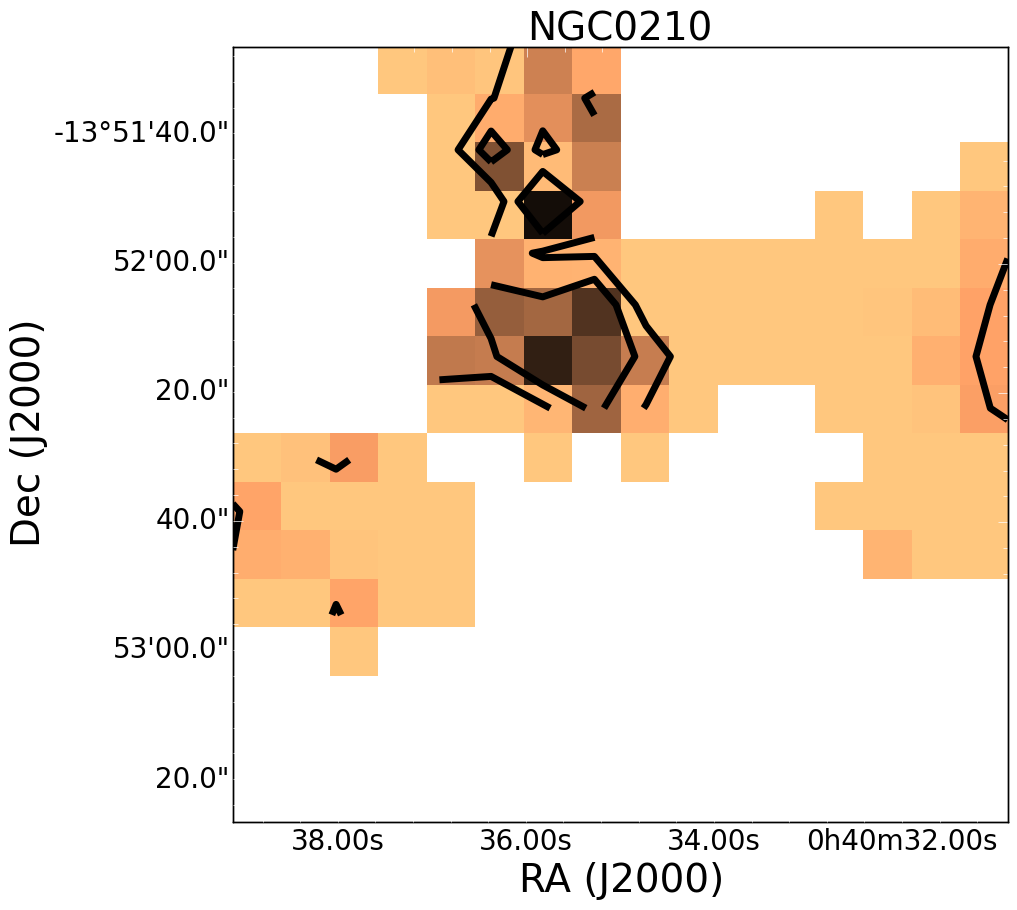}
	\includegraphics[height=3.75cm]{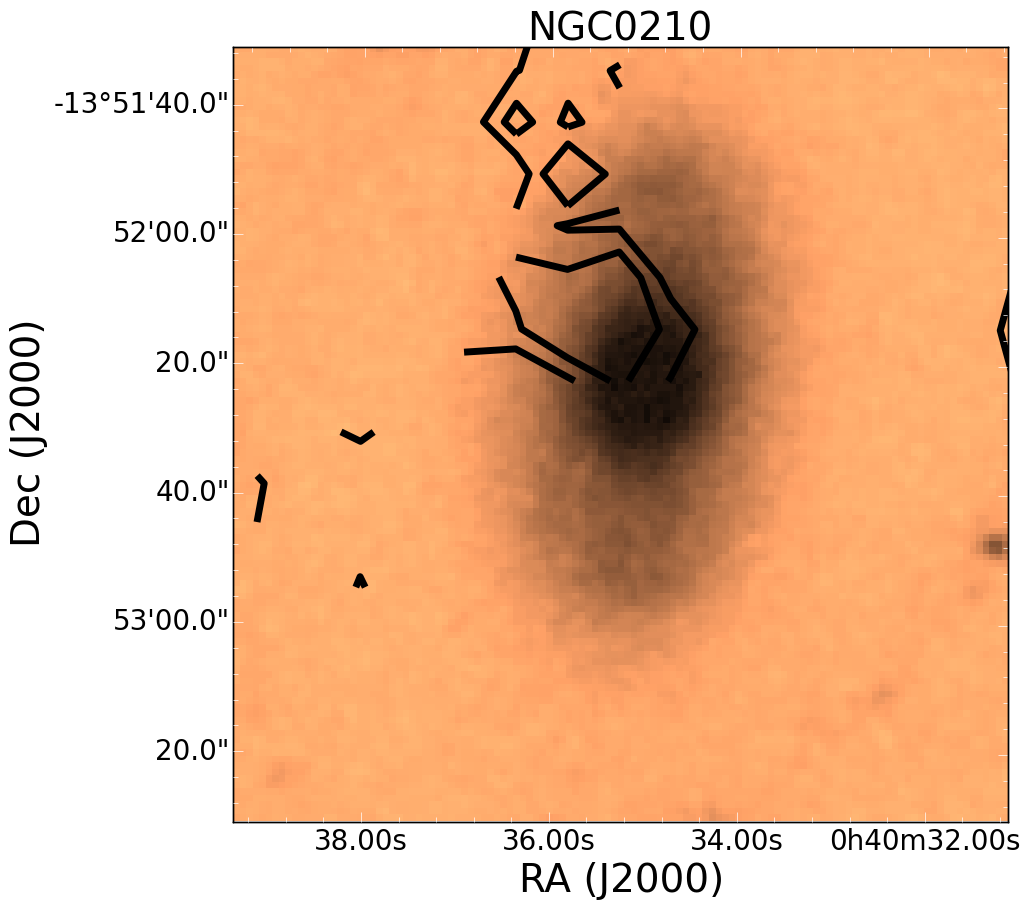}
	\includegraphics[height=3.75cm]{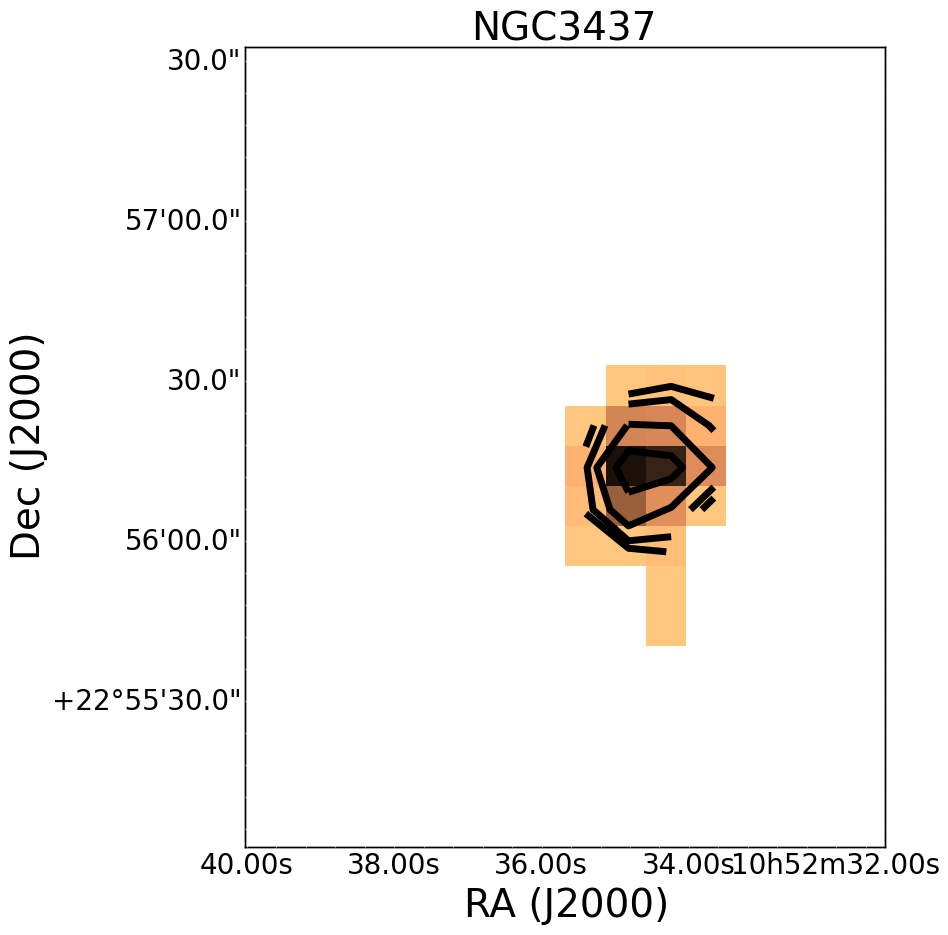}
	\includegraphics[height=3.75cm]{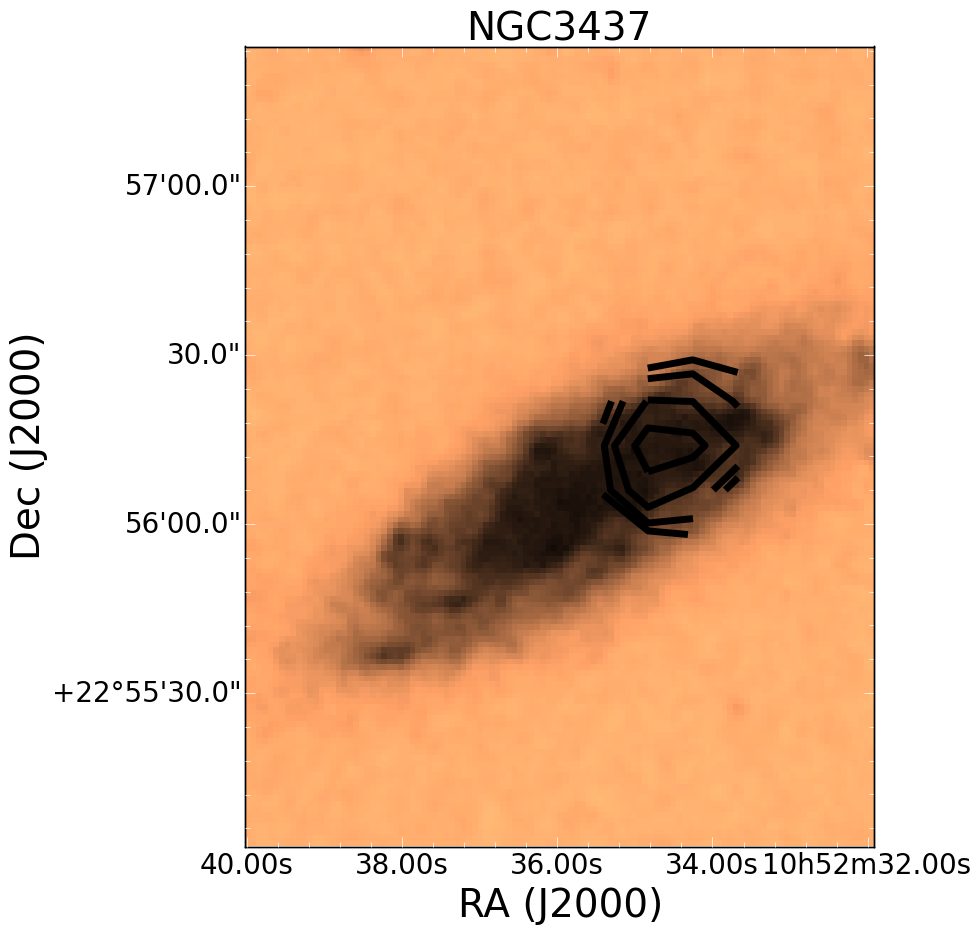}
	\includegraphics[height=3.75cm]{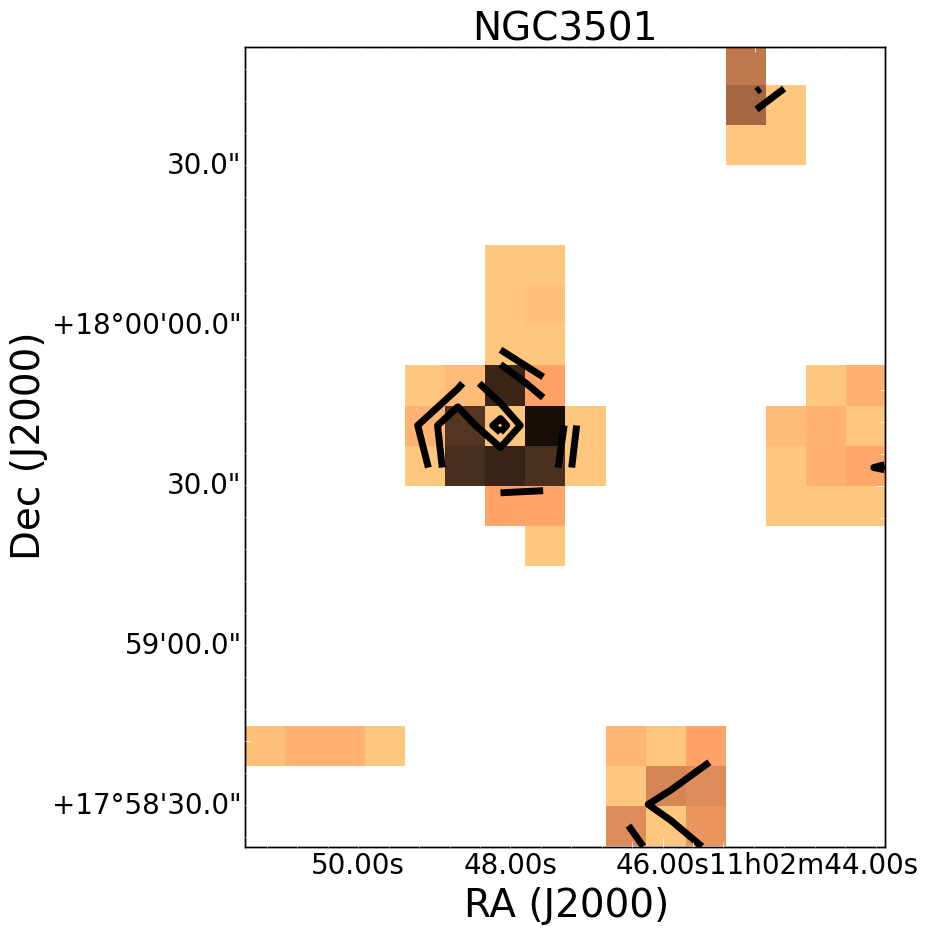}
	\includegraphics[height=3.75cm]{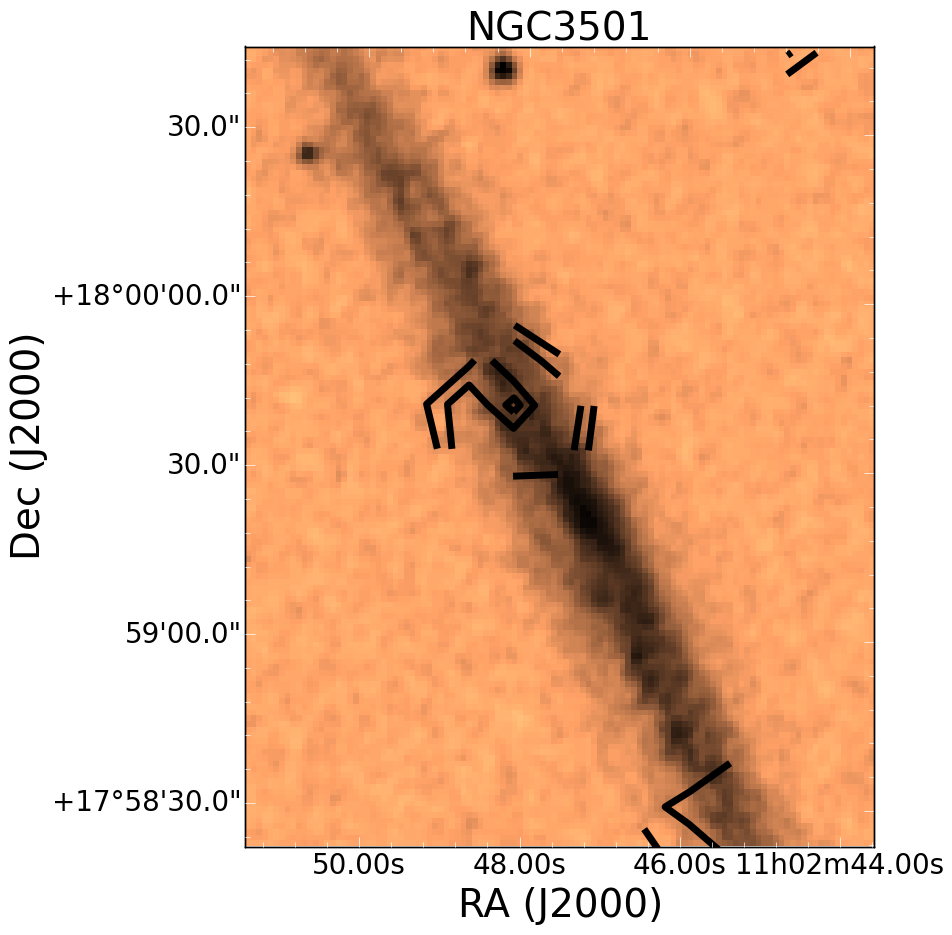}
	\includegraphics[height=3.75cm]{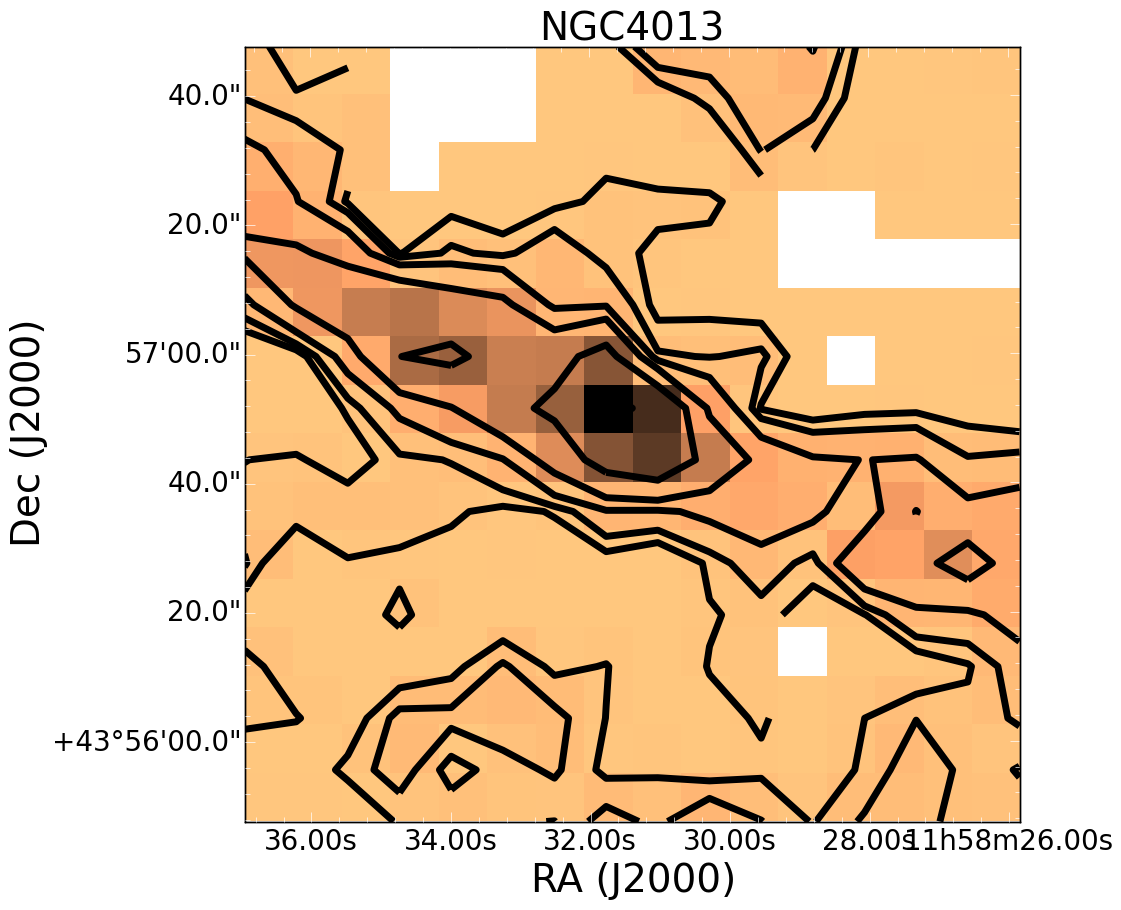}
	\includegraphics[height=3.75cm]{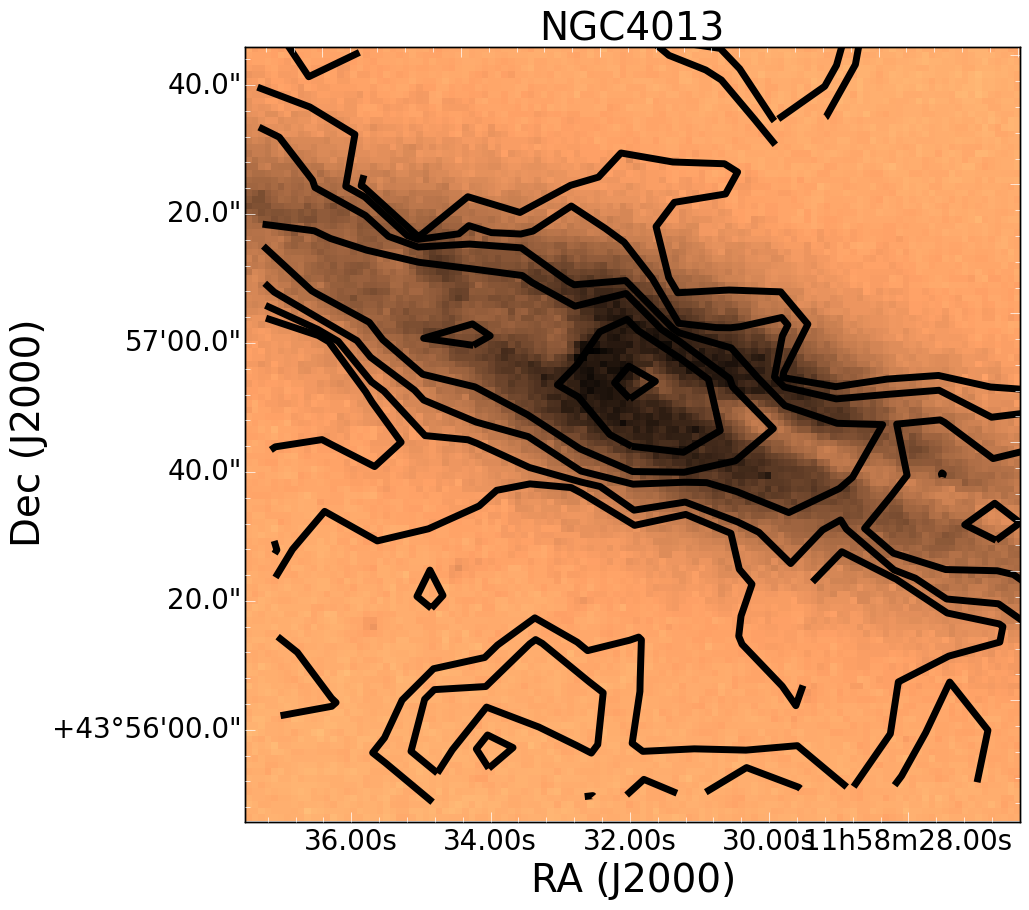}
	\includegraphics[height=3.75cm]{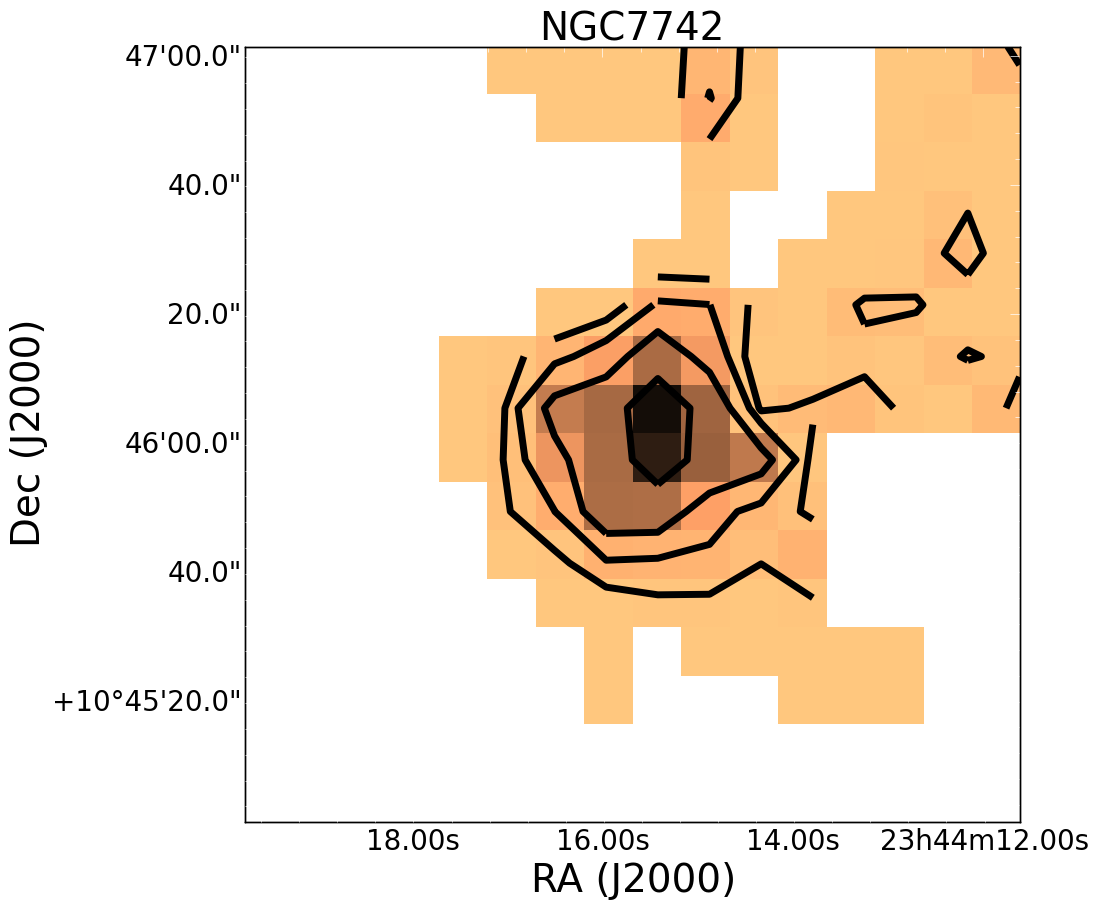}
	\includegraphics[height=3.75cm]{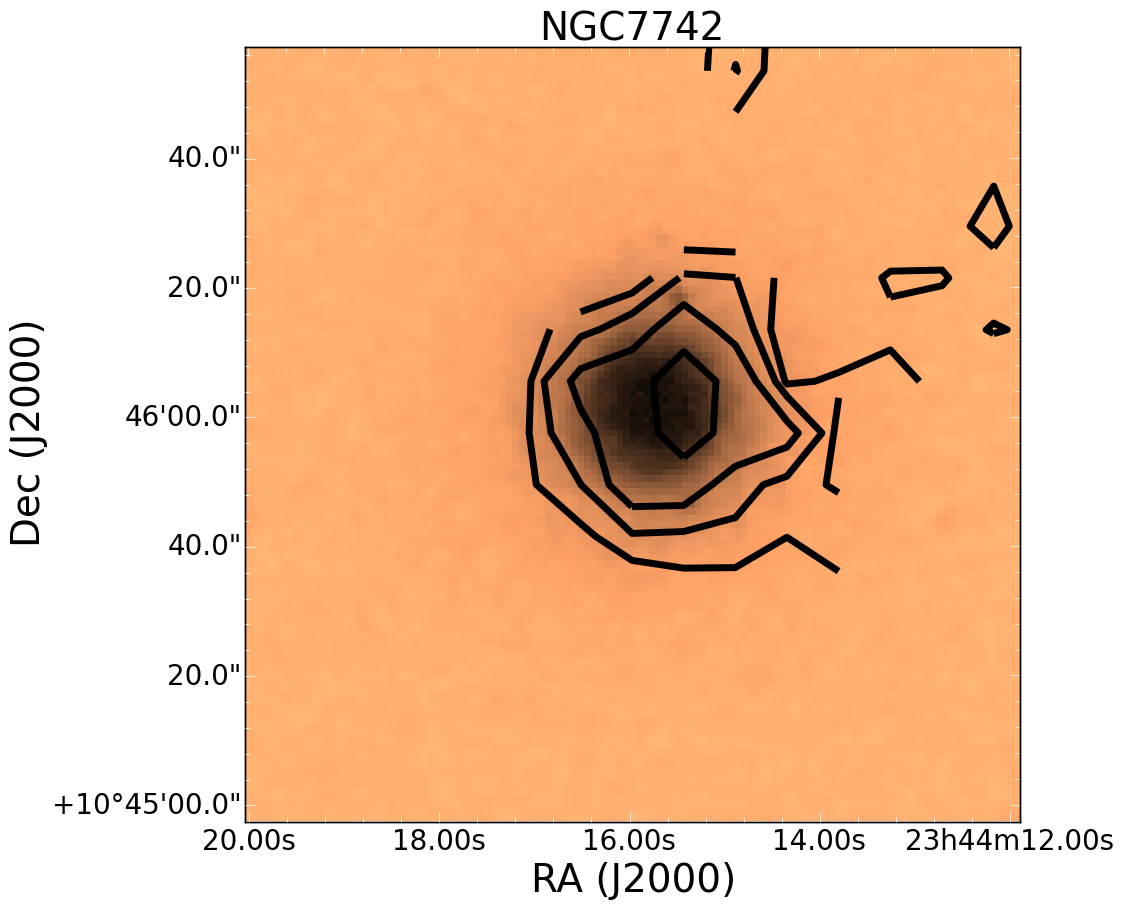}
	\caption{Images of the CO detected galaxies in the field sample. There are two plots for each galaxy. The first panel in each pair is the CO $J=3-2$ integrated galaxy map with black contour levels overlaid at 0.5, 1, and 2 K km s$^{-1}$. The second panel is the same contour levels overlaid on the optical image from the Digitized Sky Survey.}
	\label{fig-det_isolated}
\end{figure*}
\begin{figure*}
	\includegraphics[height=3.75cm]{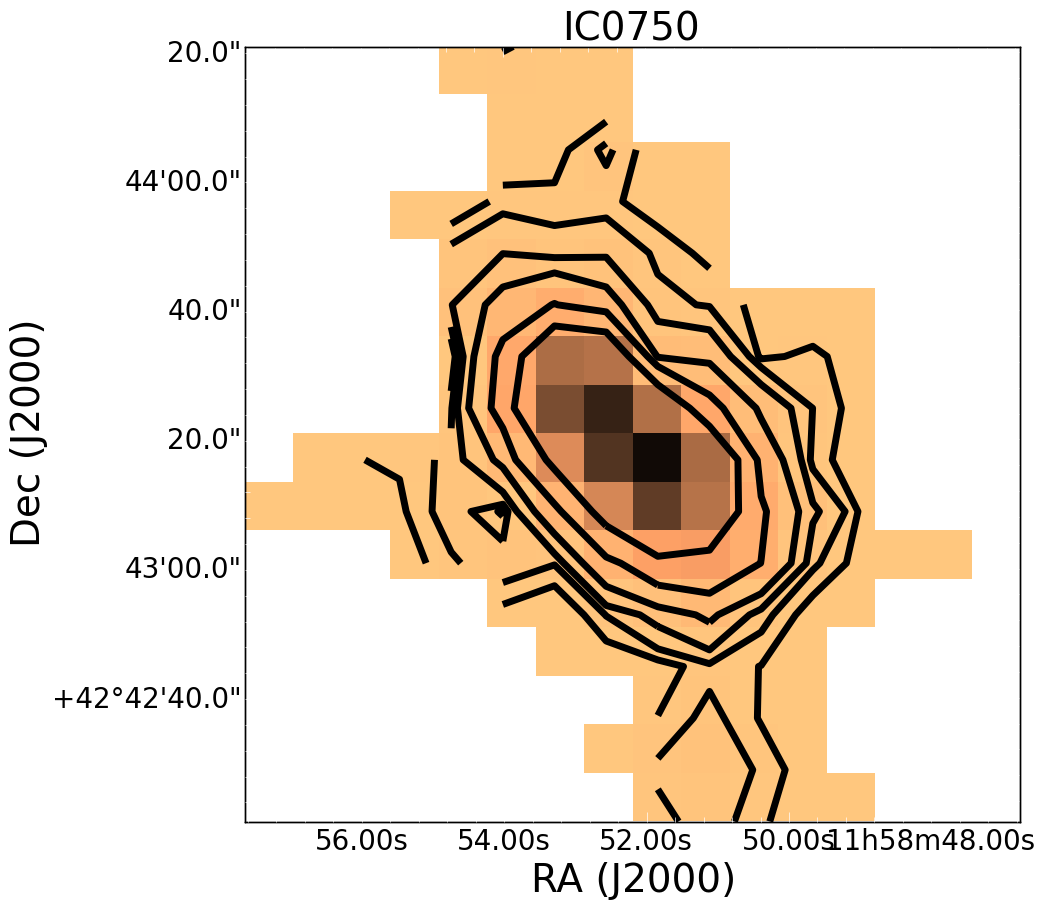}
	\includegraphics[height=3.75cm]{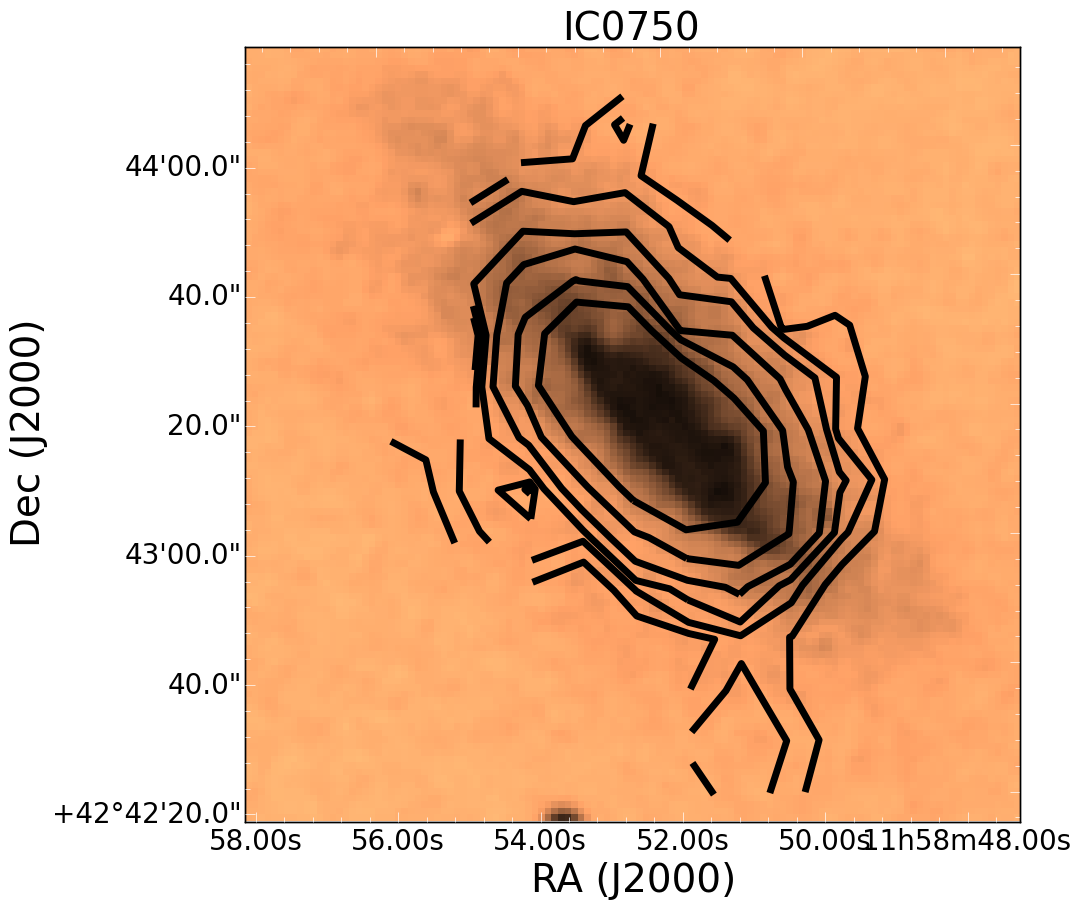}
	\includegraphics[height=3.75cm]{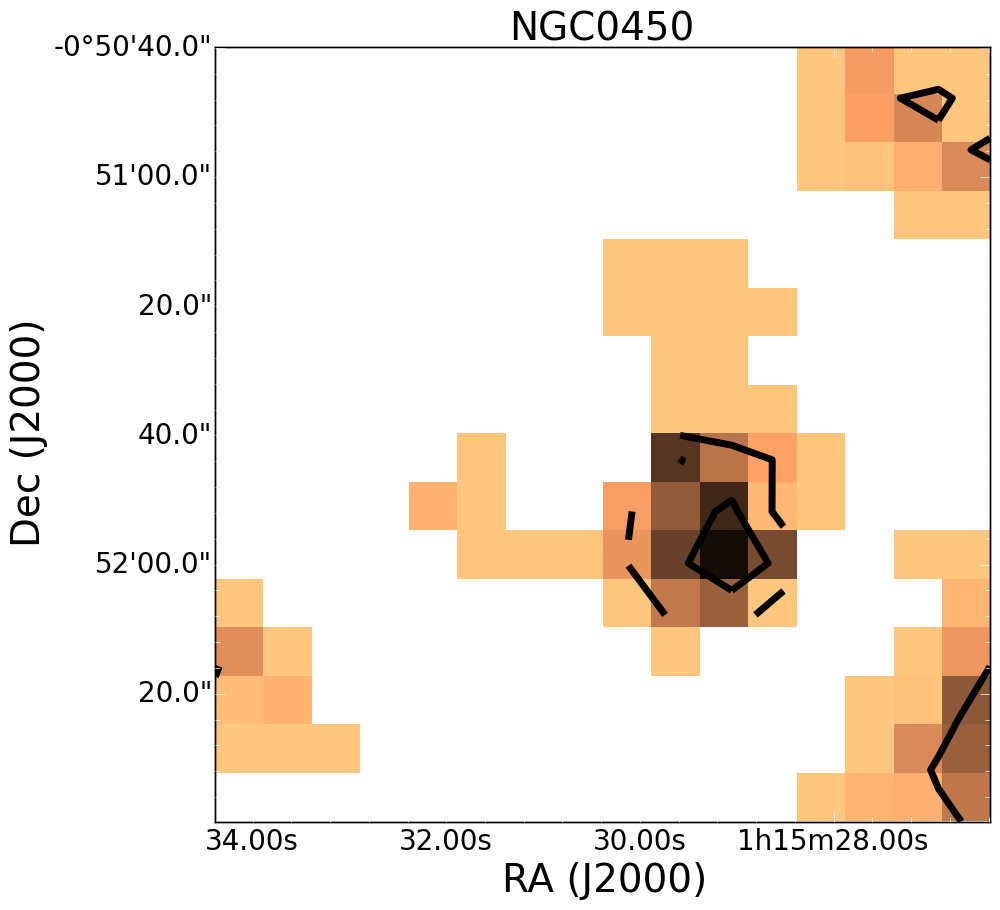}
	\includegraphics[height=3.75cm]{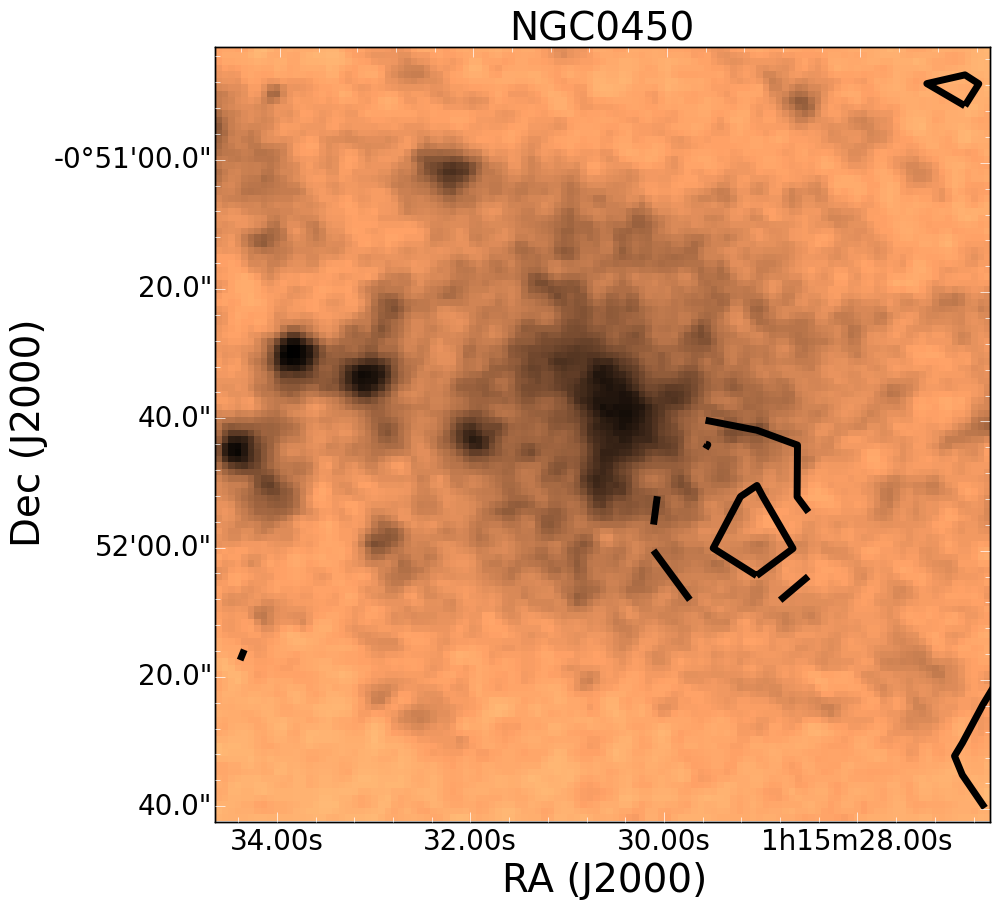}
	\includegraphics[height=3.75cm]{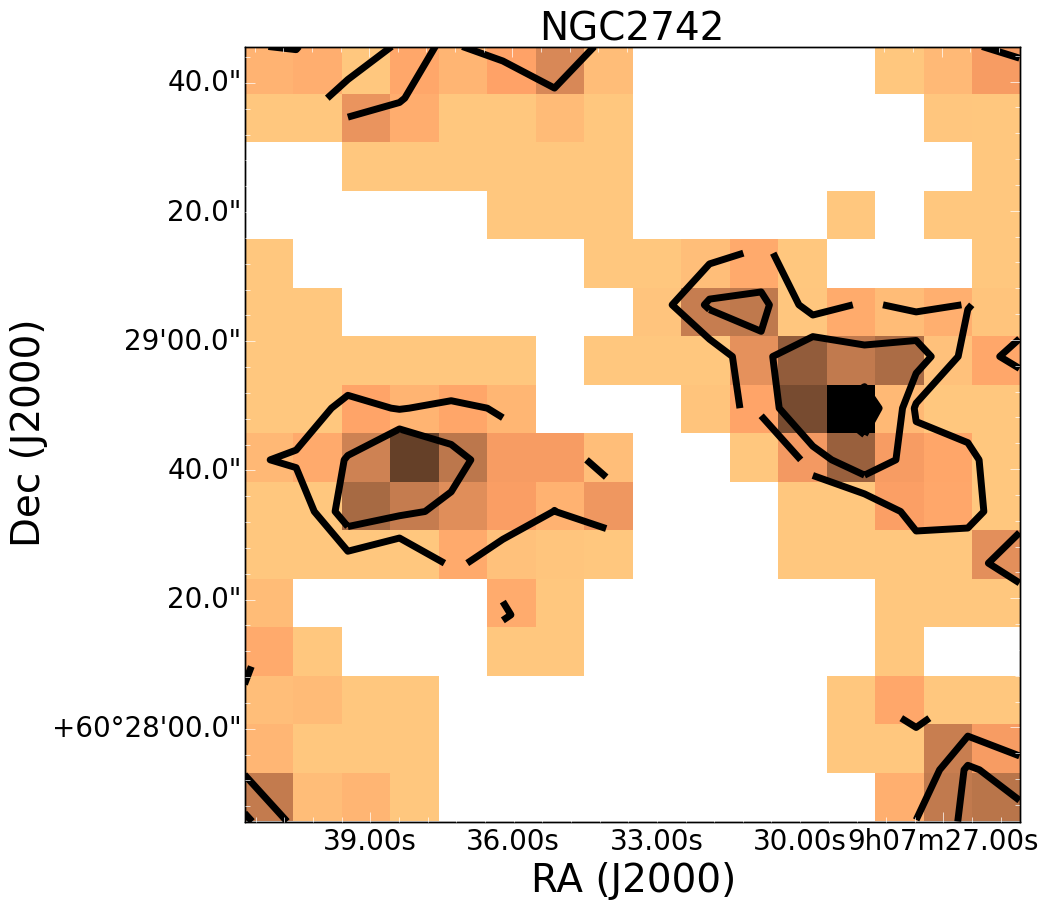}
	\includegraphics[height=3.75cm]{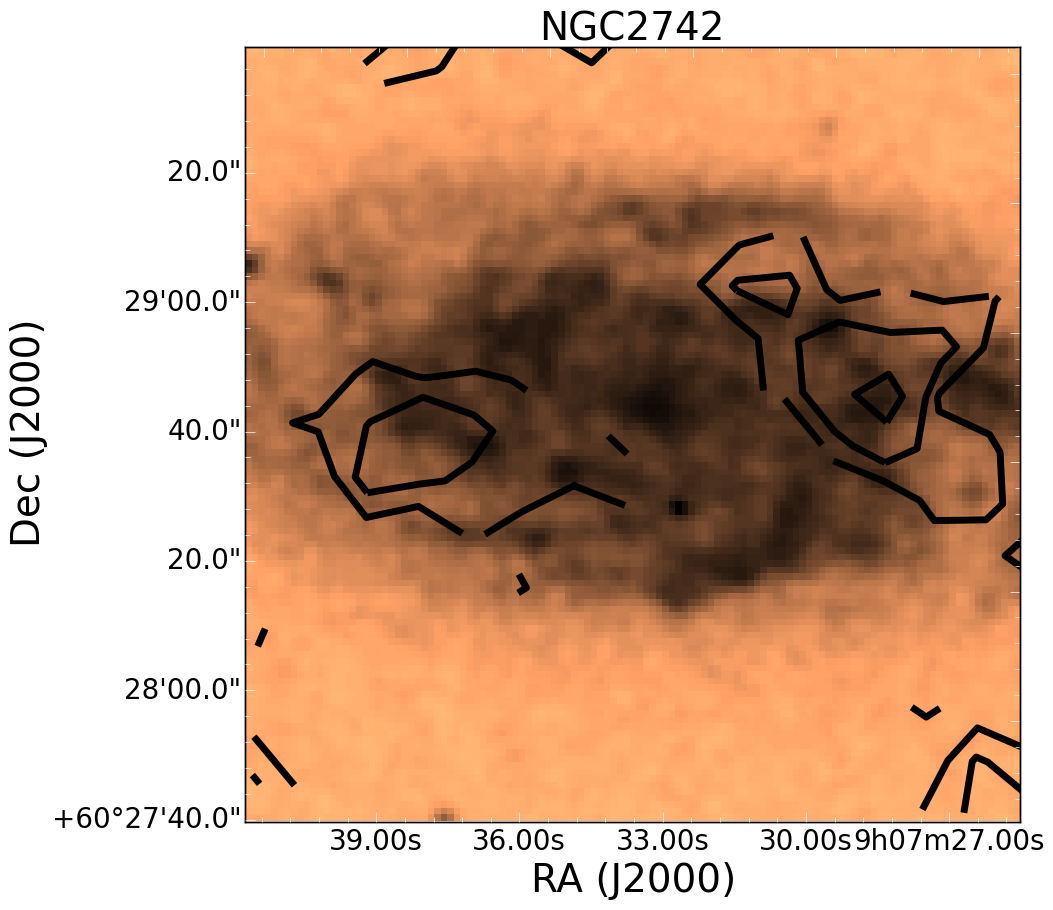}
	\includegraphics[height=3.75cm]{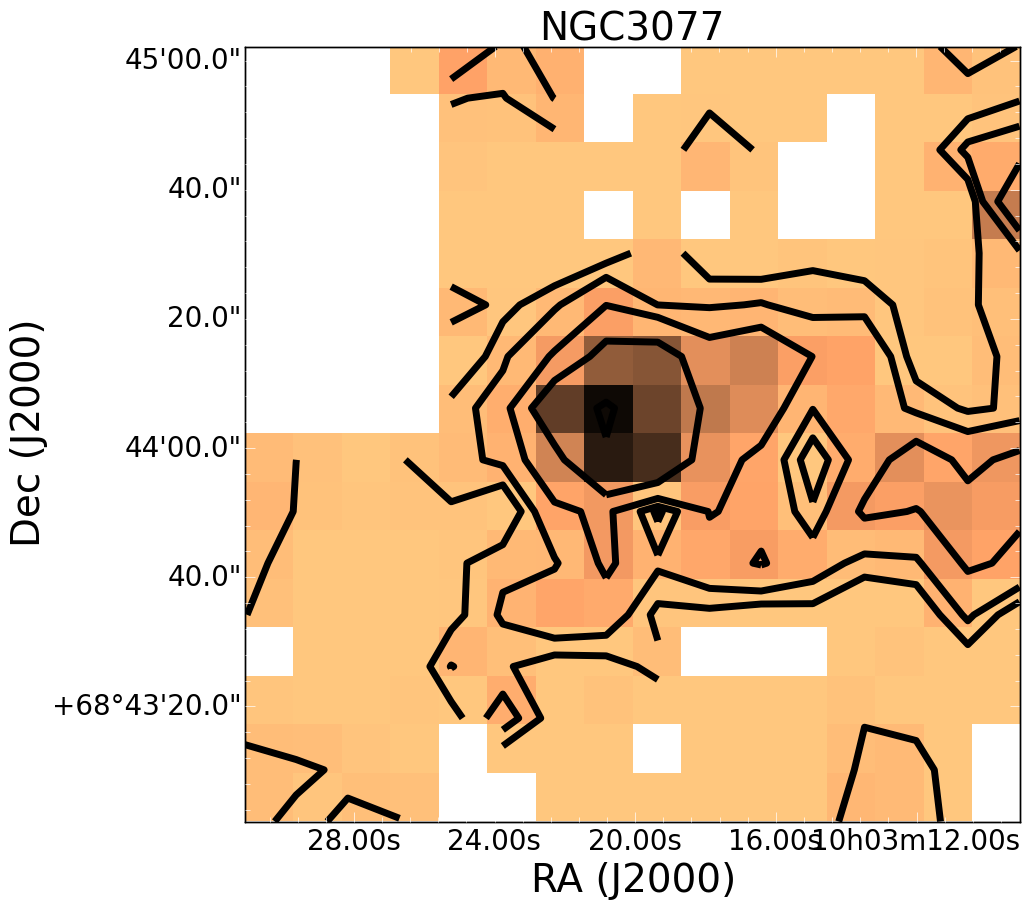}
	\includegraphics[height=3.75cm]{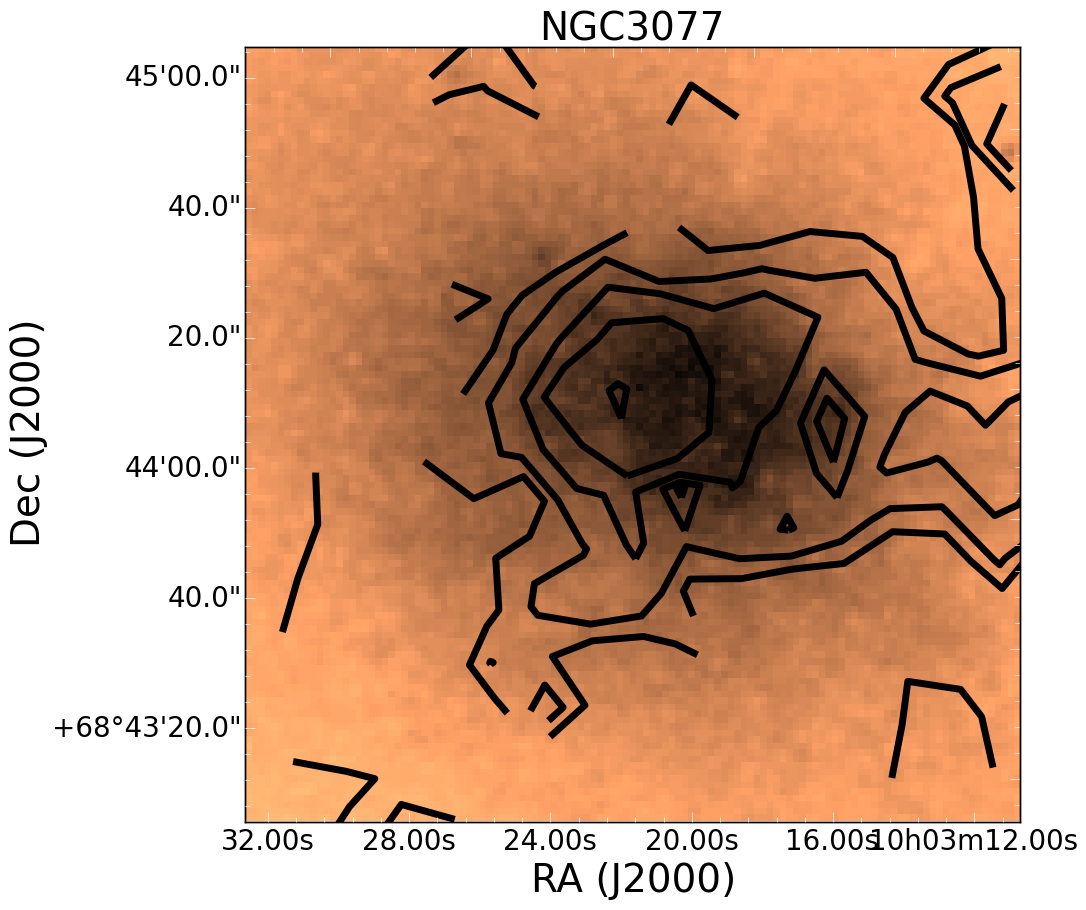}
	\includegraphics[height=3.75cm]{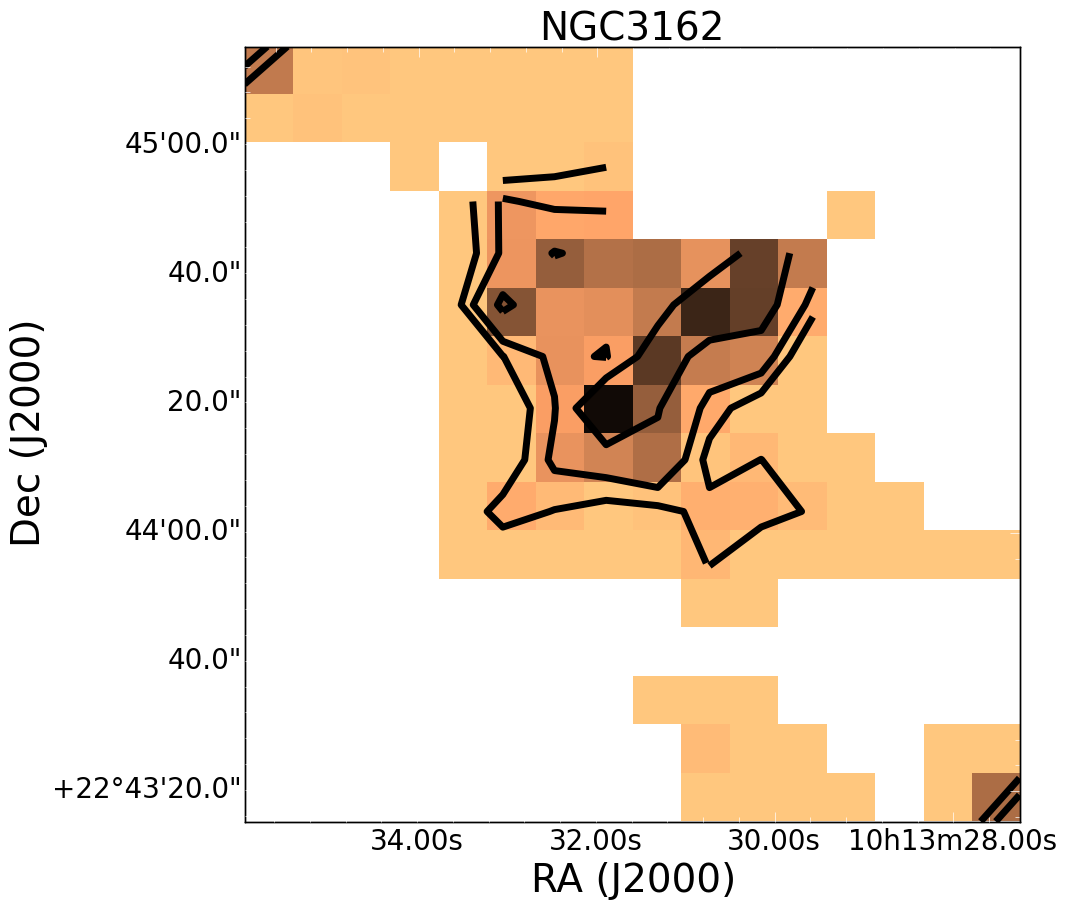}
	\includegraphics[height=3.75cm]{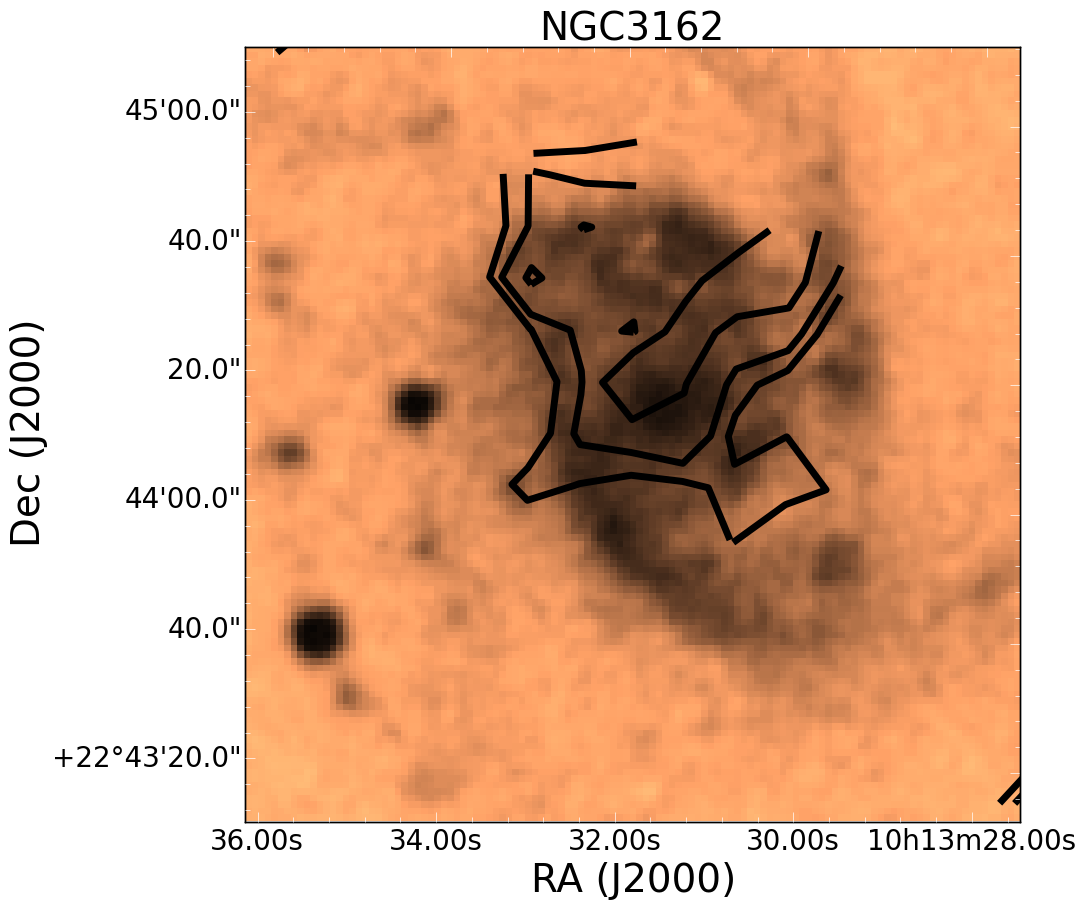}
	\includegraphics[height=3.75cm]{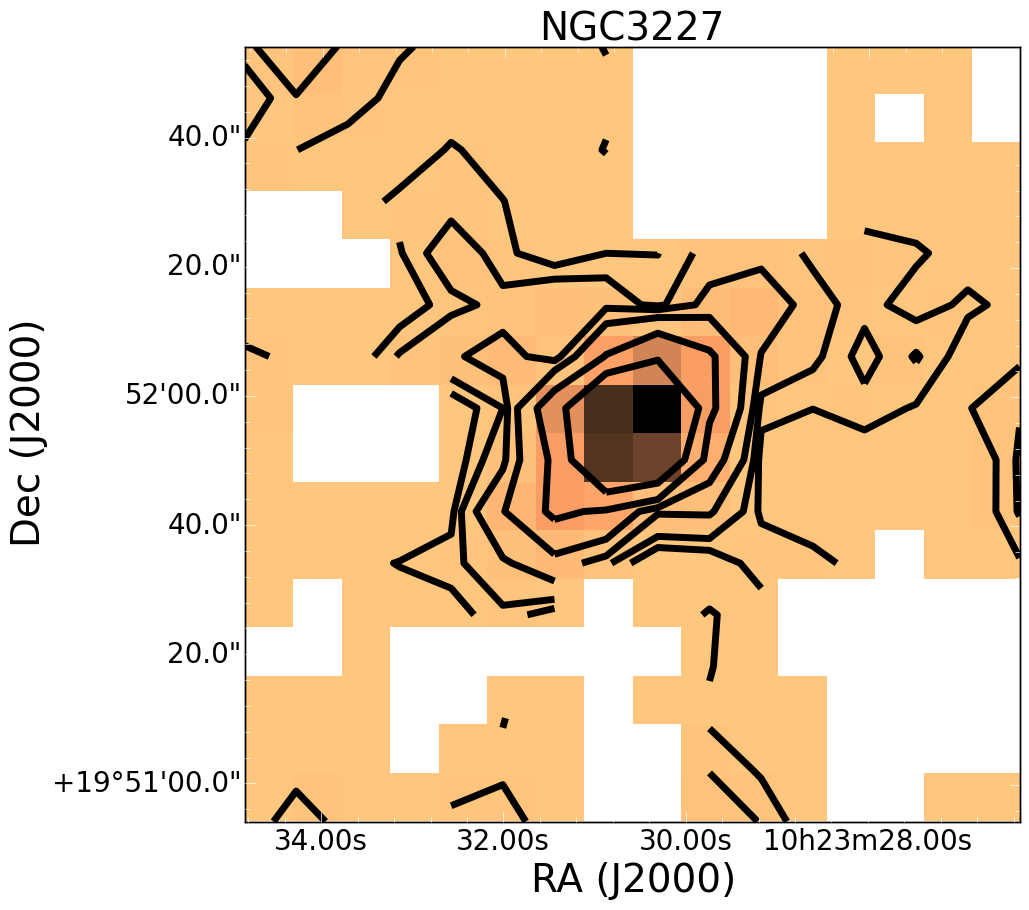}
	\includegraphics[height=3.75cm]{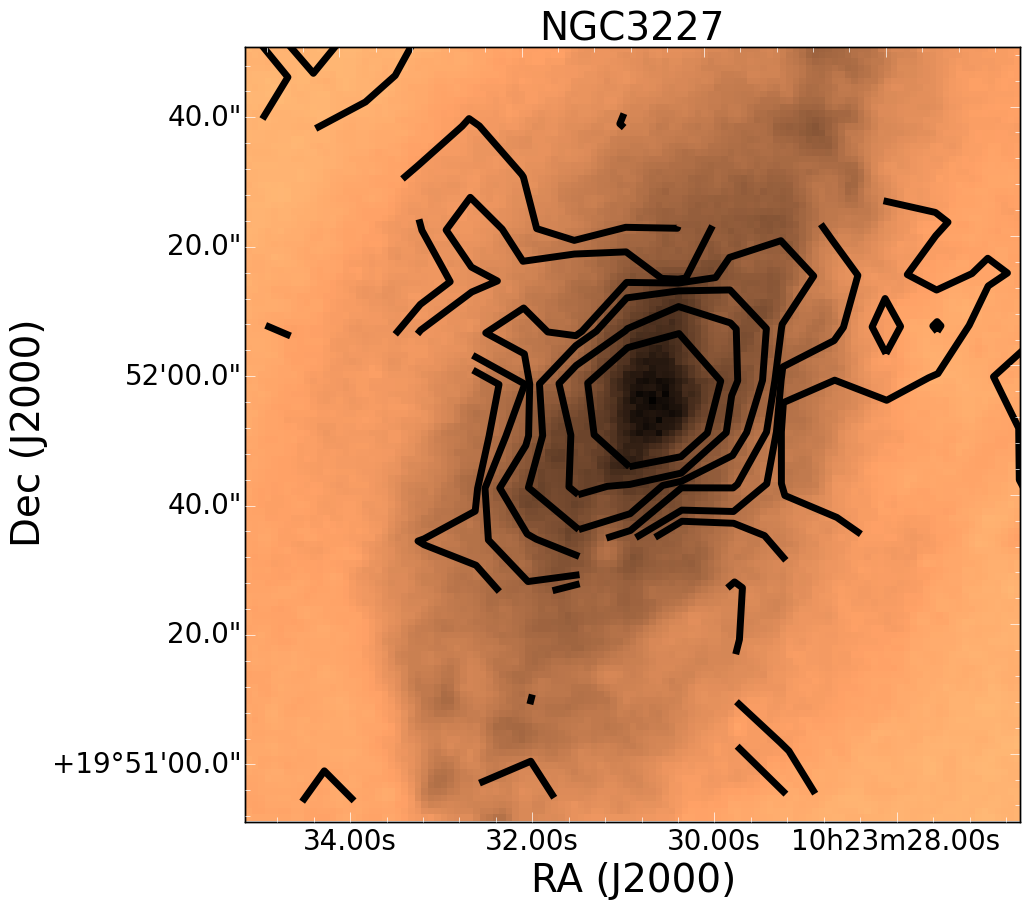}
	\includegraphics[height=3.75cm]{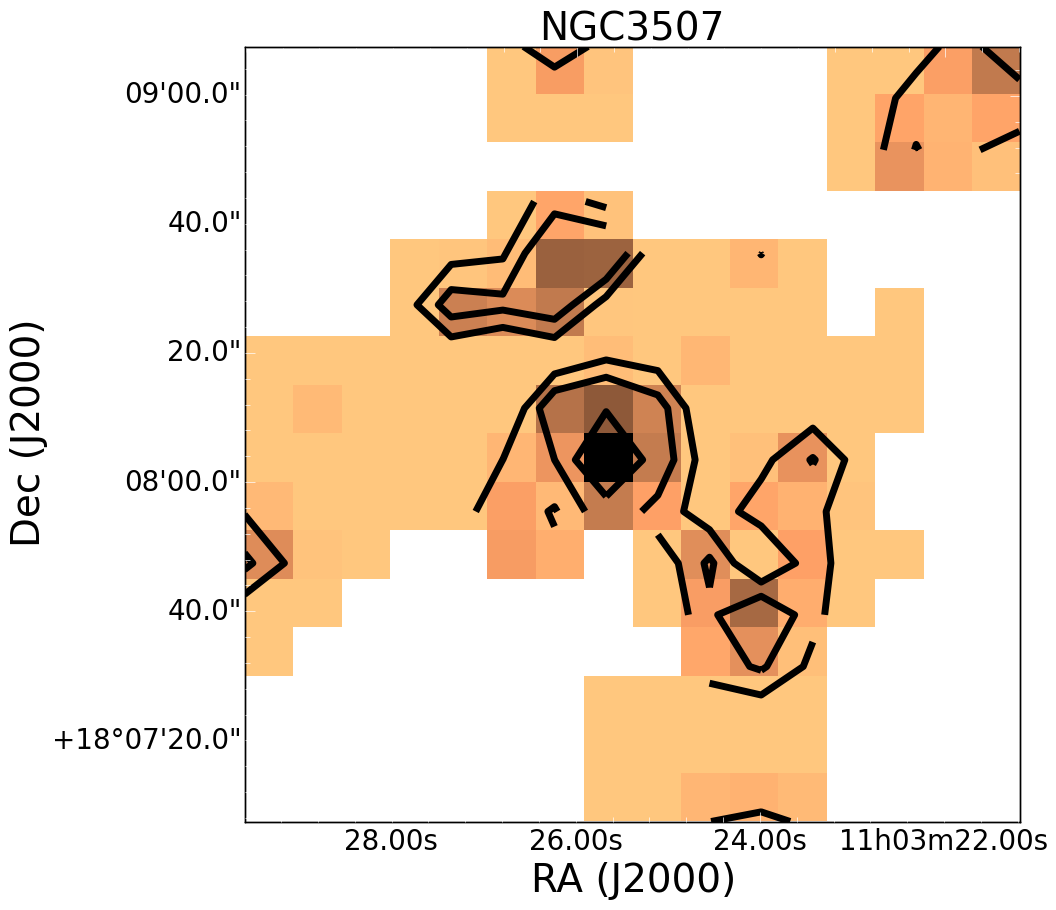}
	\includegraphics[height=3.75cm]{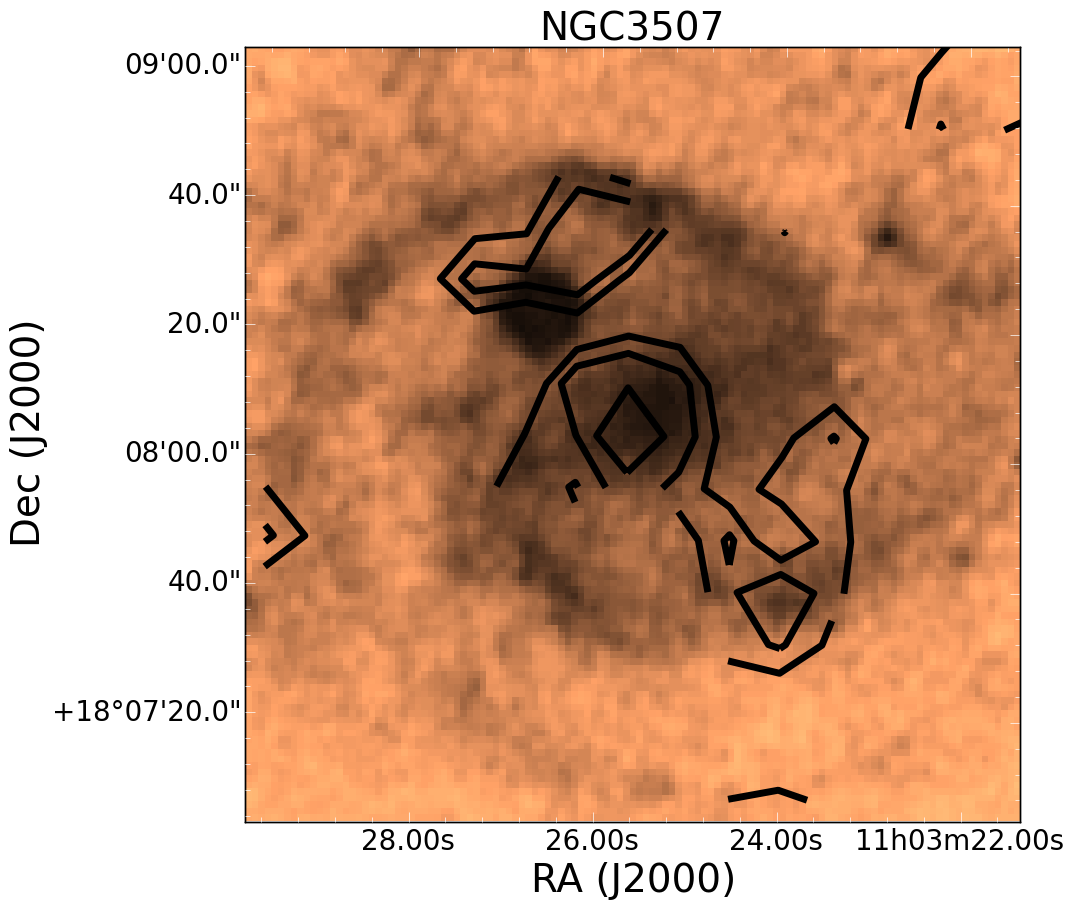}
	\includegraphics[height=3.75cm]{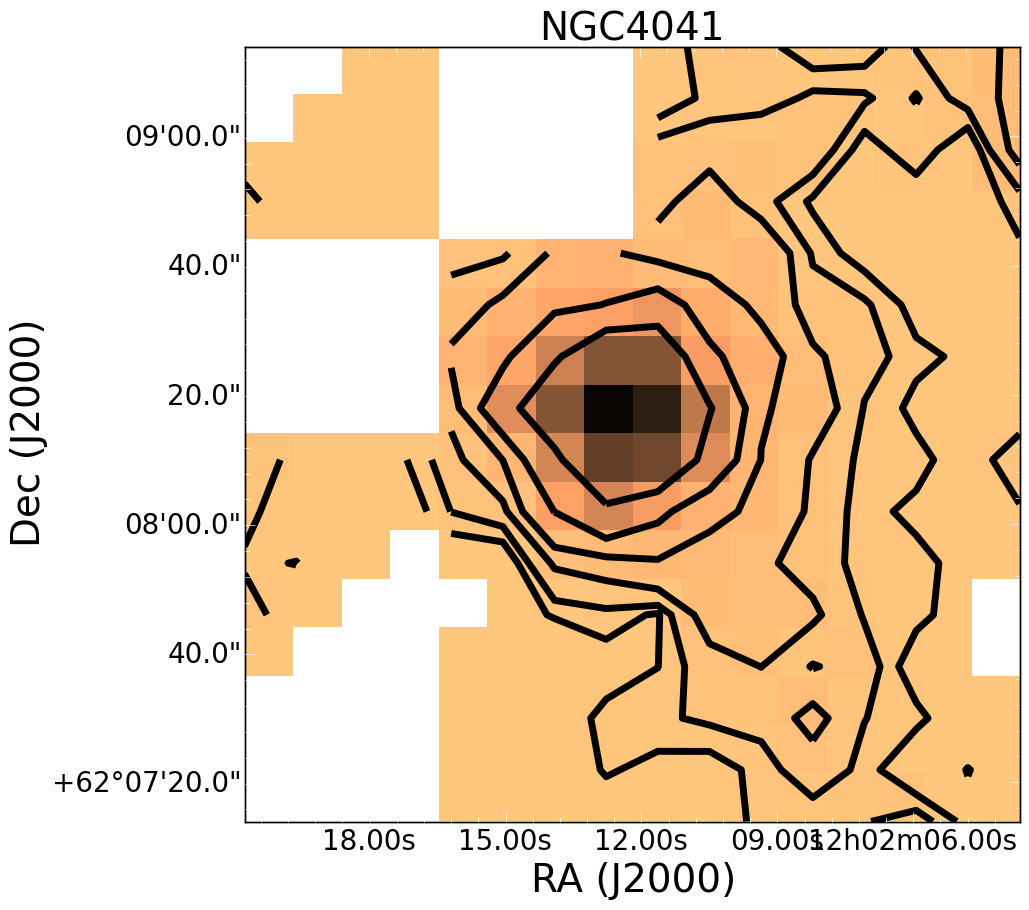}
	\includegraphics[height=3.75cm]{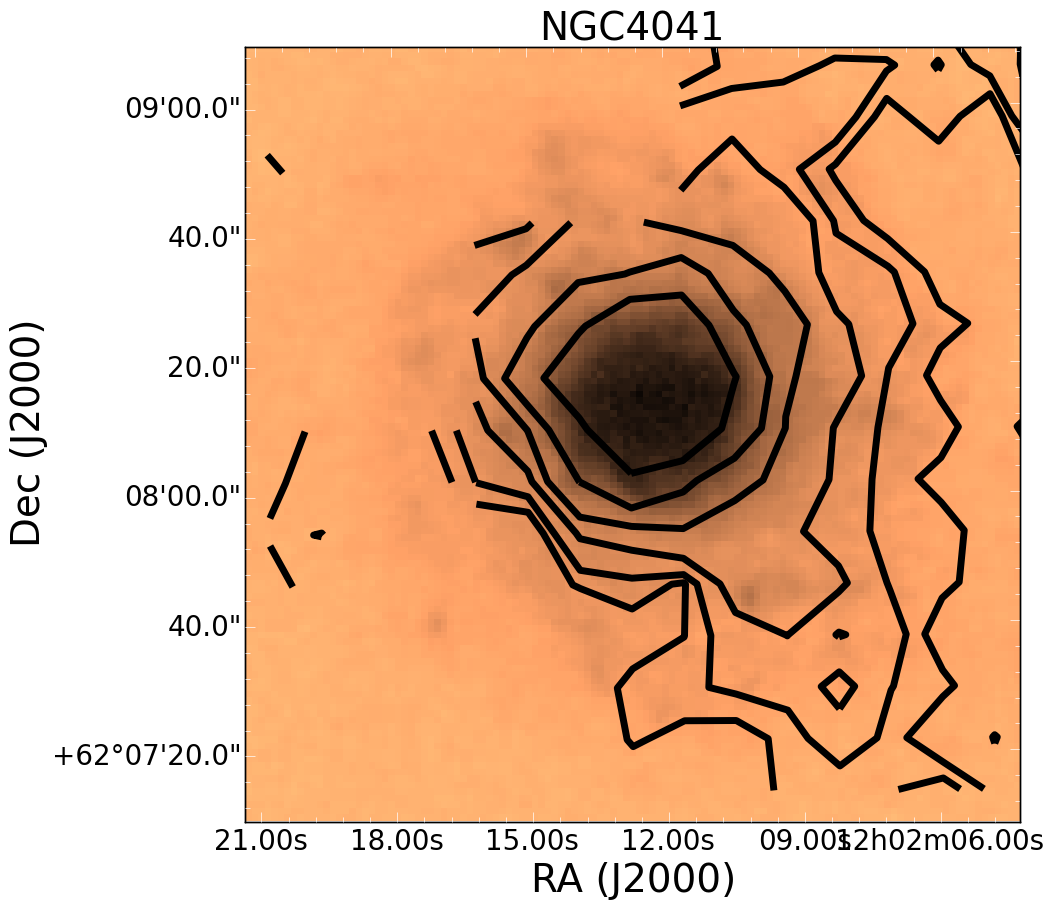}
	\caption{Images of the CO detected galaxies in the group sample. The first panel in each pair is the CO $J=3-2$ integrated galaxy map with black contour levels overlaid at 0.5, 1, 2, 4, 8, and 16 K km s$^{-1}$. The second panel is the same contour levels overlaid on the optical image from the Digitized Sky Survey.}
	\label{fig-det_group1}
\end{figure*}
\begin{figure*}
	\includegraphics[height=3.75cm]{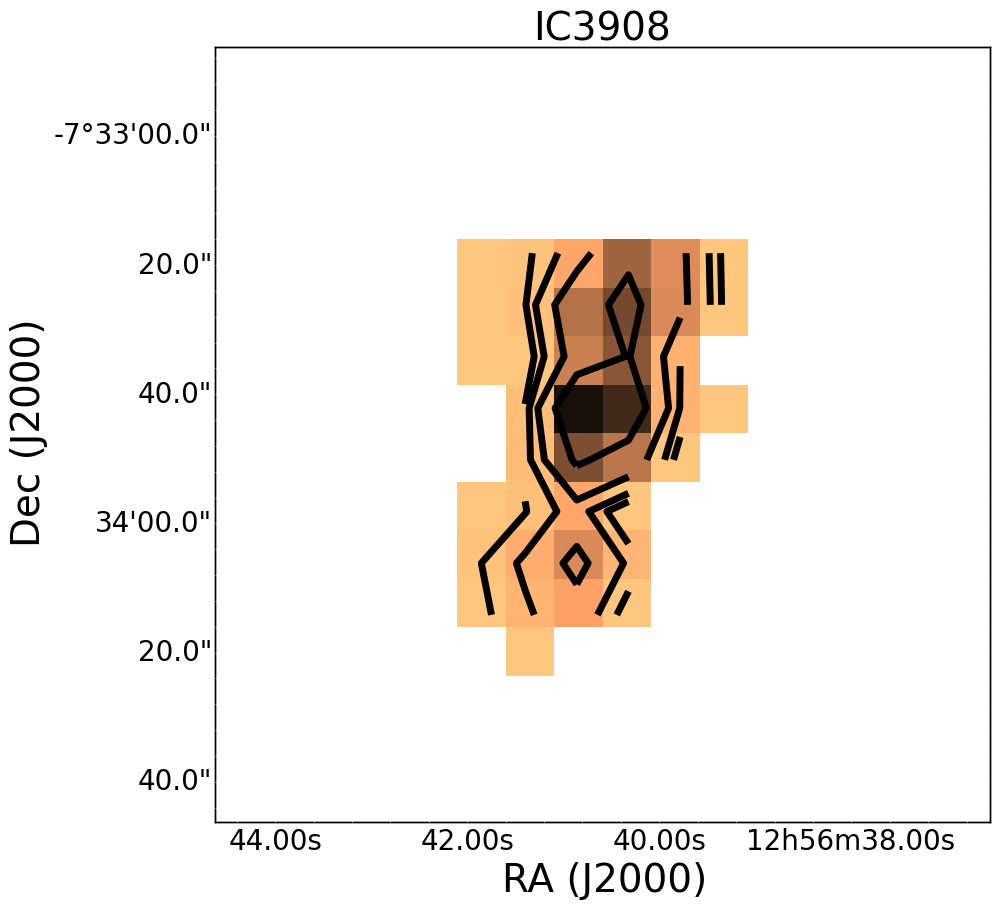}
	\includegraphics[height=3.75cm]{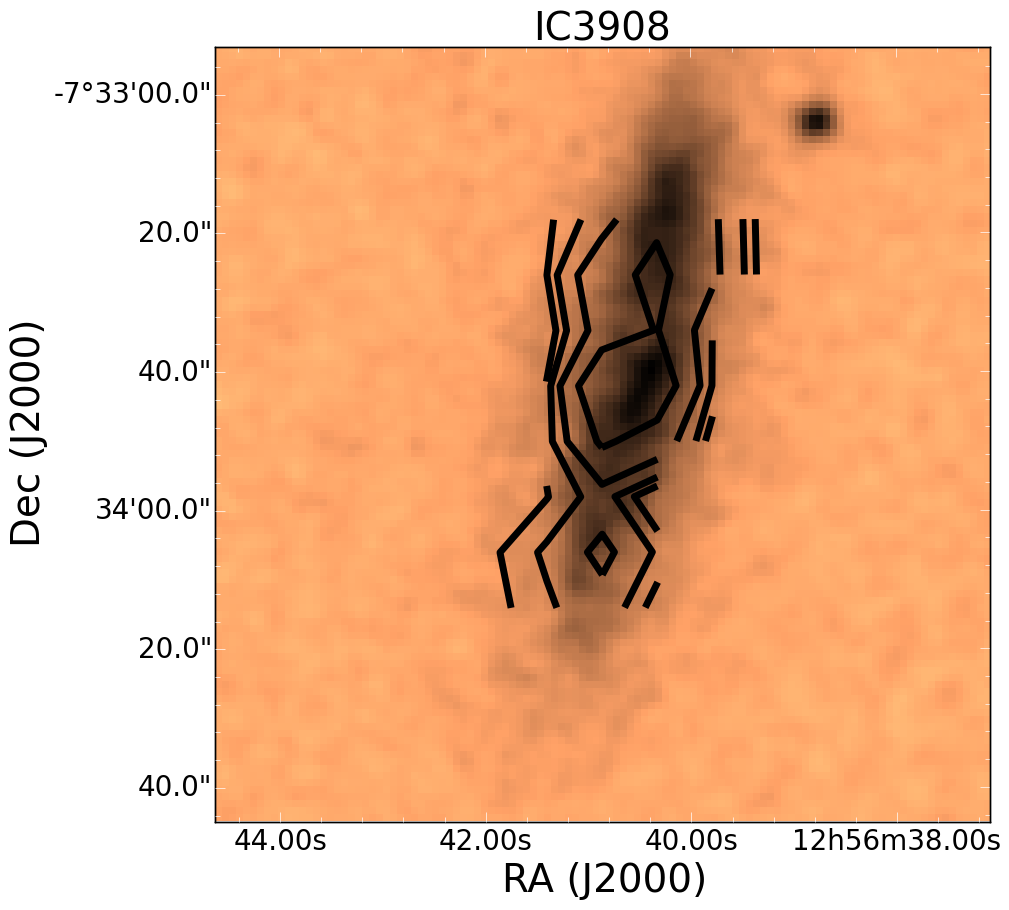}
	\includegraphics[height=3.75cm]{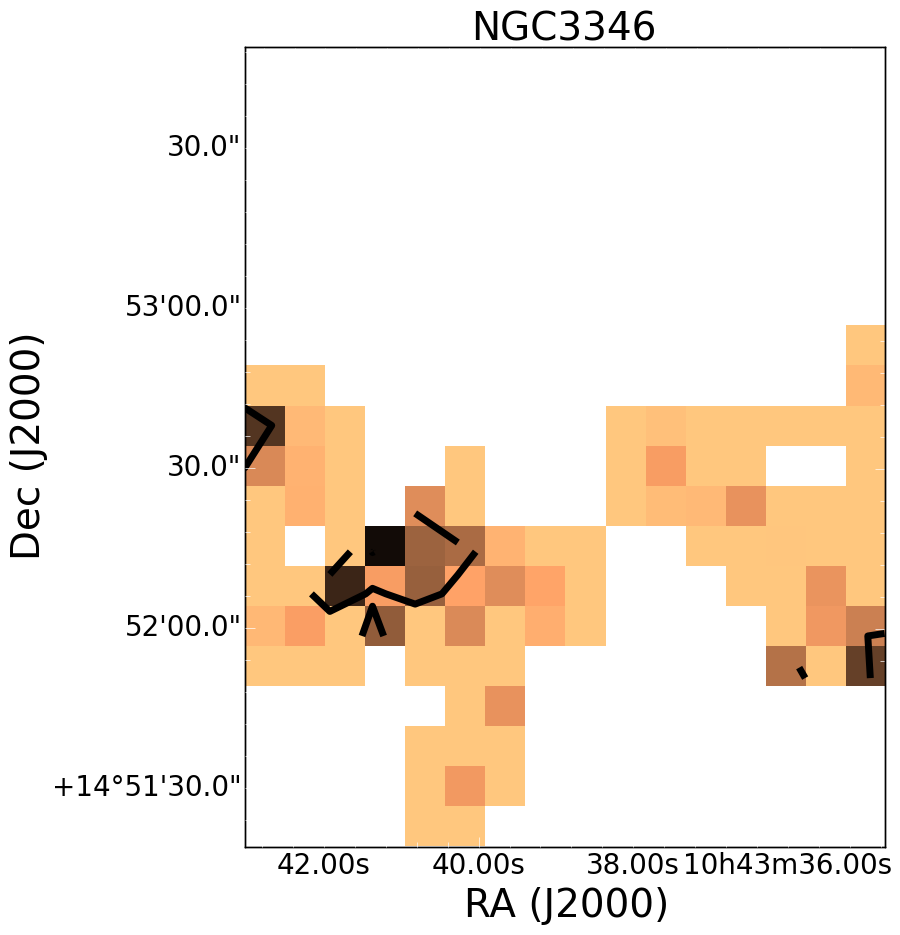}
	\includegraphics[height=3.75cm]{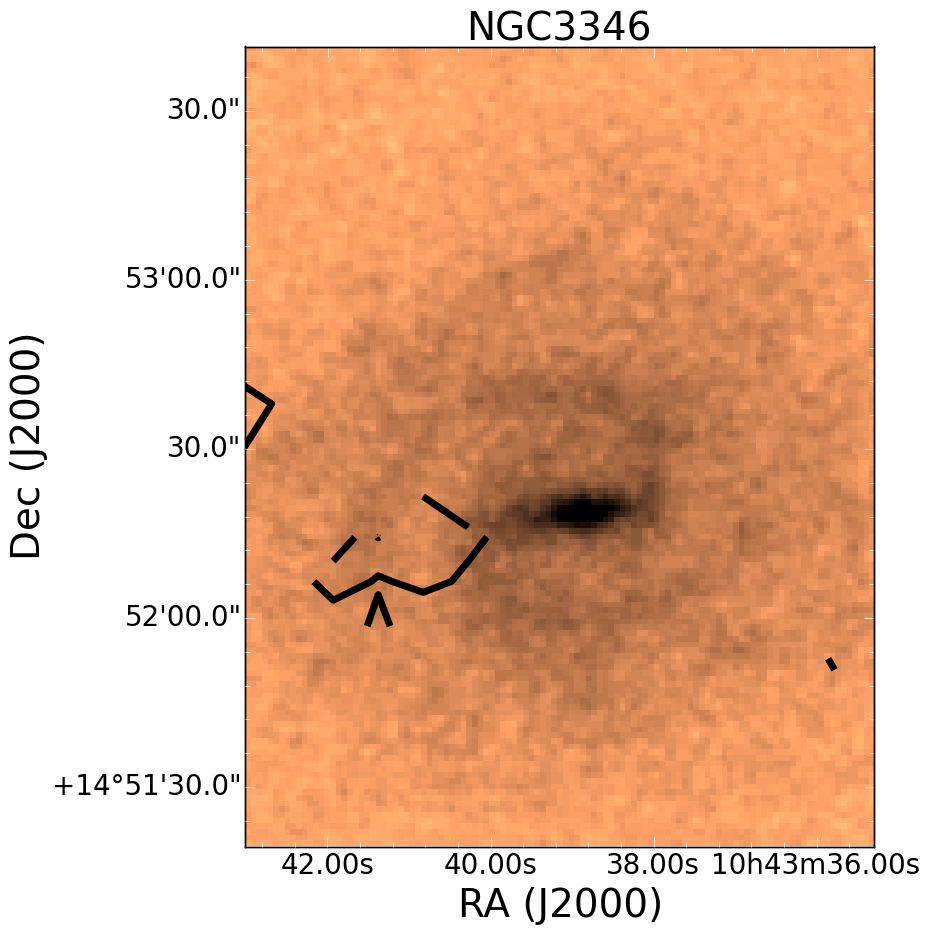}
	\includegraphics[height=3.75cm]{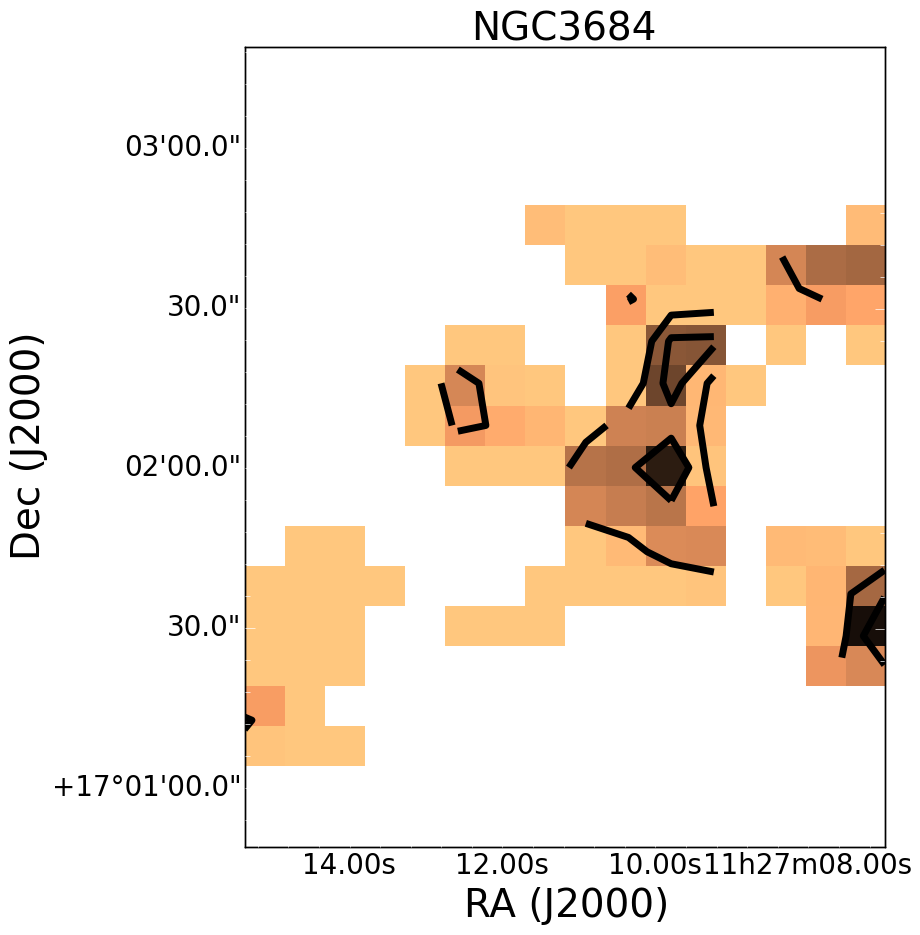}
	\includegraphics[height=3.75cm]{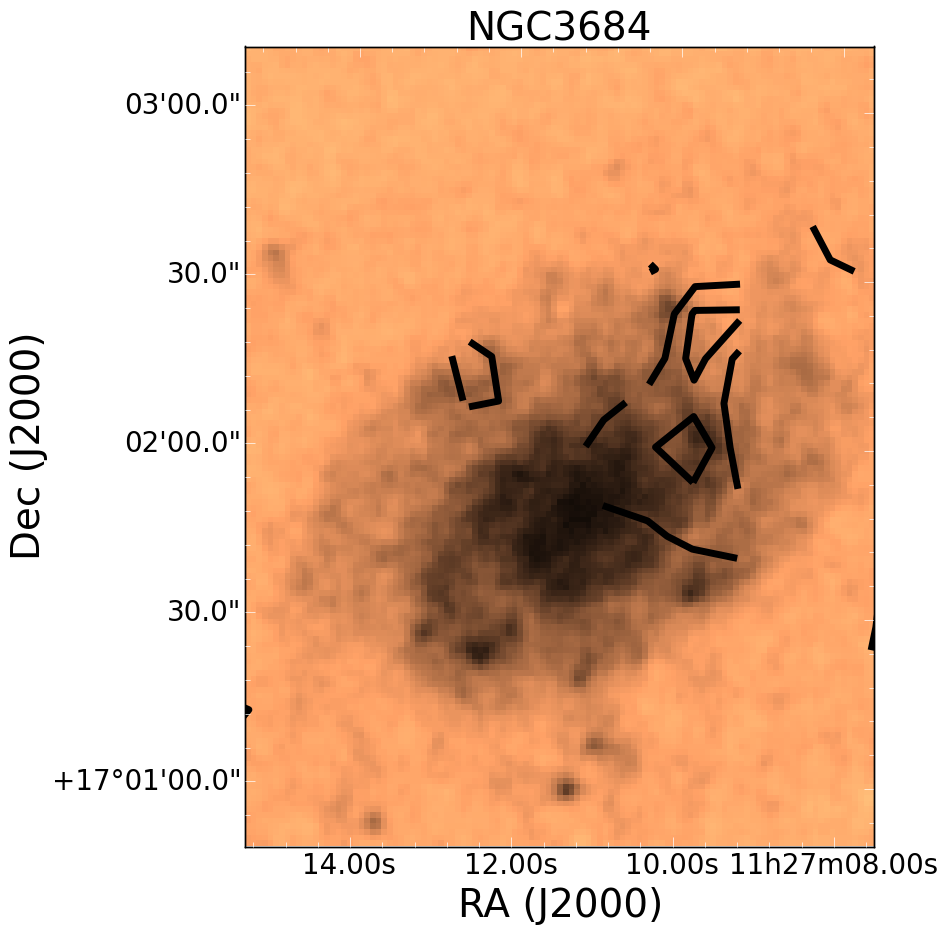}
	\includegraphics[height=3.75cm]{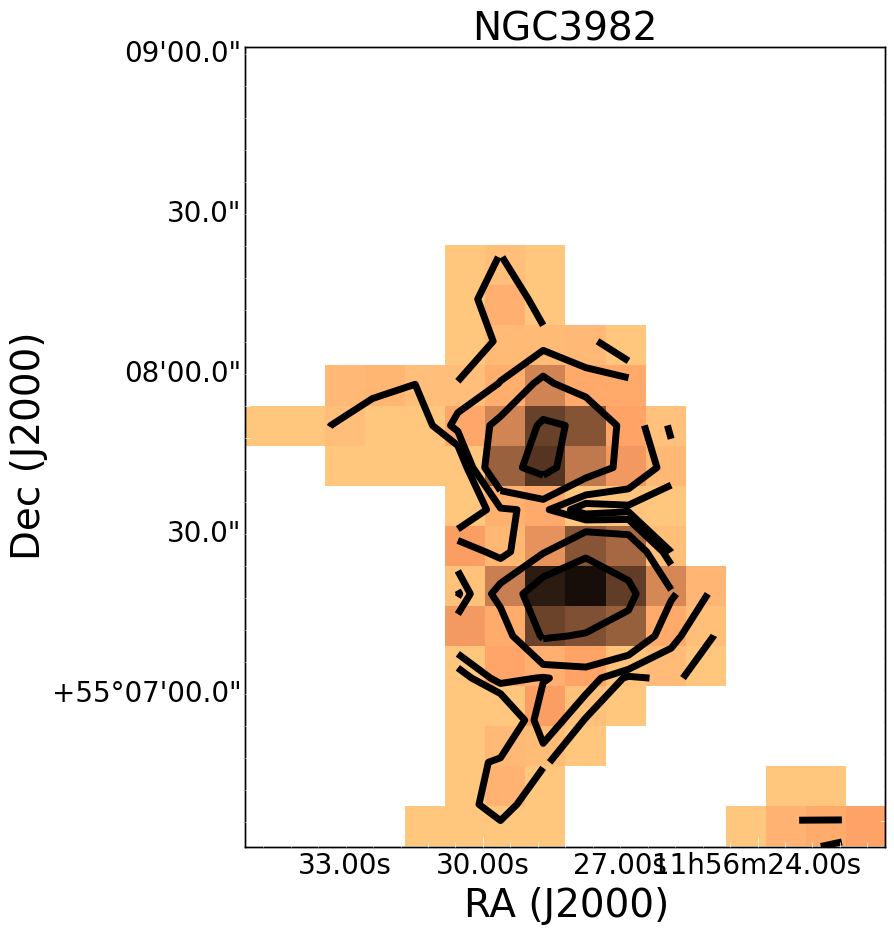}
	\includegraphics[height=3.75cm]{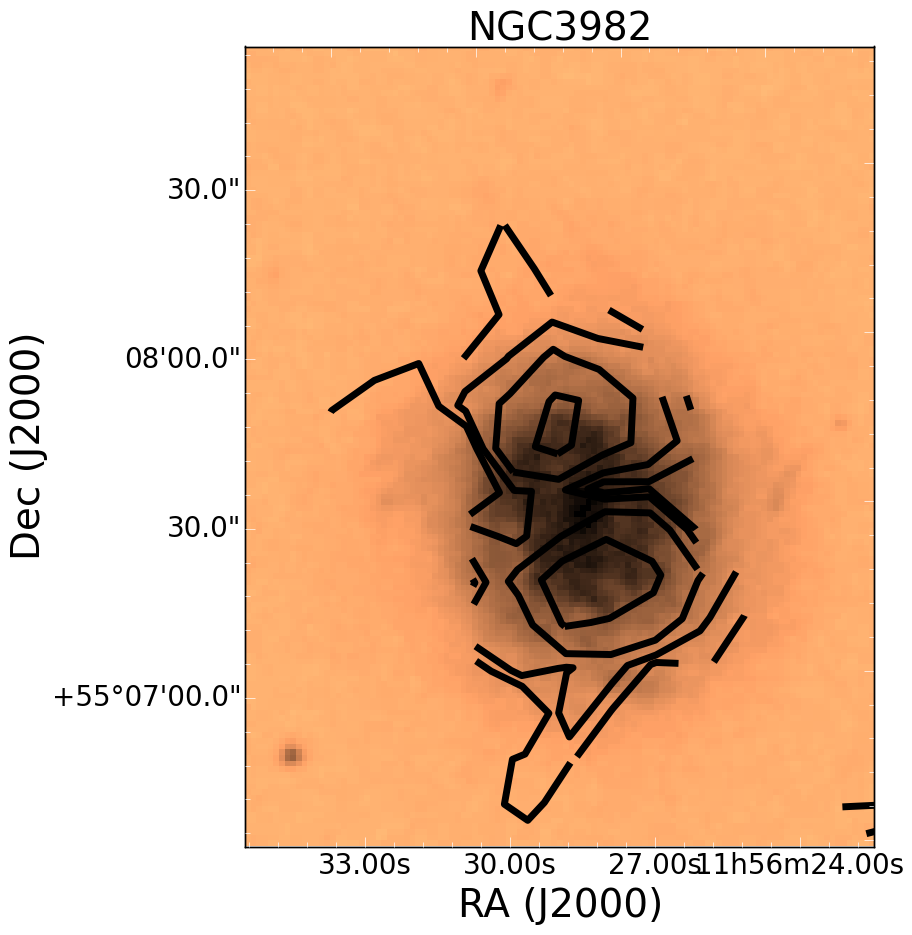}
	\includegraphics[height=3.75cm]{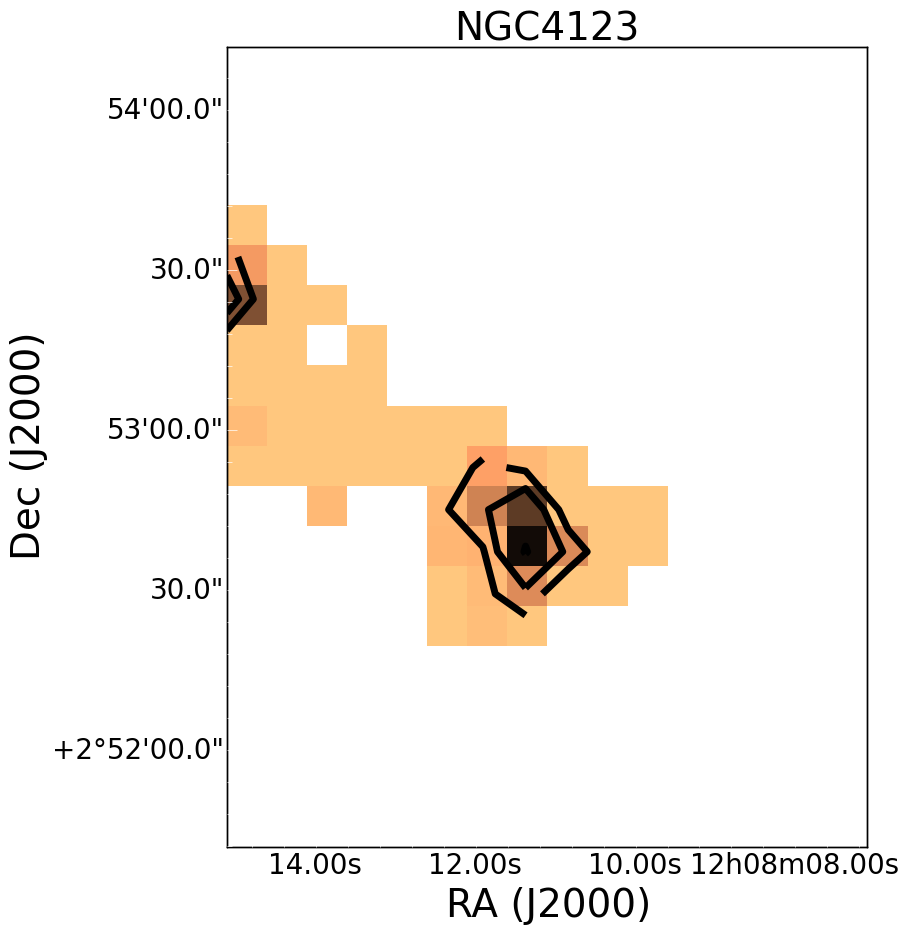}
	\includegraphics[height=3.75cm]{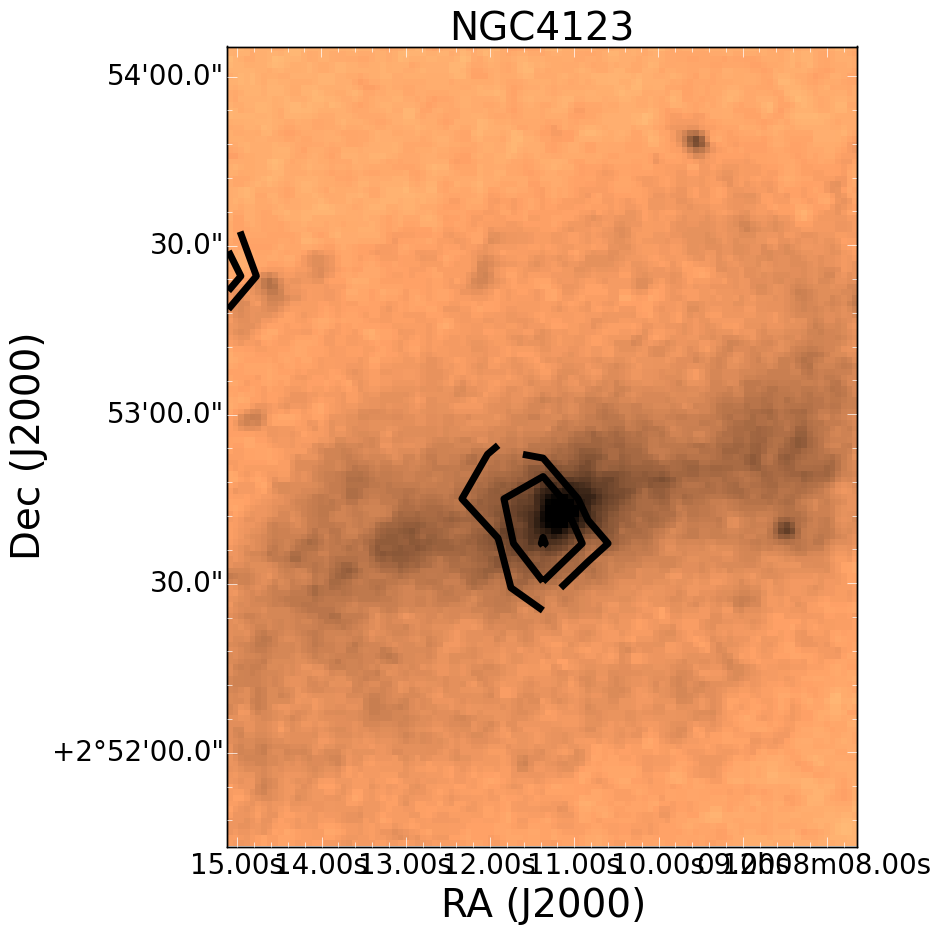}
	\includegraphics[height=3.75cm]{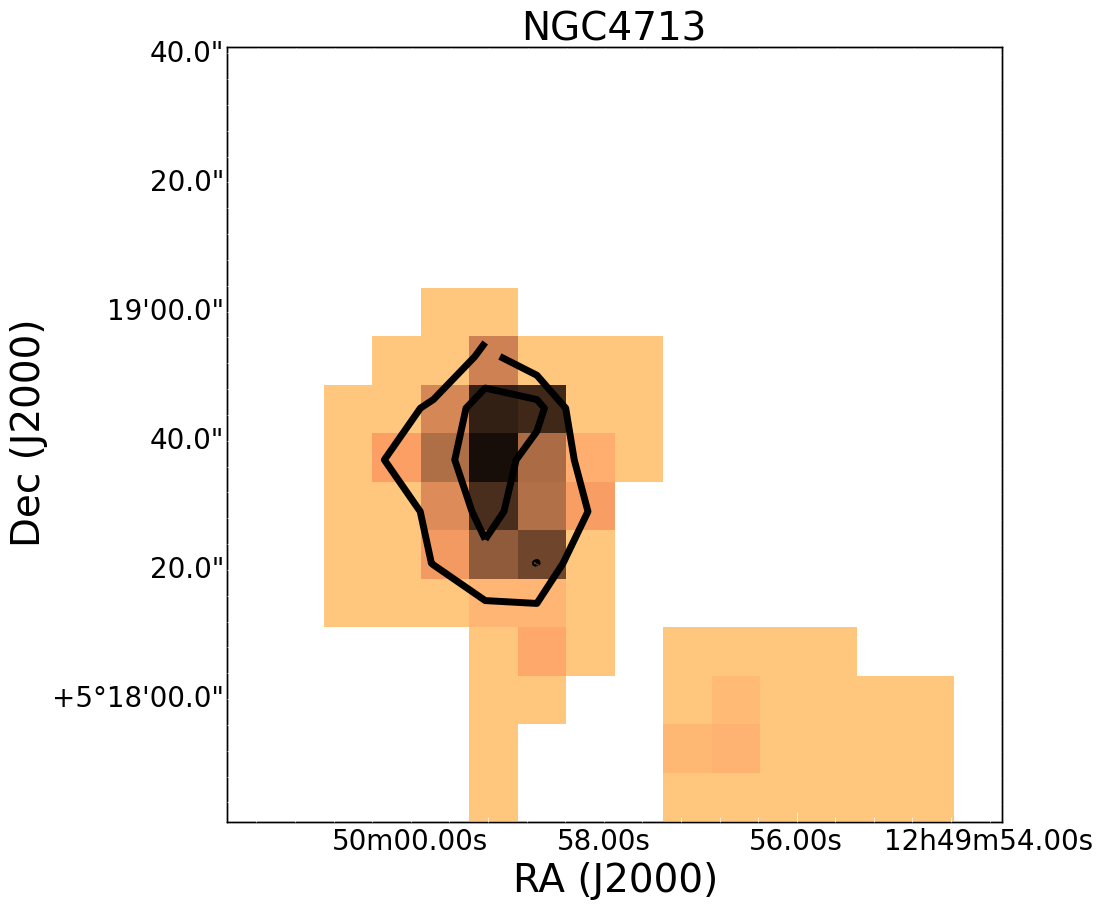}
	\includegraphics[height=3.75cm]{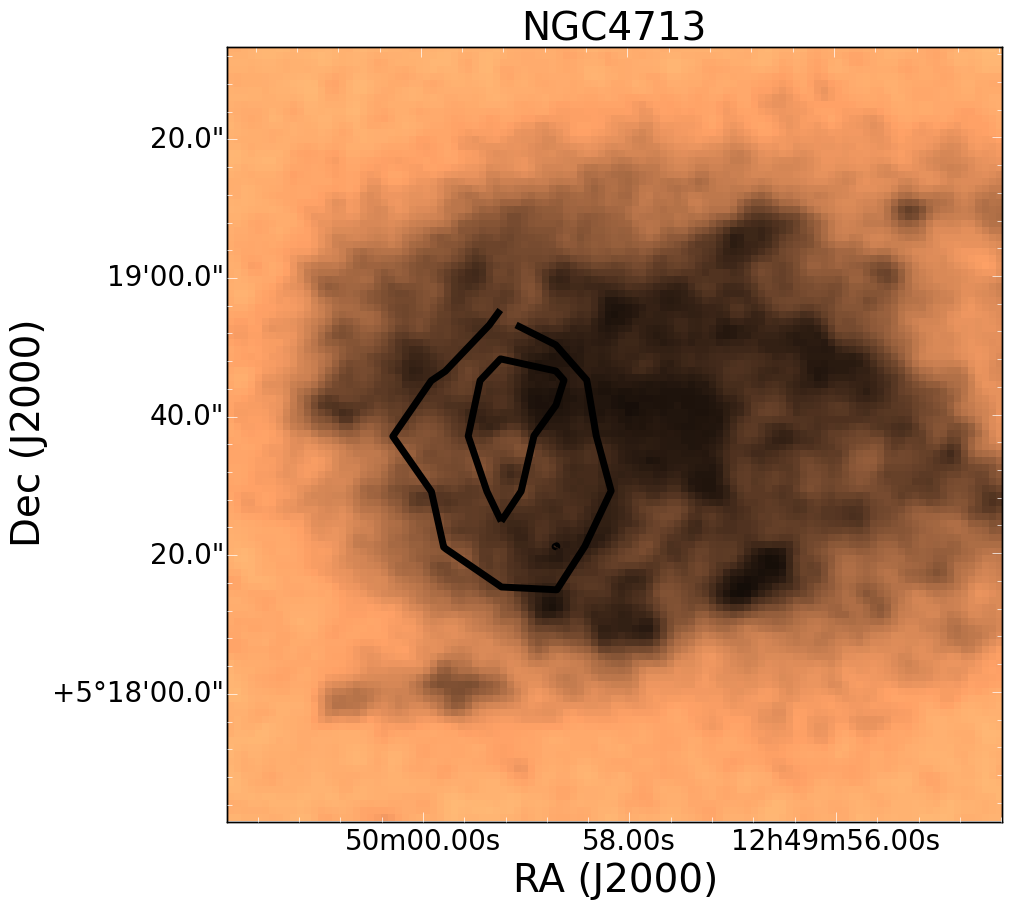}
	\includegraphics[height=3.75cm]{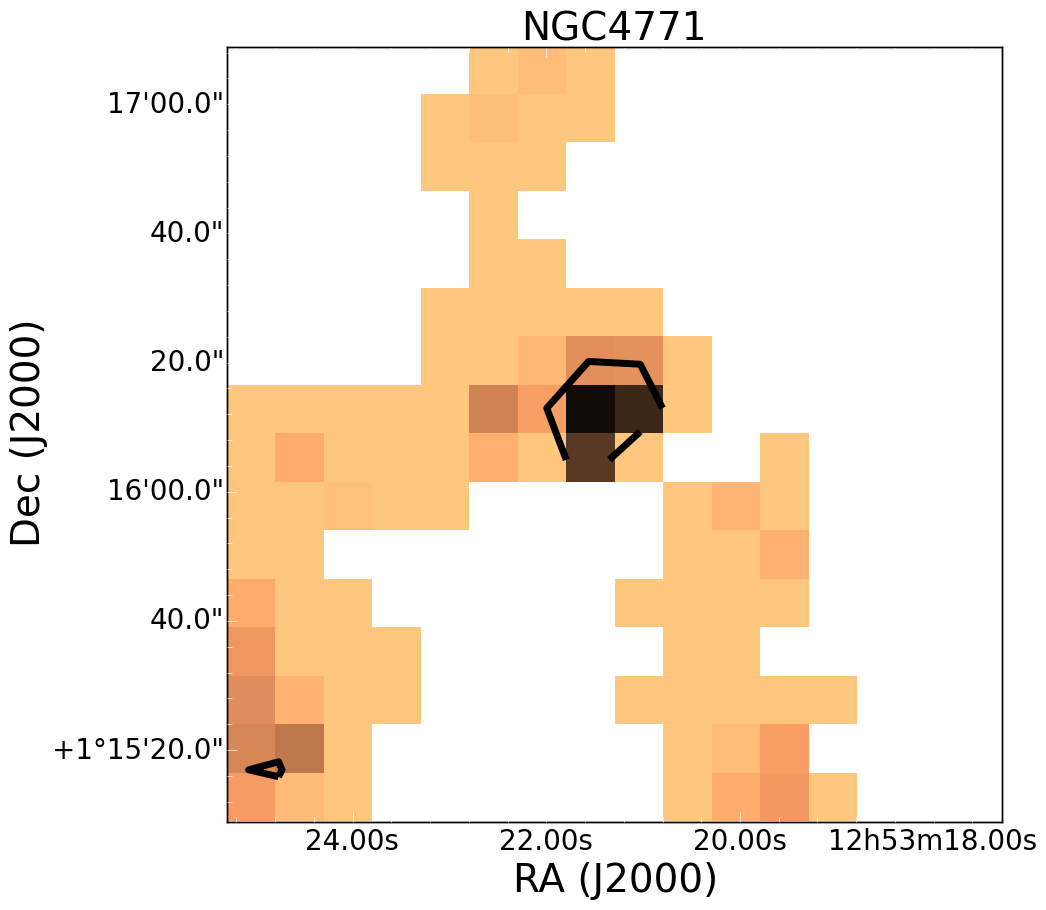}
	\includegraphics[height=3.75cm]{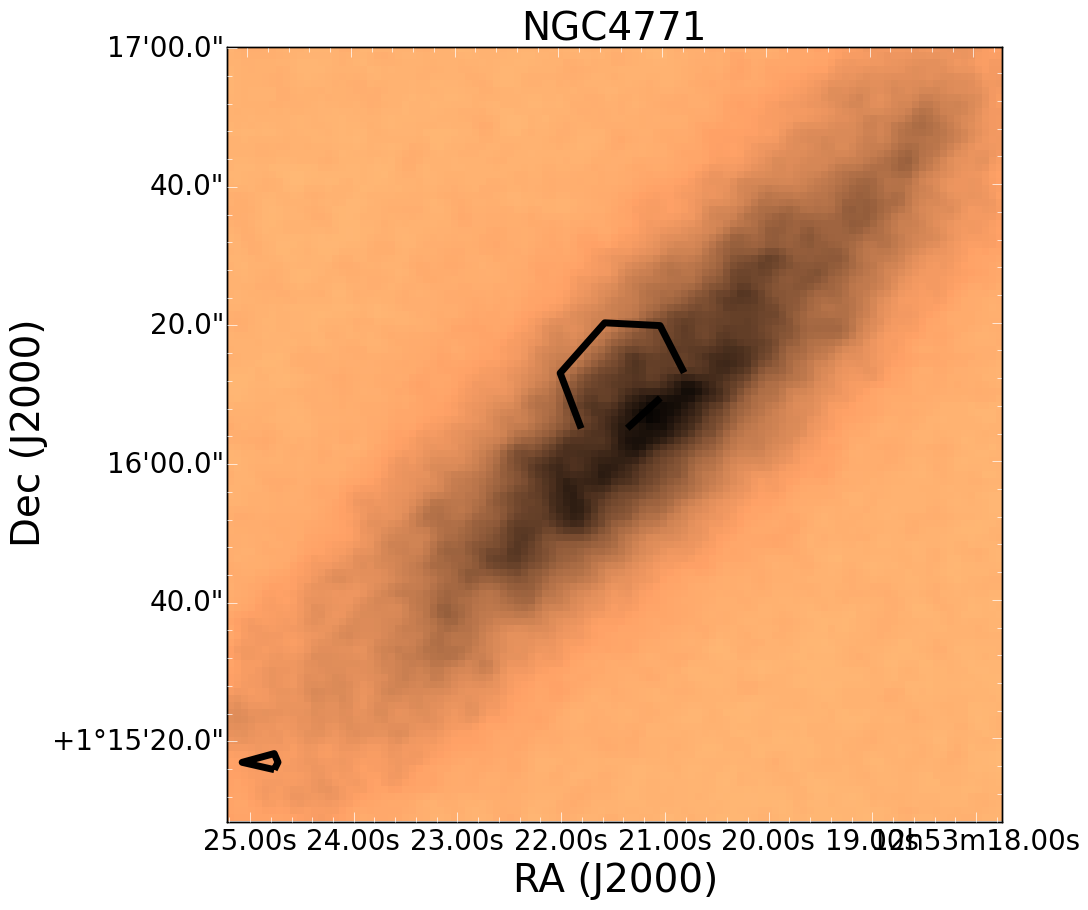}
	\includegraphics[height=3.75cm]{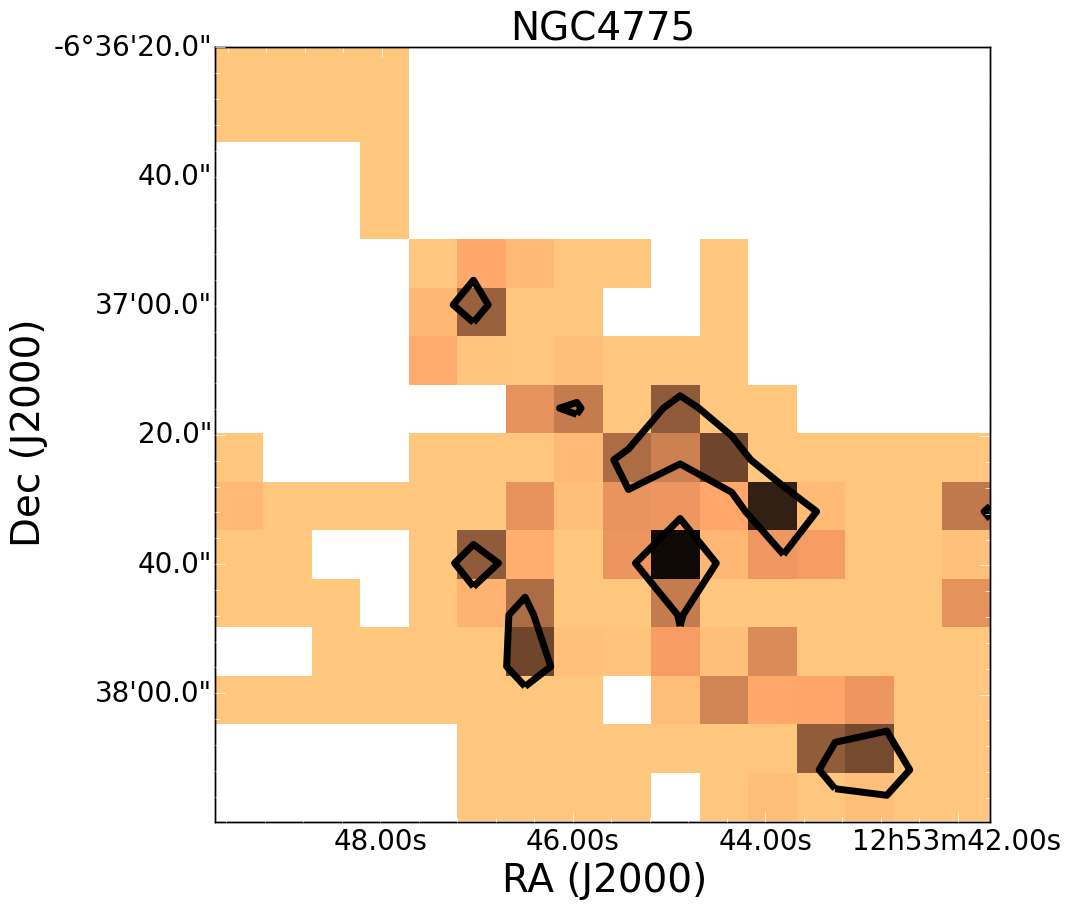}
	\includegraphics[height=3.75cm]{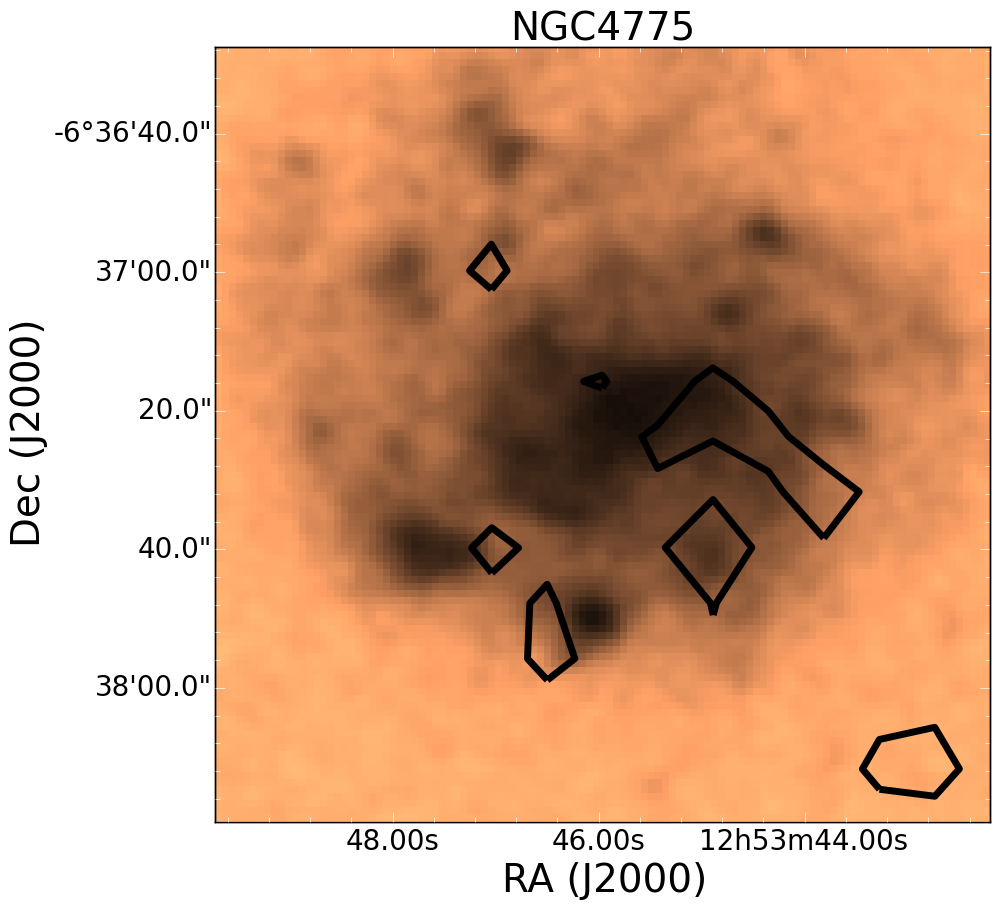}
	\includegraphics[height=3.75cm]{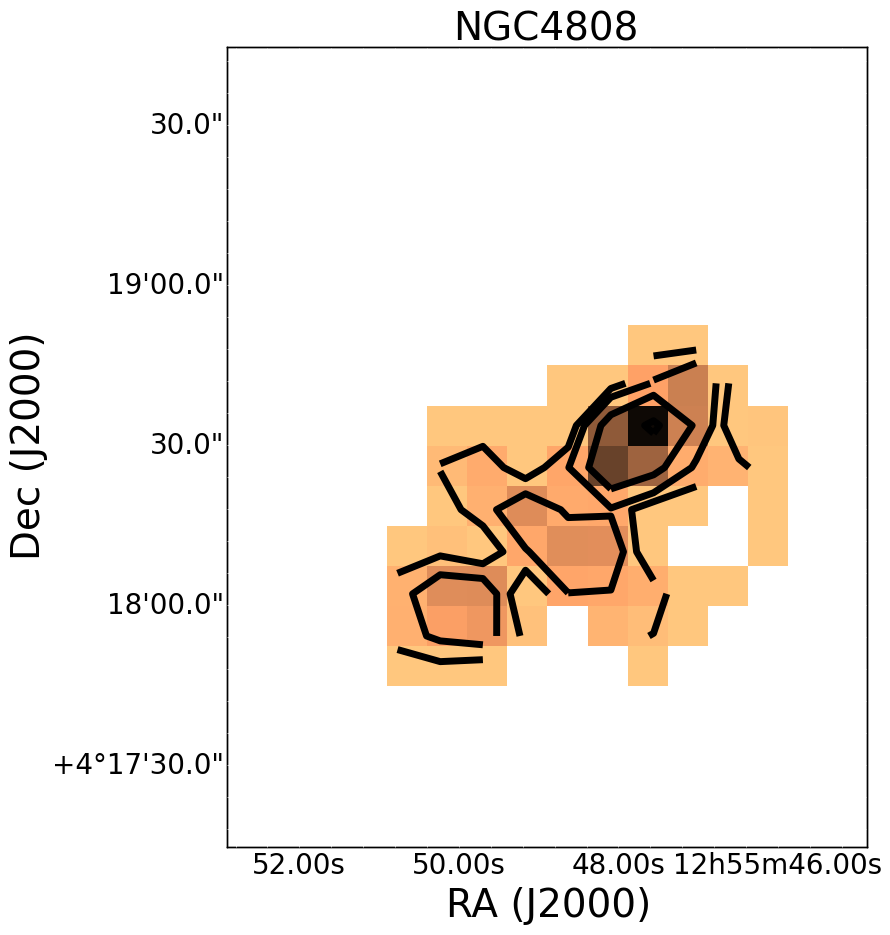}
	\includegraphics[height=3.75cm]{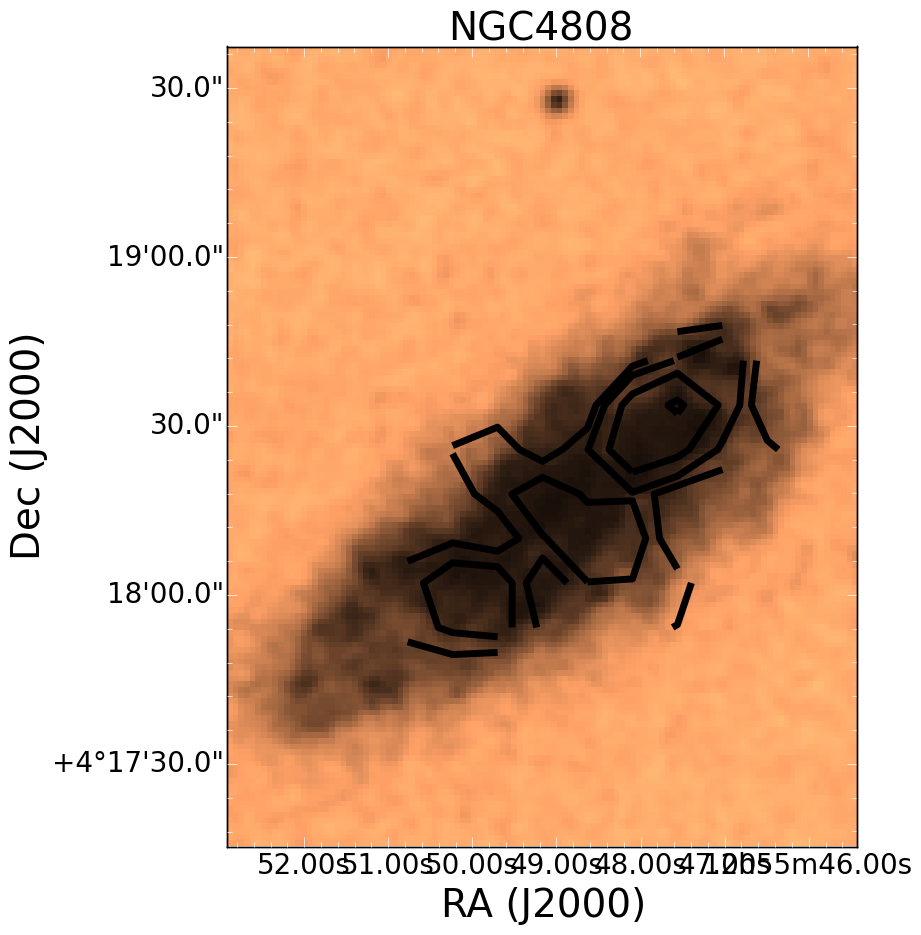}
	\caption{Additional images of the CO detected galaxies in the group sample. The first panel in each pair is the CO $J=3-2$ integrated galaxy map with black contour levels overlaid at 0.5, 1, 2, 4, 8, and 16 K km s$^{-1}$. The second panel is the same contour levels overlaid on the optical image from the Digitized Sky Survey.}
	\label{fig-det_group2}
\end{figure*}
\begin{figure*}
	\includegraphics[height=3.75cm]{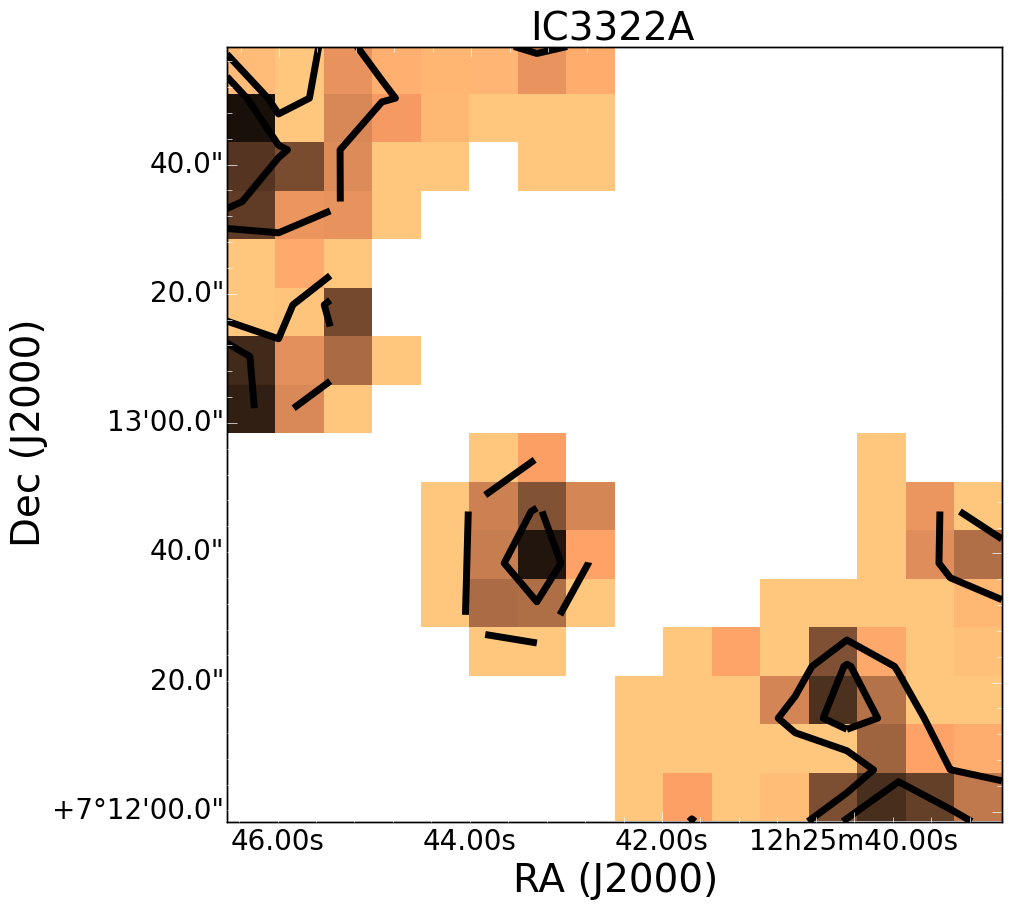}
	\includegraphics[height=3.75cm]{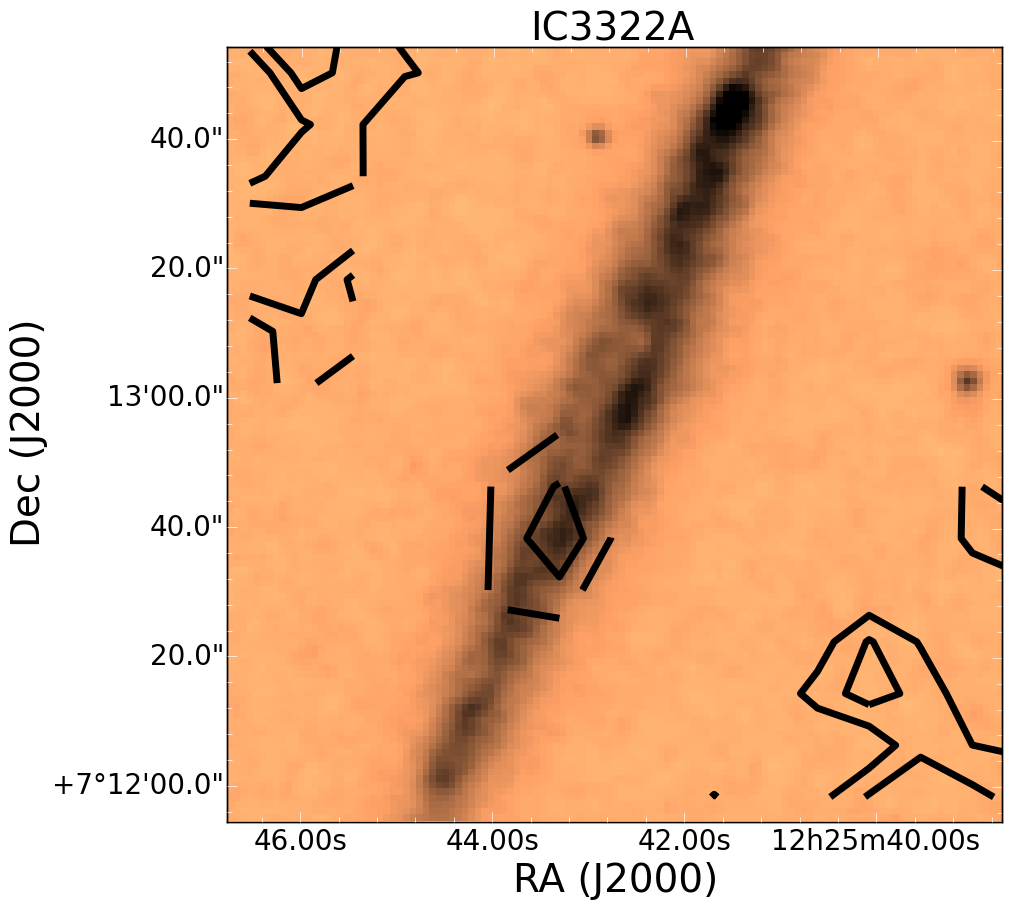}
	\includegraphics[height=3.75cm]{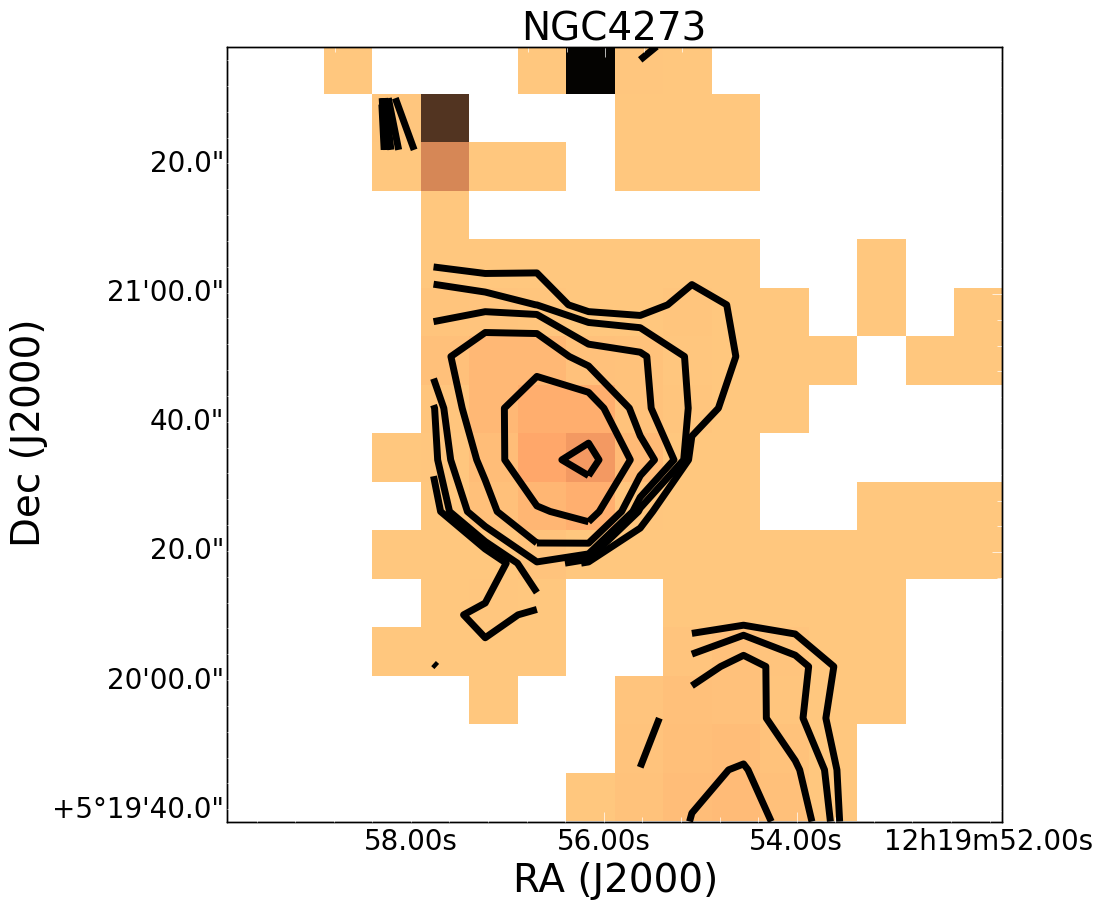}
	\includegraphics[height=3.75cm]{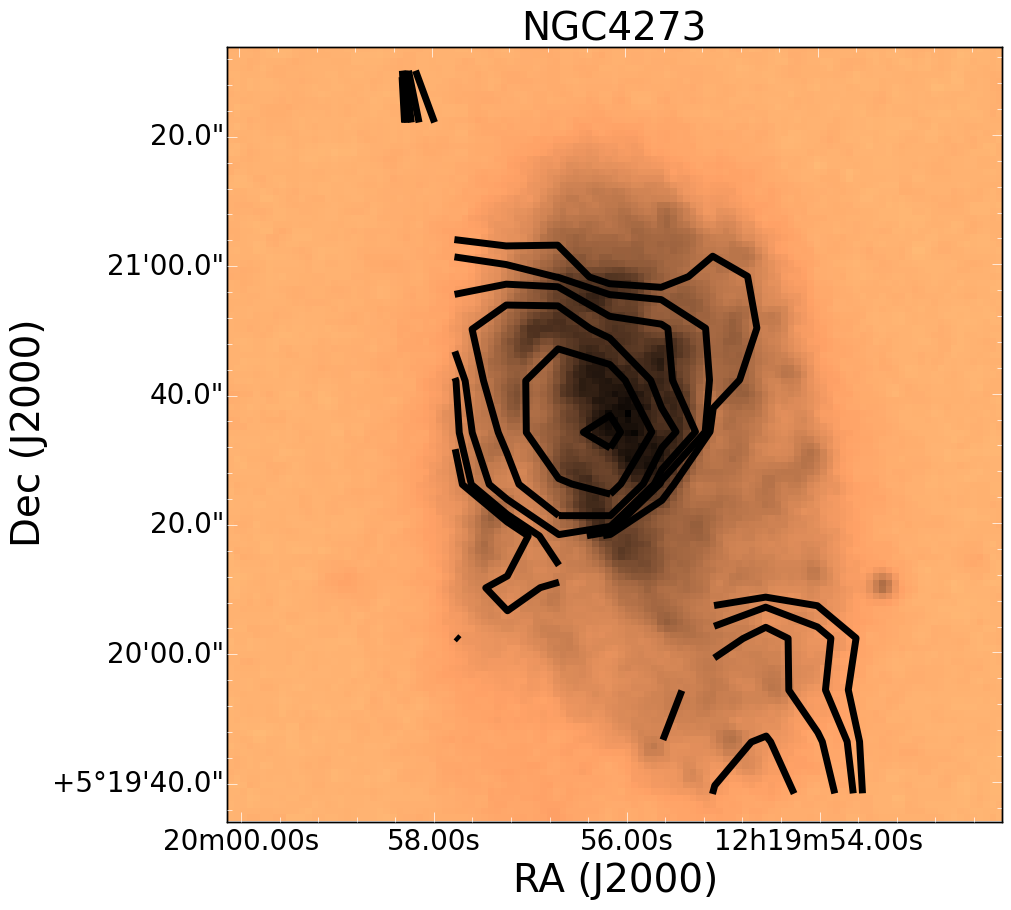}
	\includegraphics[height=3.75cm]{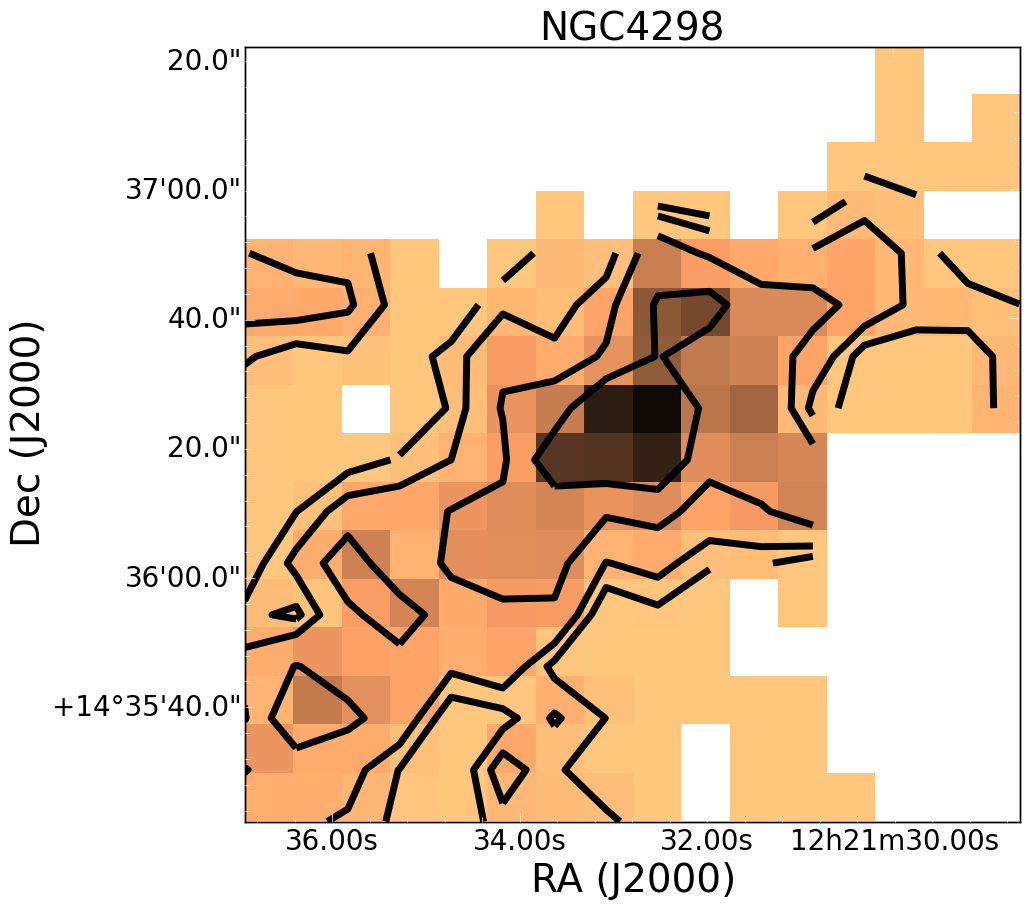}
	\includegraphics[height=3.75cm]{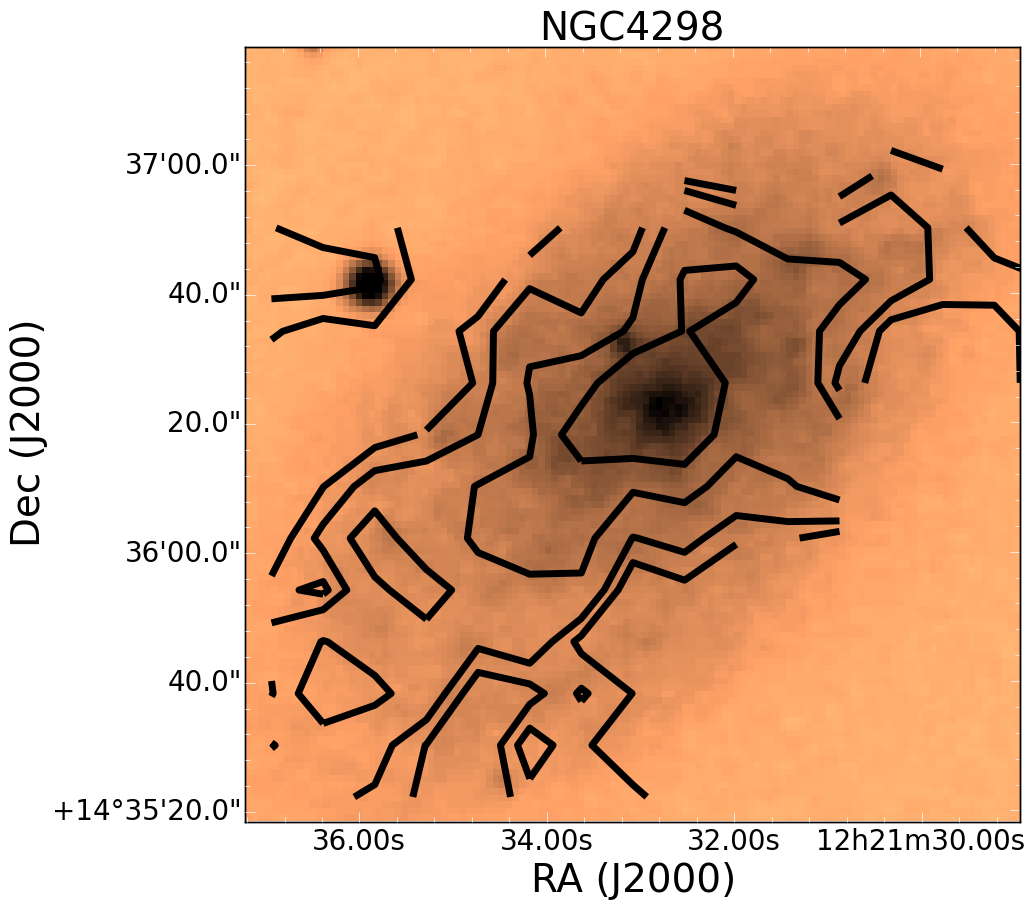}
	\includegraphics[height=3.75cm]{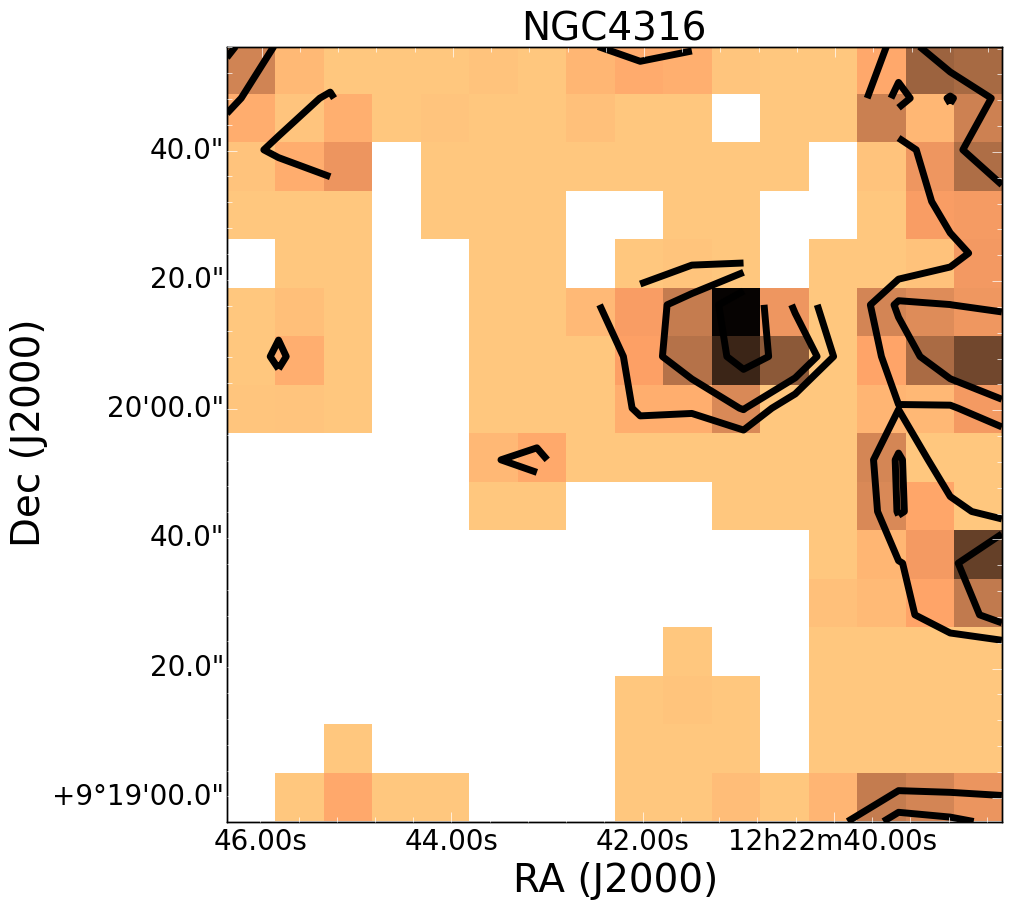}
	\includegraphics[height=3.75cm]{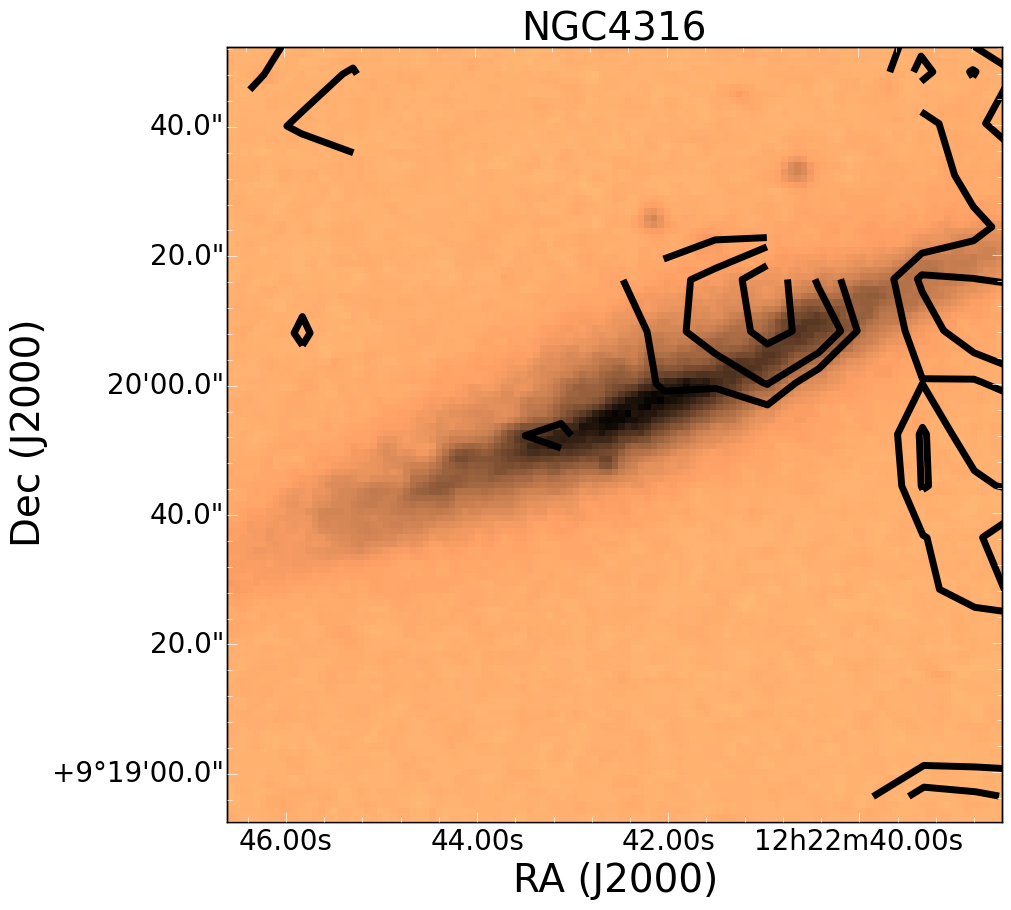}
	\includegraphics[height=3.75cm]{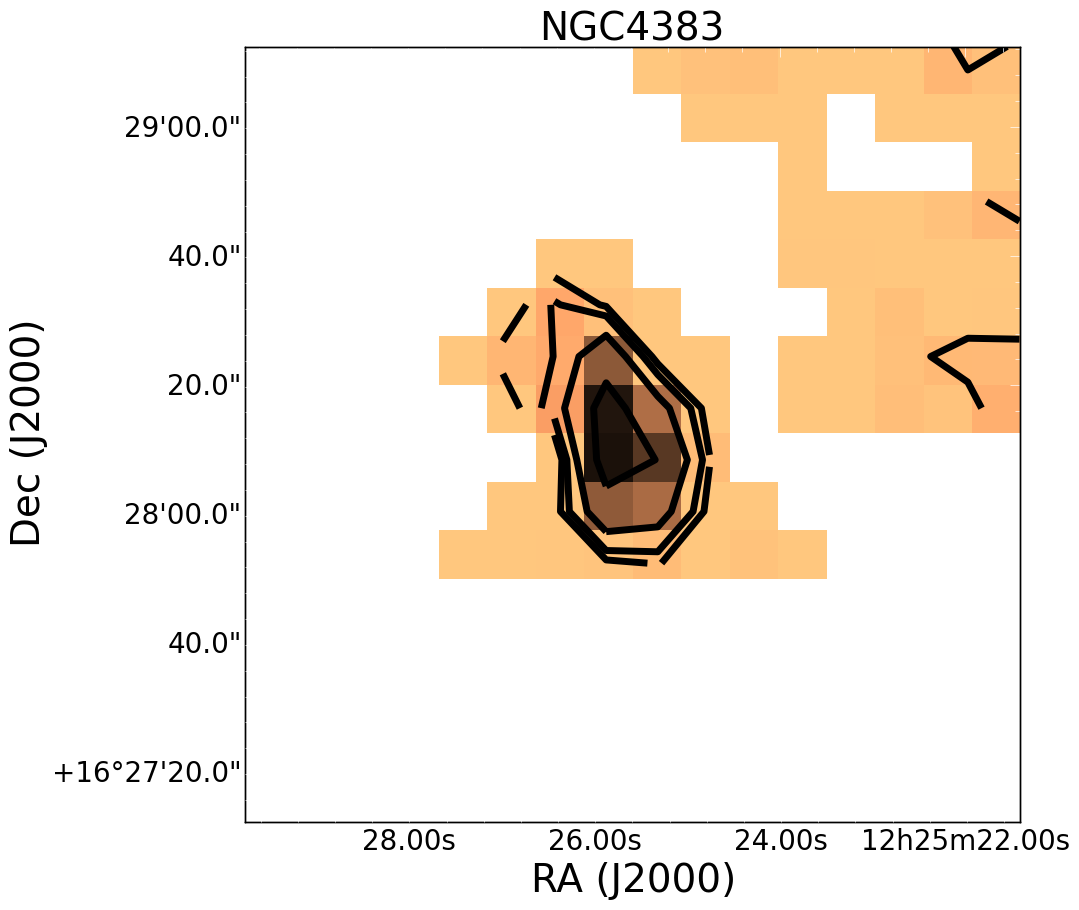}
	\includegraphics[height=3.75cm]{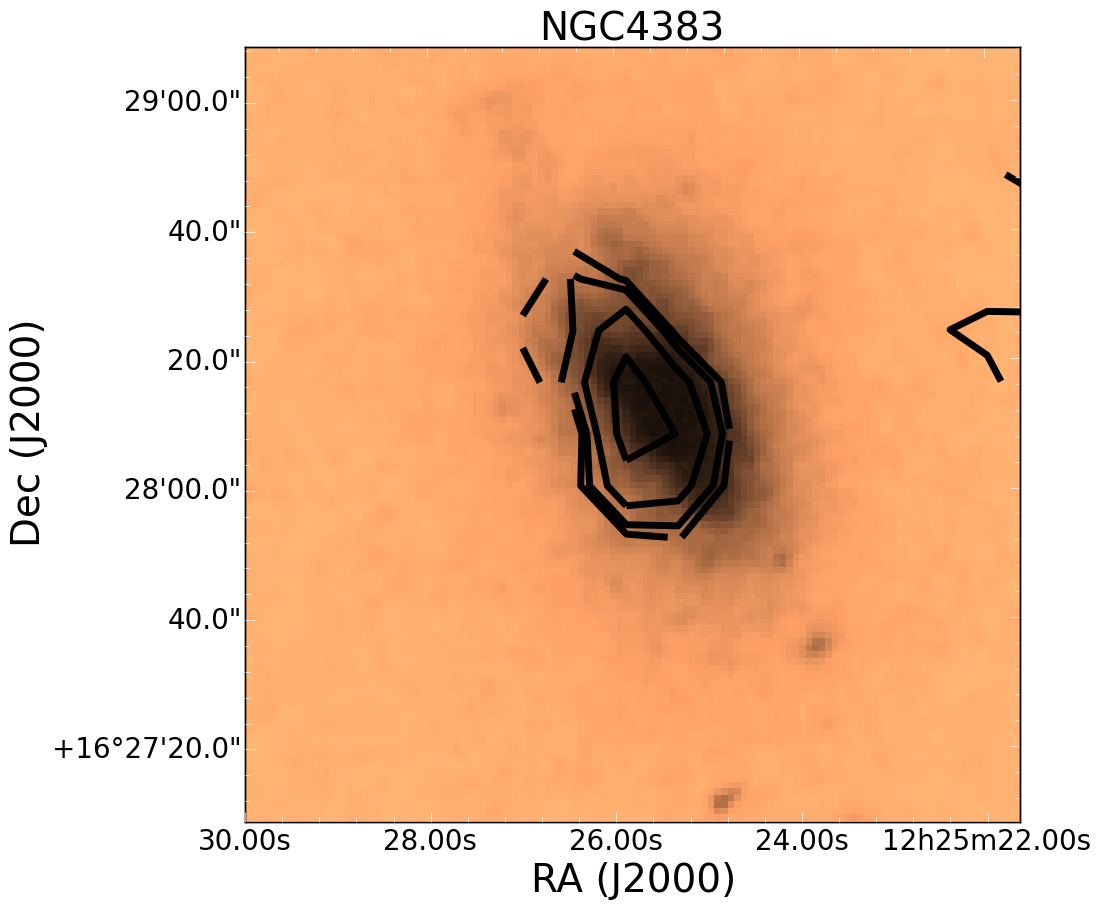}
	\includegraphics[height=3.75cm]{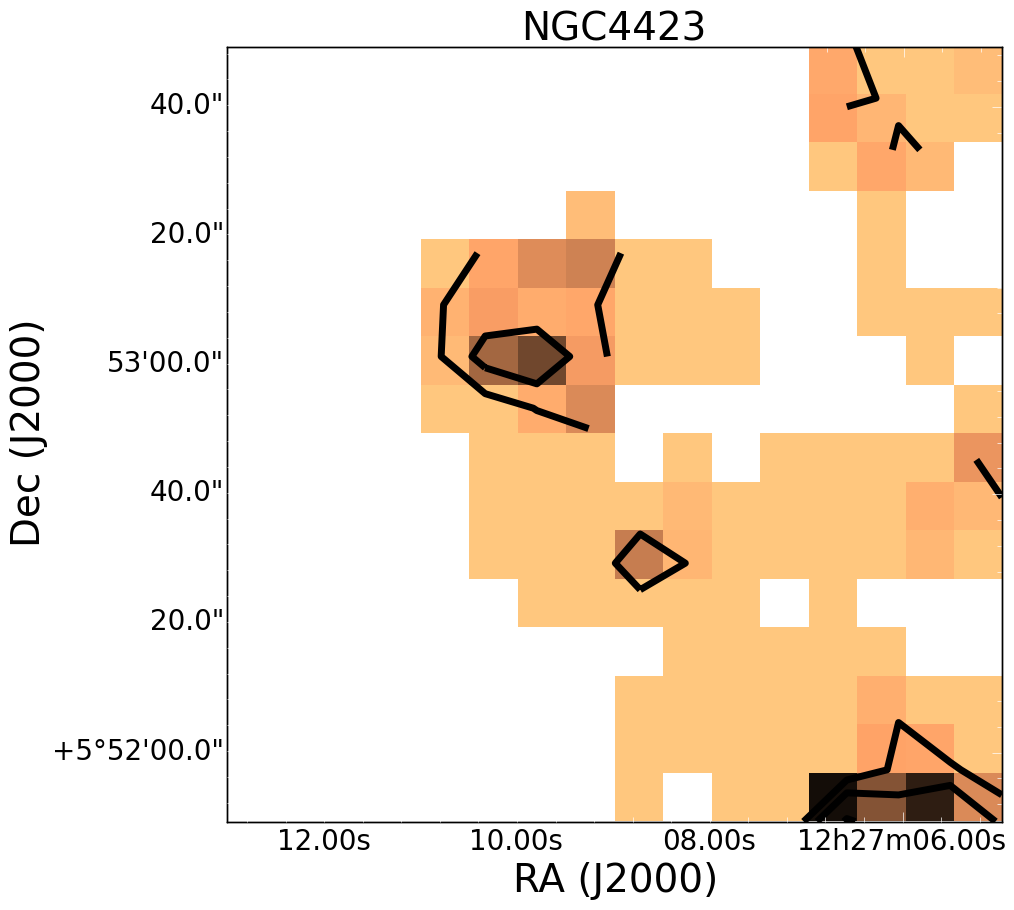}
	\includegraphics[height=3.75cm]{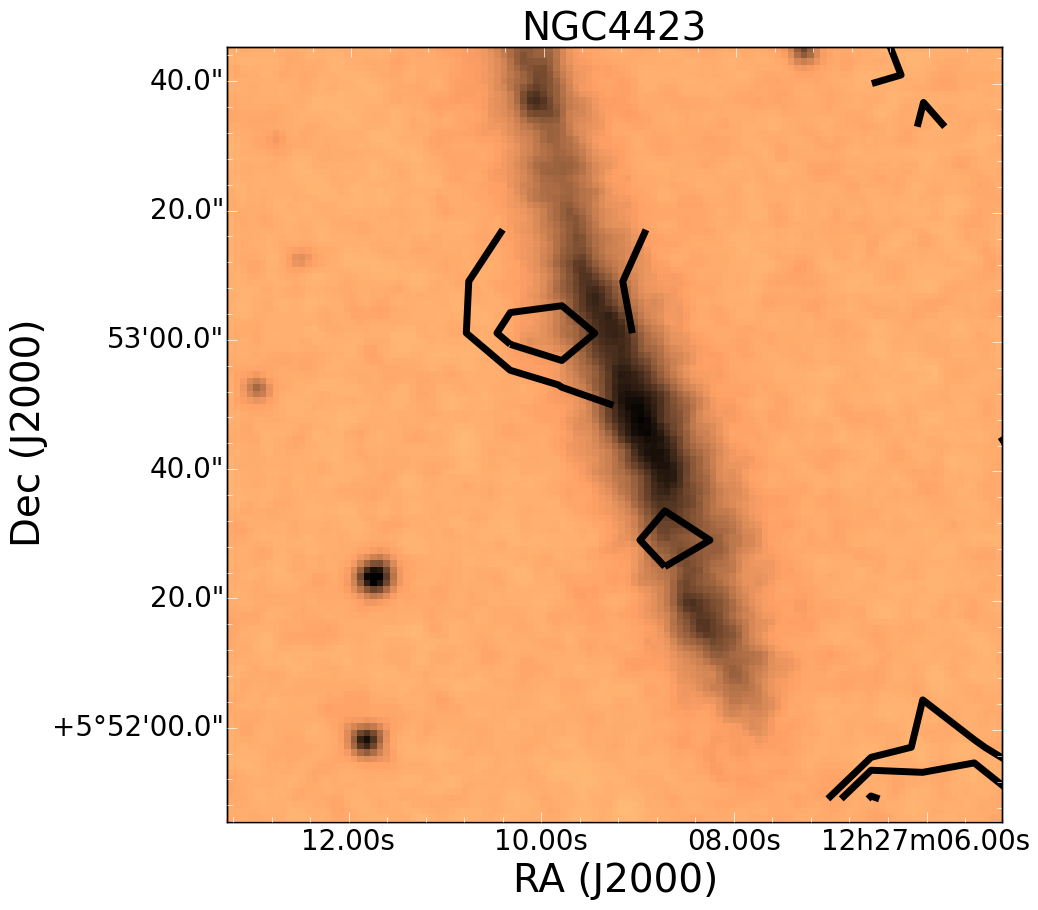}
	\includegraphics[height=3.75cm]{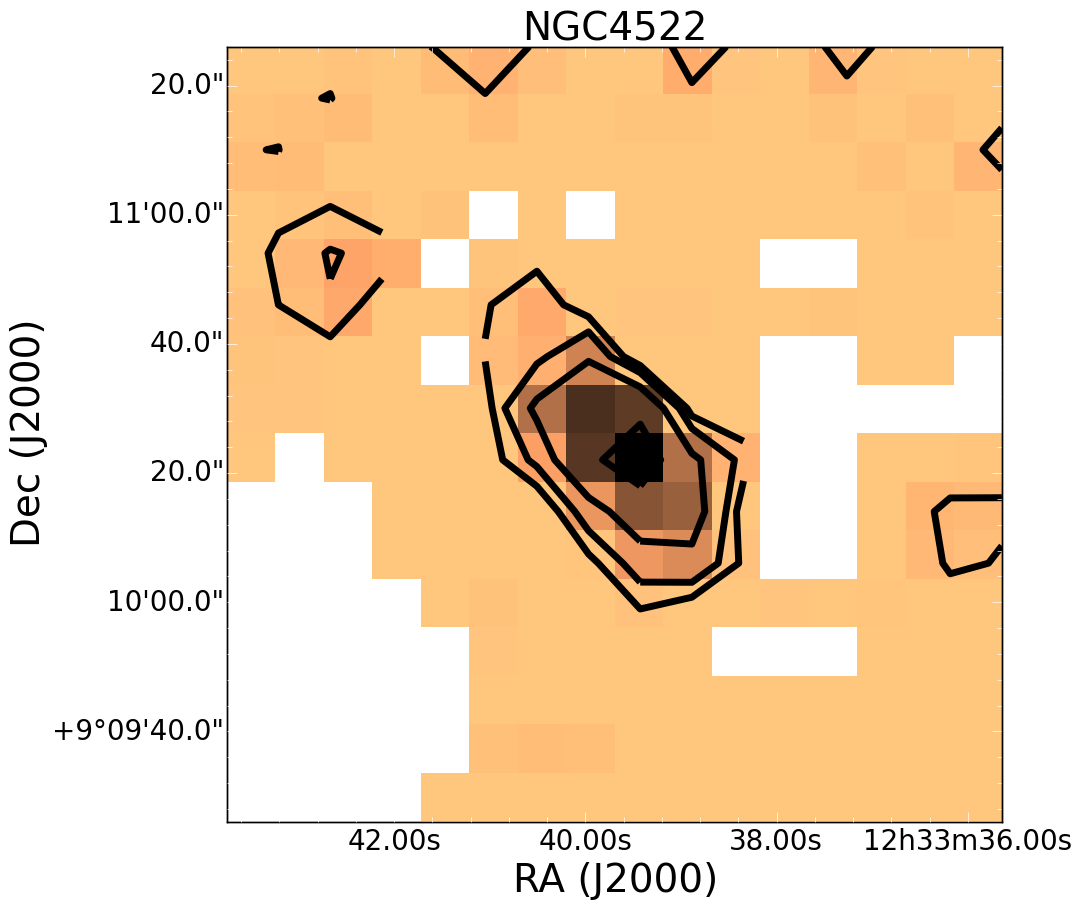}
	\includegraphics[height=3.75cm]{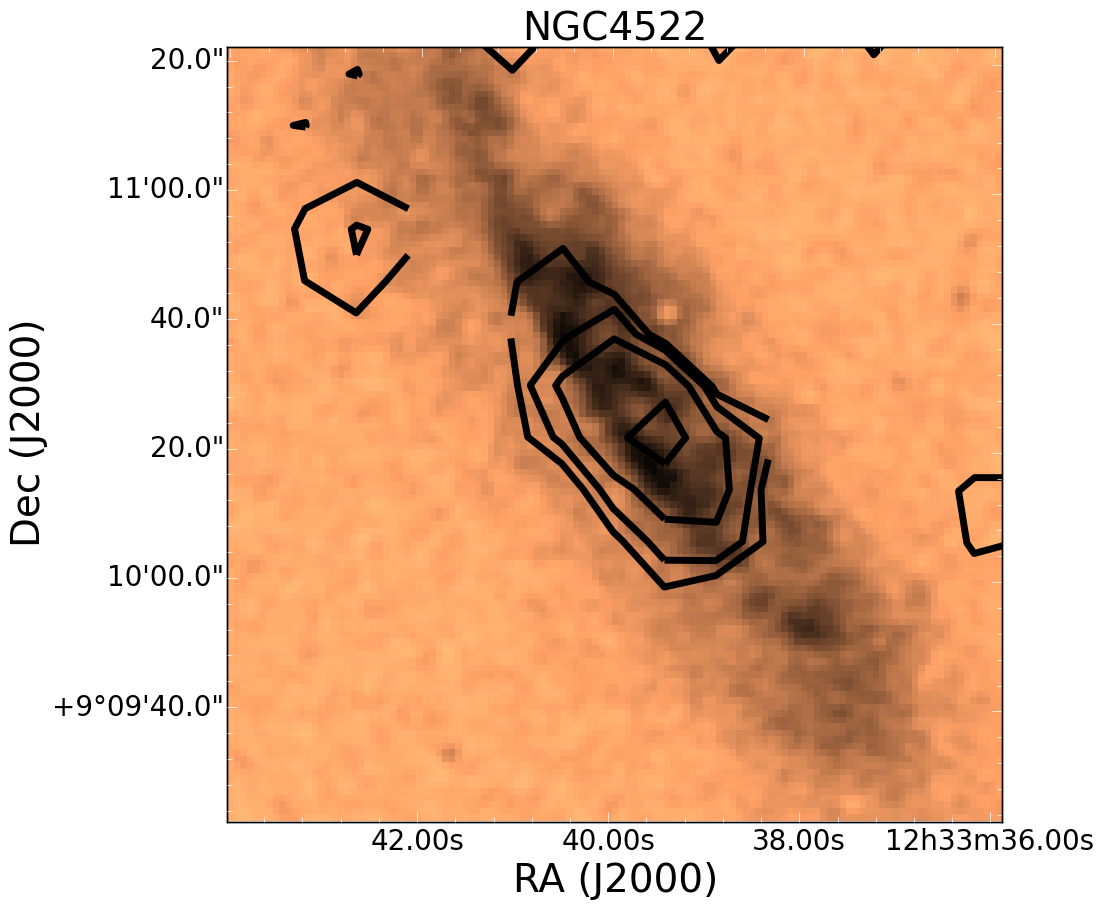}
	\includegraphics[height=3.75cm]{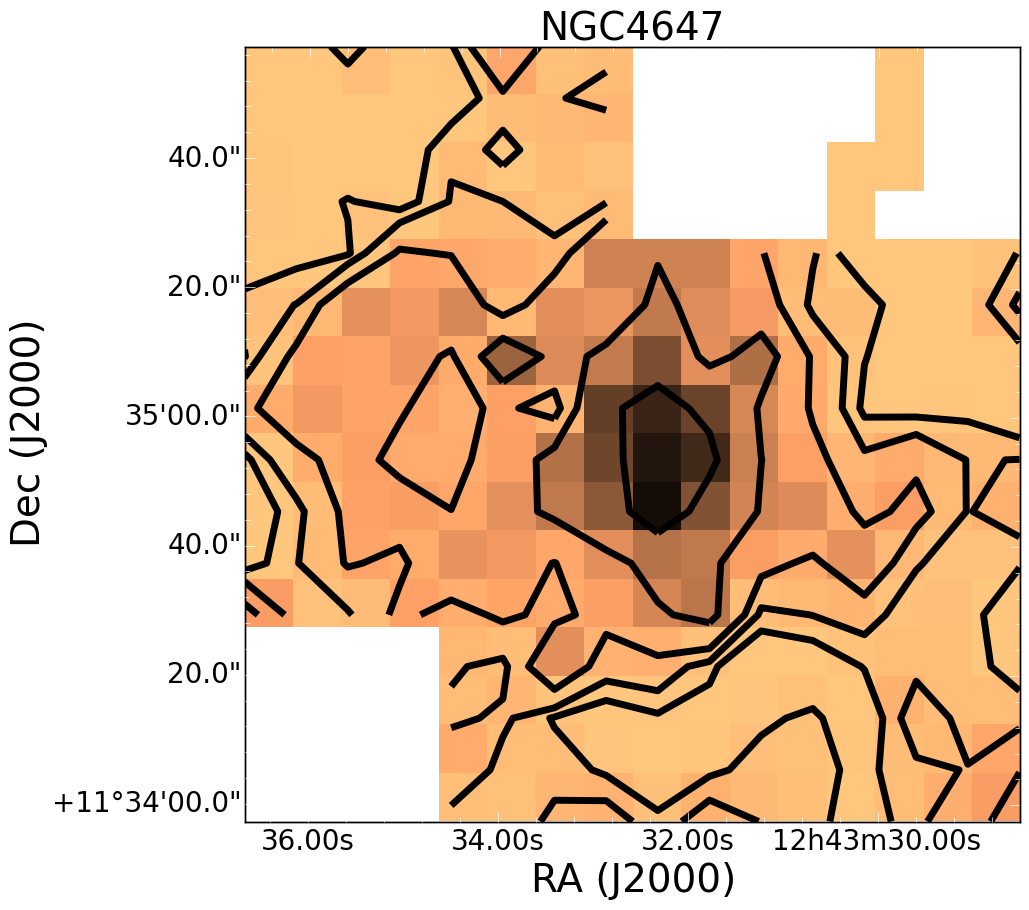}
	\includegraphics[height=3.75cm]{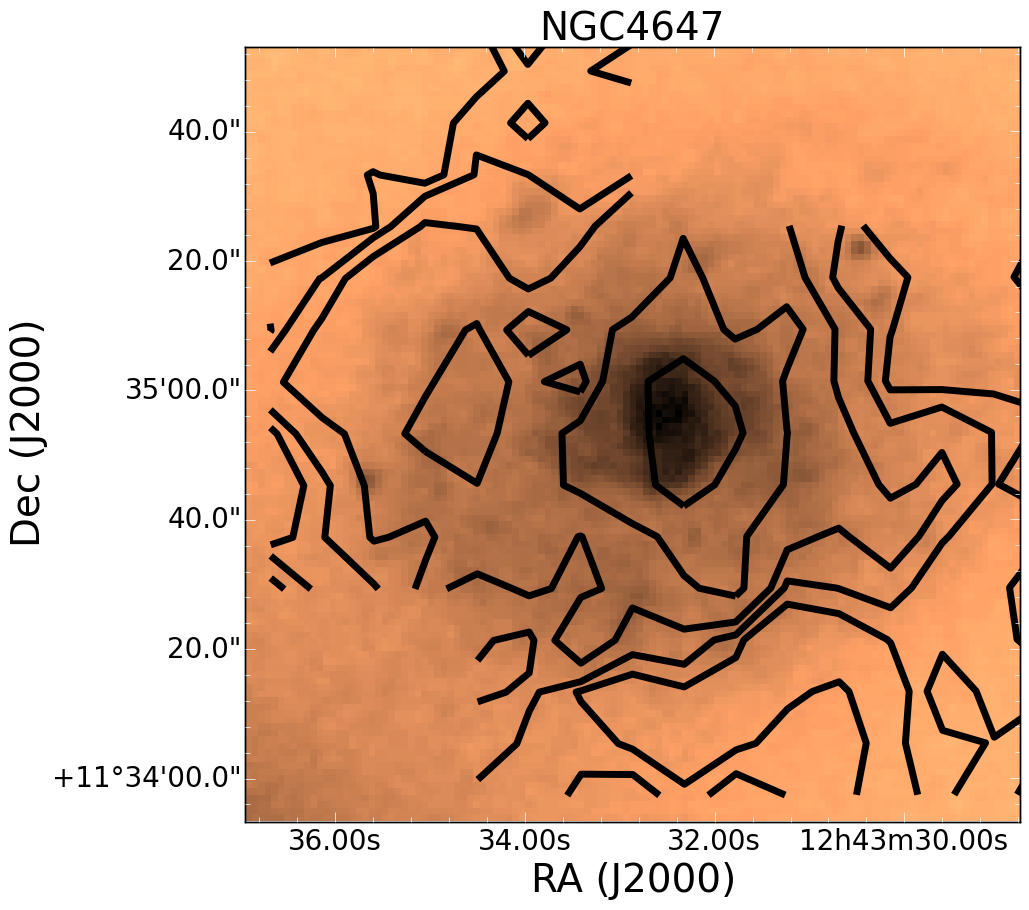}
	\includegraphics[height=3.75cm]{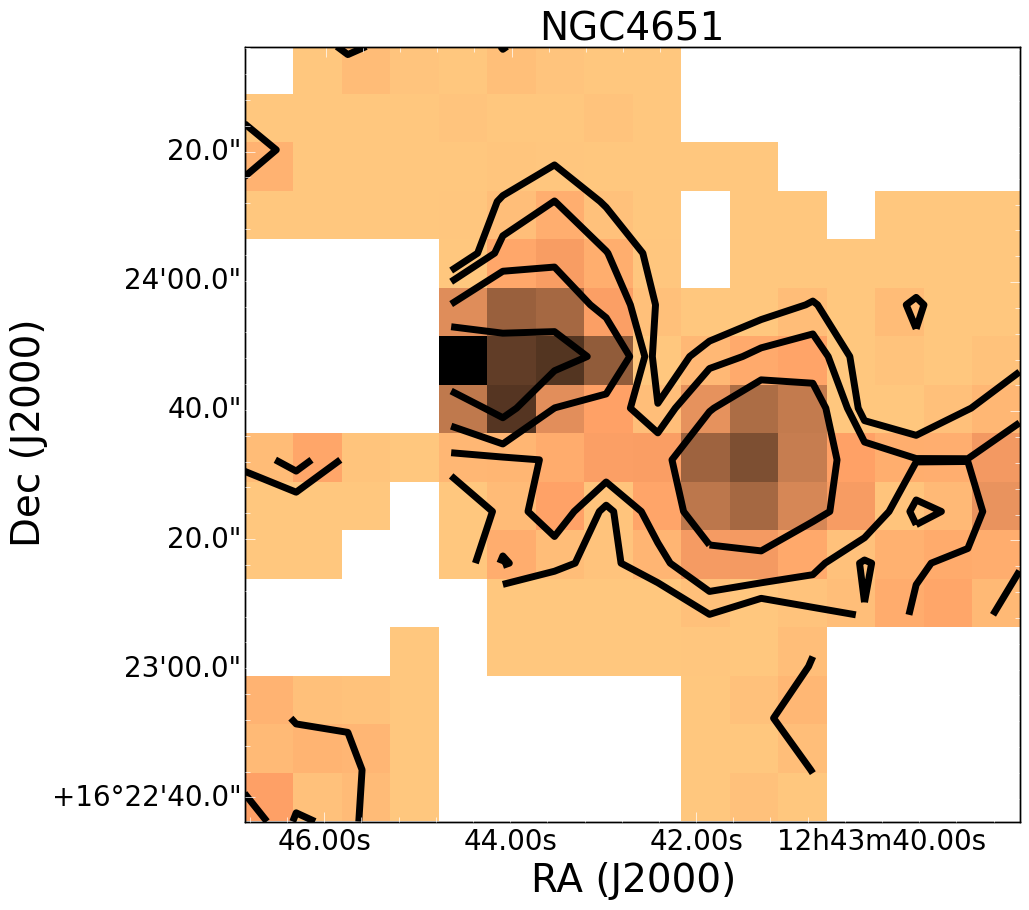}
	\includegraphics[height=3.75cm]{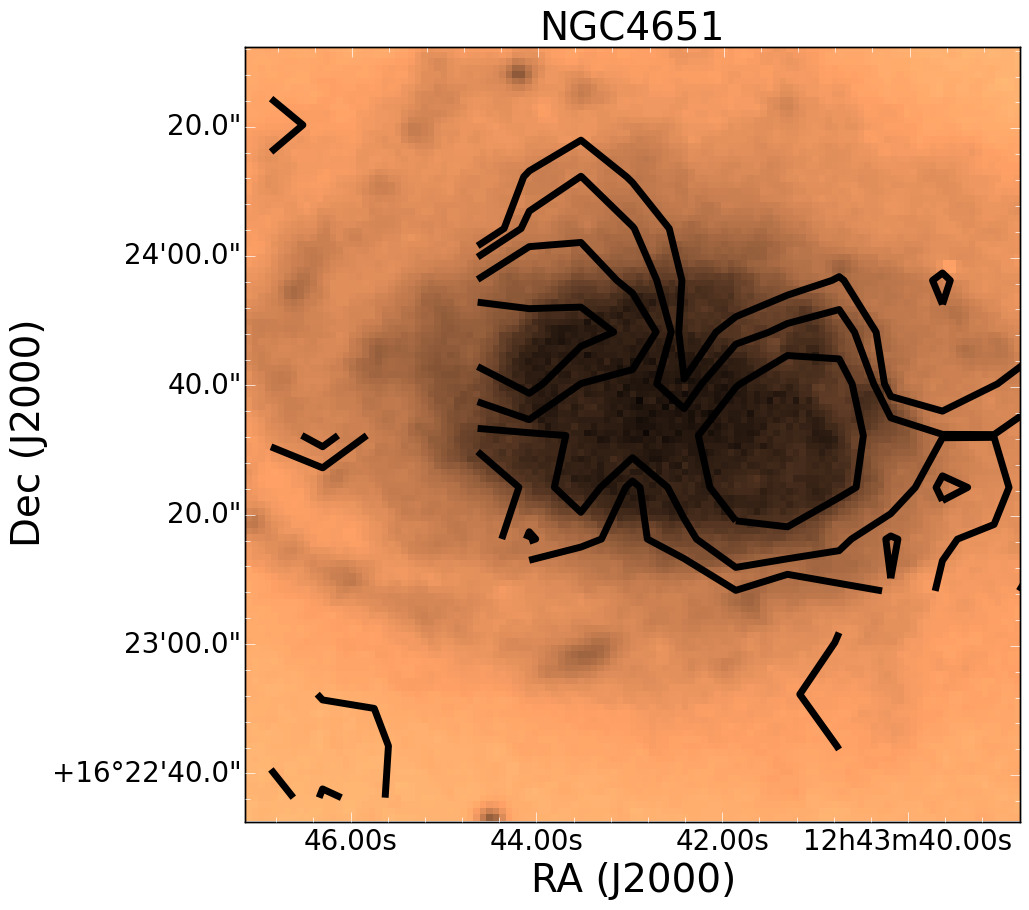}
	\includegraphics[width=4.5cm]{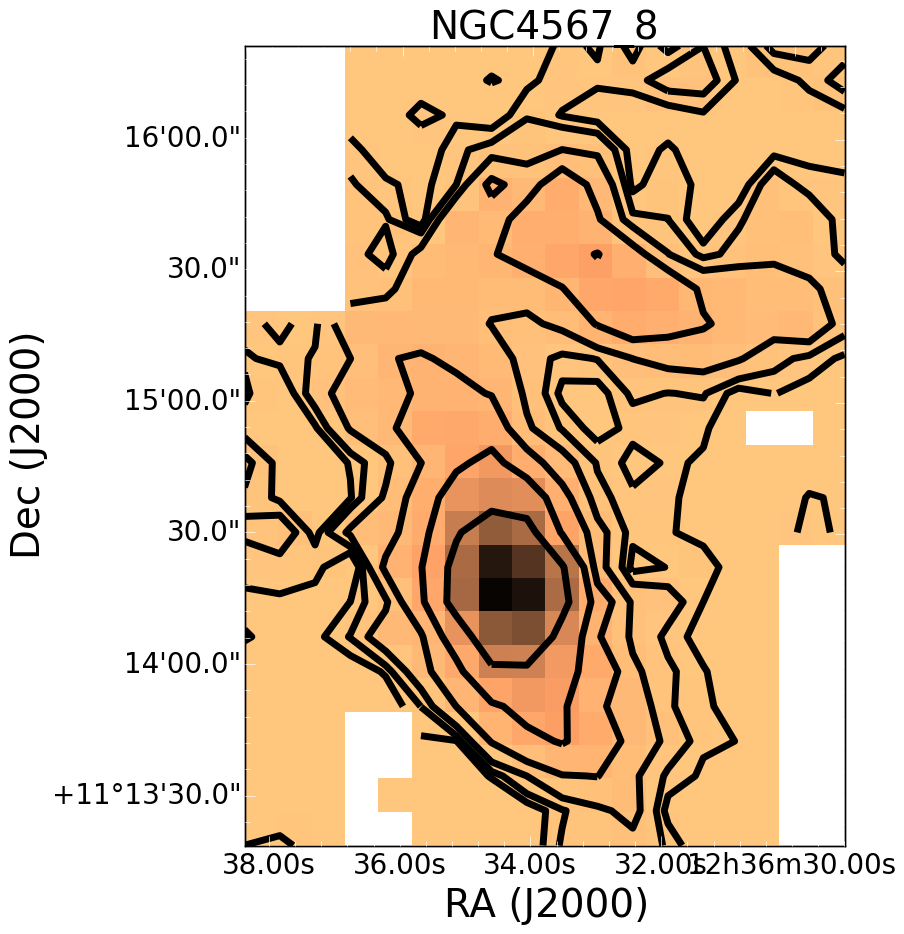}
	\includegraphics[width=4.5cm]{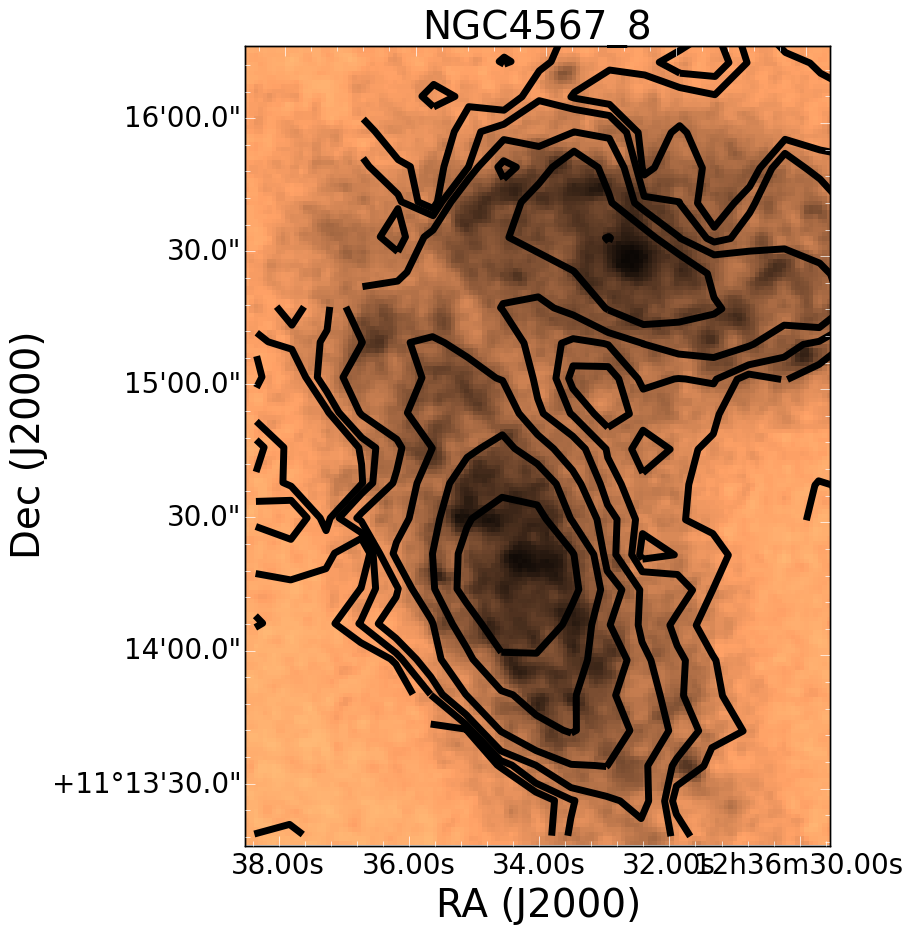}
	\caption{Images of the CO detected galaxies in the Virgo sample. The first panel in each pair is the CO $J=3-2$ integrated galaxy map with black contour levels overlaid at 0.5, 1, 2, 4, 8, and 16 K km s$^{-1}$. The second panel is the same contour levels overlaid on the optical image from the Digitized Sky Survey. Note that NGC 4567 and 4568 were combined into one image due to their close proximity.}
	\label{fig-det_virgo}
\end{figure*}
\begin{figure*}
	\includegraphics[width=6.5cm]{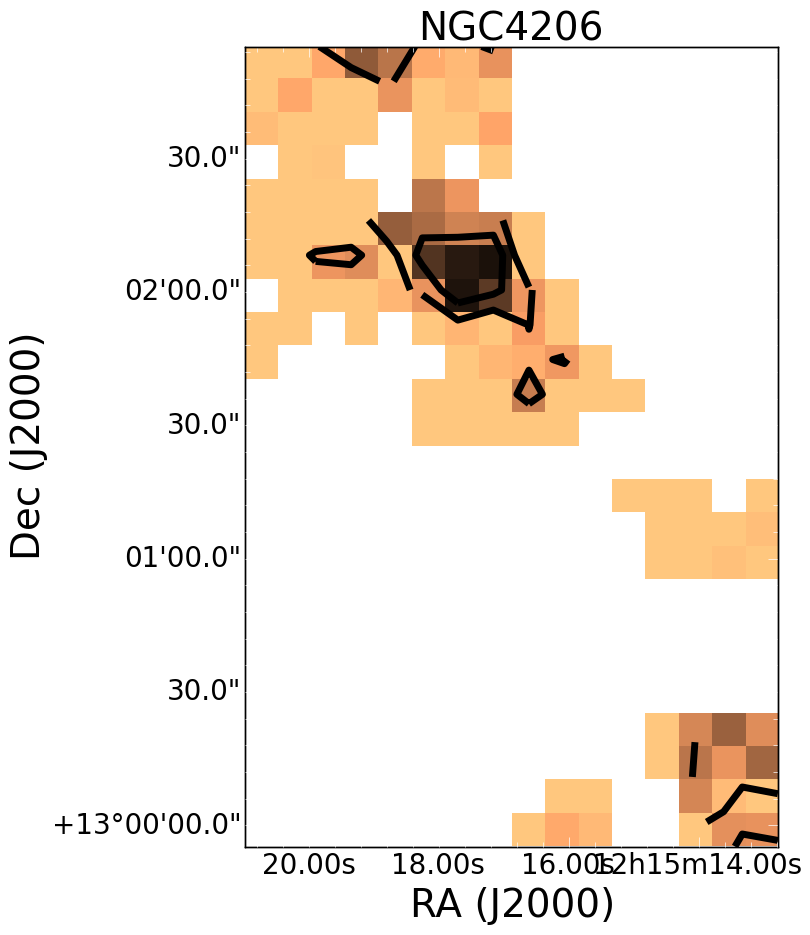}
	\includegraphics[width=6.5cm]{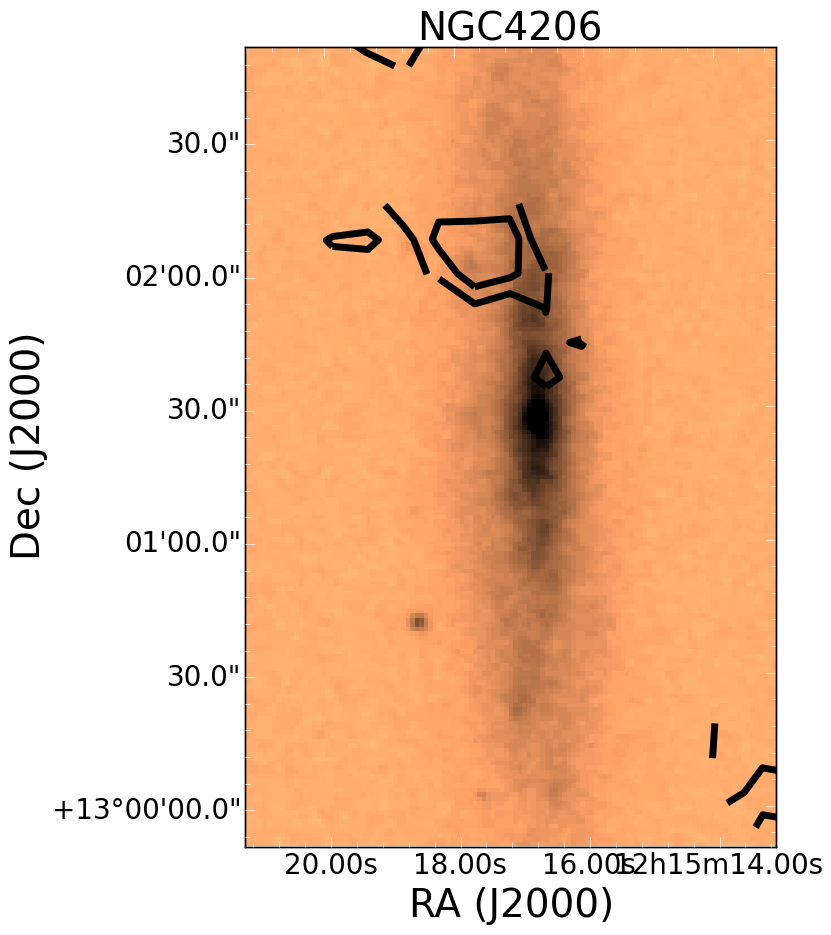}
	\includegraphics[width=6.5cm]{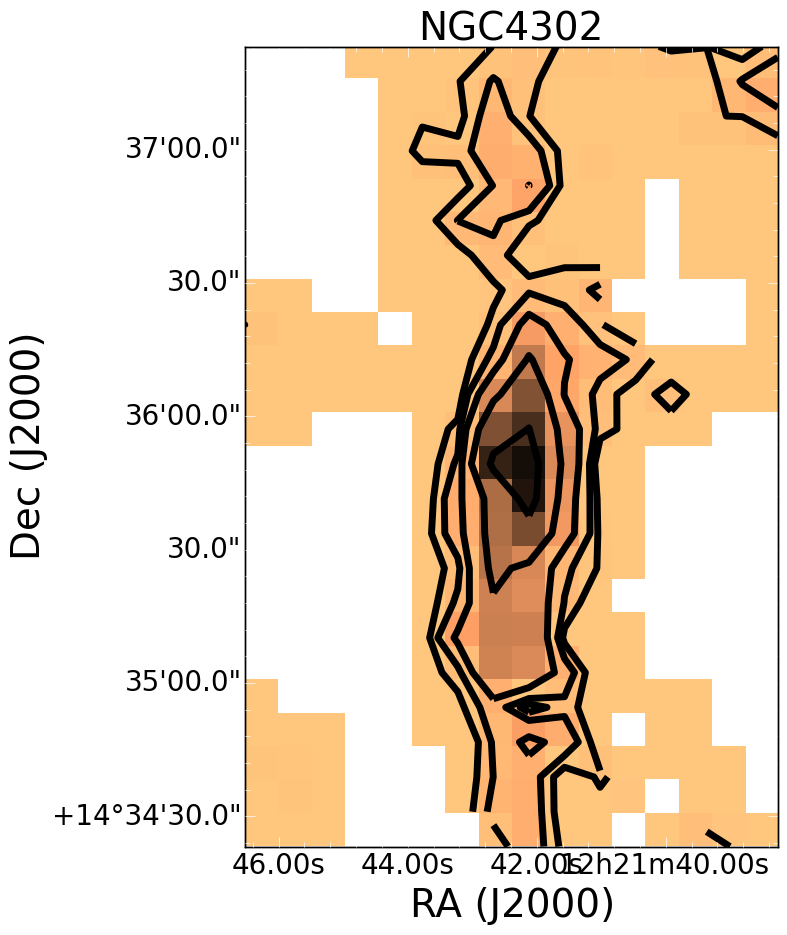}
	\includegraphics[width=6.5cm]{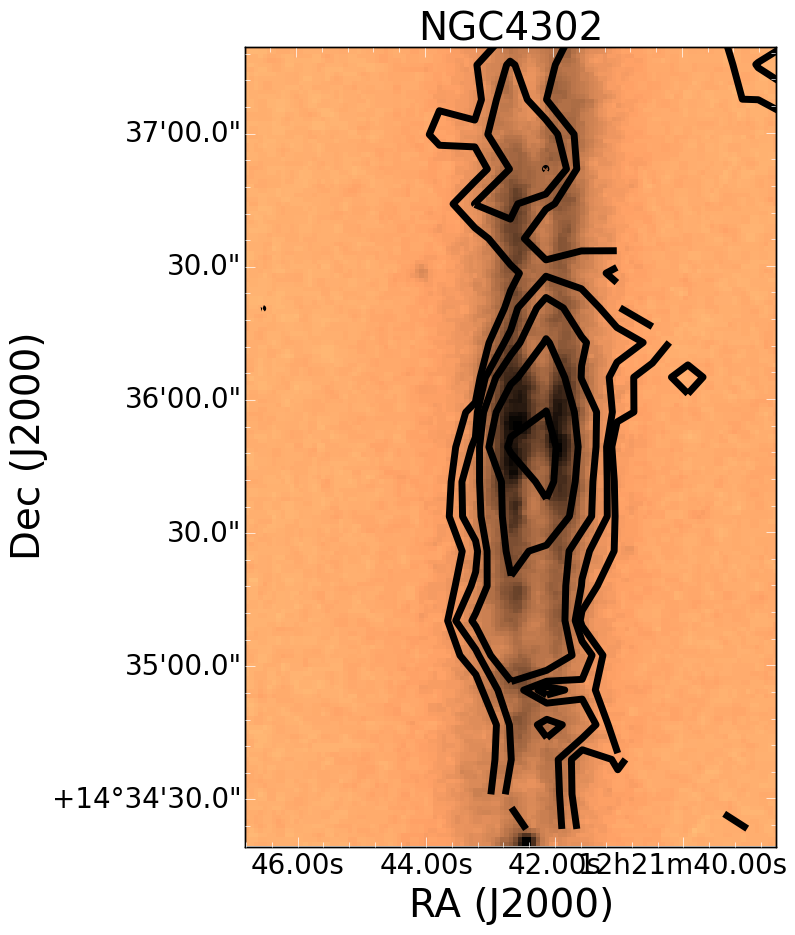}
	\includegraphics[width=7.5cm]{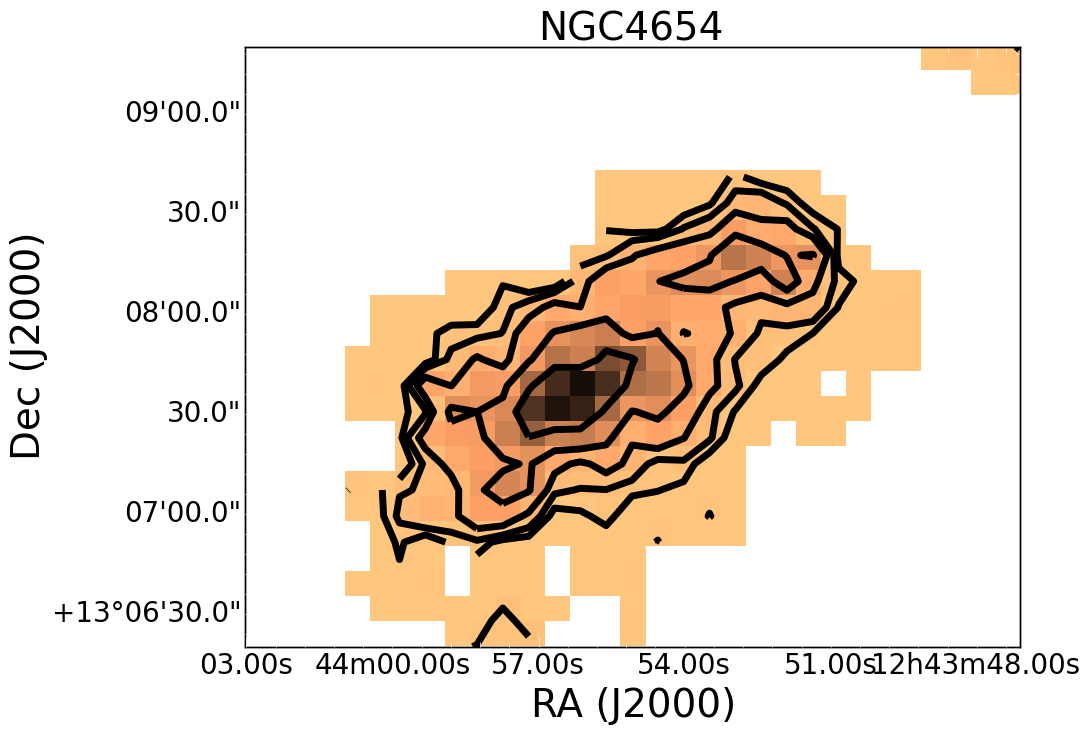}
	\includegraphics[width=7.5cm]{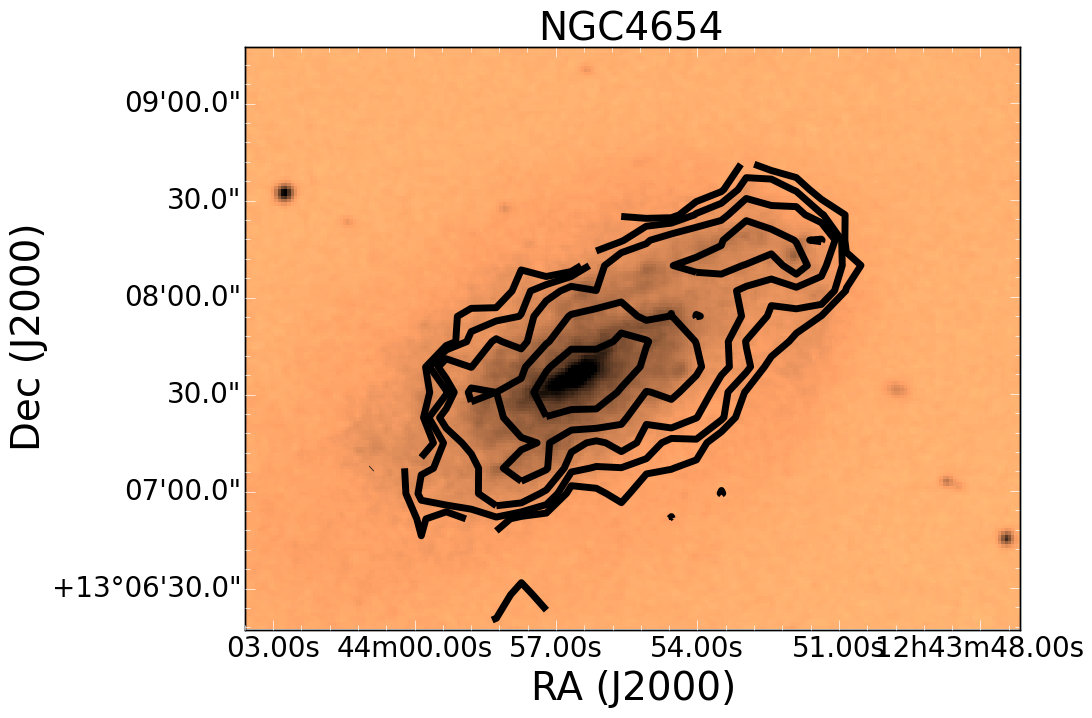}
	\caption{Images of the CO detected galaxies in the Virgo sample, observed with two overlapping fields. The first panel in each pair is the CO $J=3-2$ integrated galaxy map with black contour levels overlaid at 0.5, 1, 2, 4, and 16 K km s$^{-1}$. The second panel is the same contour levels overlaid on the optical image from the Digitized Sky Survey.}
	\label{fig-det_virgo_raster}
\end{figure*}
\begin{figure*}
	\includegraphics[width=9cm]{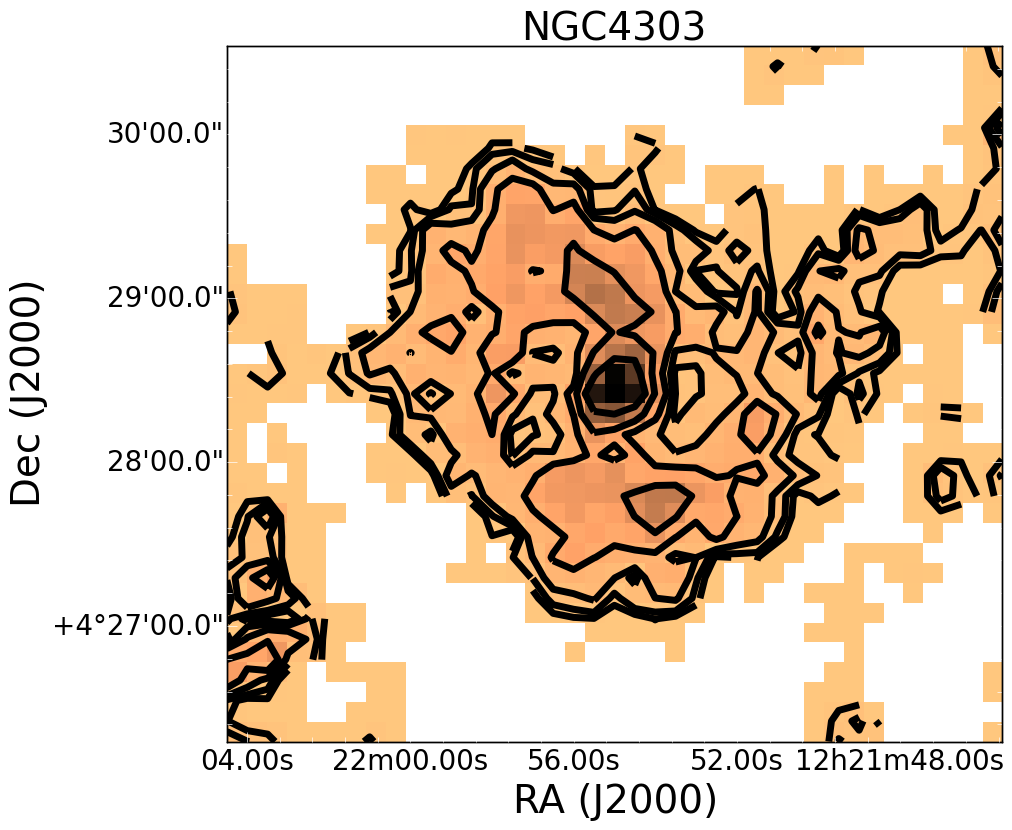}
	\includegraphics[width=9cm]{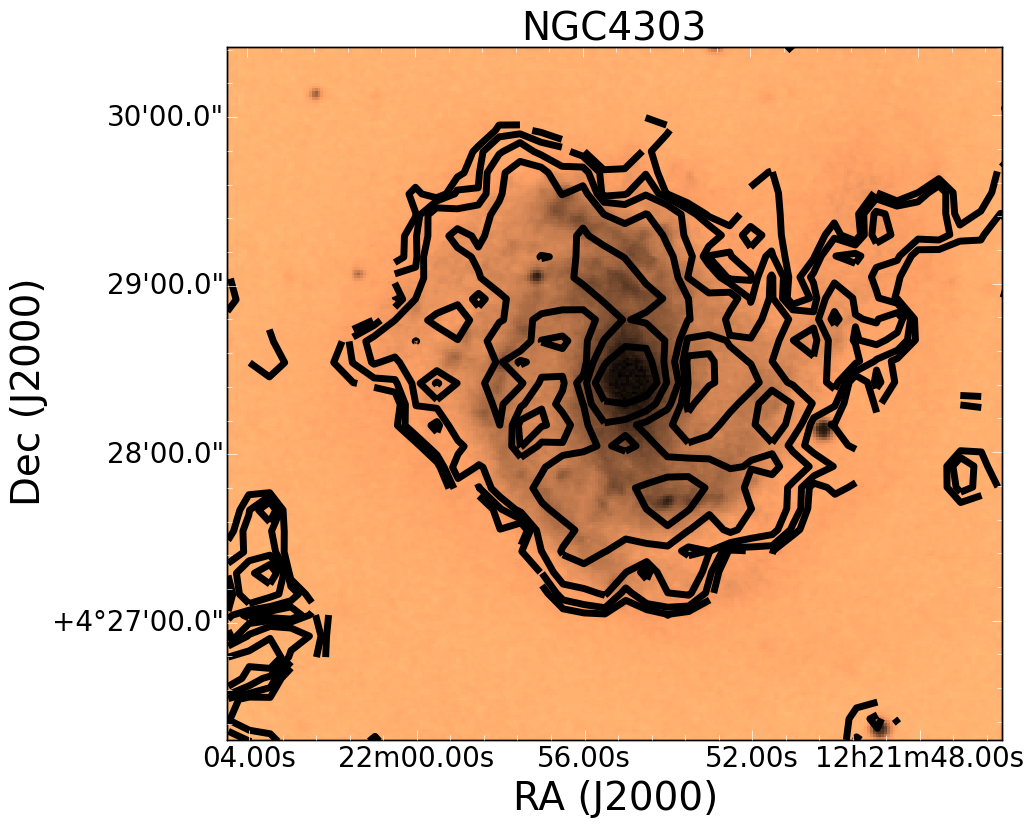}
	\caption{Images of NGC 4303 in the Virgo sample, observed using the raster method. The first panel in each pair is the CO $J=3-2$ integrated galaxy map with black contour levels overlaid at 0.5, 1, 2, 4, 8, and 16 K km s$^{-1}$. The second panel is the same contour levels overlaid on the optical image from the Digitized Sky Survey.}
	\label{fig-det_virgo_double}
\end{figure*}

\end{document}